\documentstyle[12pt,aps,eqsecnum]{revtex}

\newcommand{\sign}{\mathop{\rm sign}\nolimits}
\newcommand{\dive}{\mathop{\rm div}\nolimits}

\newcommand{\Real}{\mathop{\rm Re}\nolimits}
\newcommand{\Reyn}{\mathop{\rm R}\nolimits}

\def\ab{{\mbox{\boldmath $a$}}}
\def\bb{{\mbox{\boldmath $b$}}}
\def\cb{{\mbox{\boldmath $c$}}}
\def\eb{{\mbox{\boldmath $e$}}}
\def\fb{{\mbox{\boldmath $f$}}}

\def\kb{{\mbox{\boldmath $k$}}}
\def\lb{{\mbox{\boldmath $l$}}}
\def\nb{{\mbox{\boldmath $n$}}}
\def\nablab{{\mbox{\boldmath $\nabla$}}}
\def\rb{{\mbox{\boldmath $r$}}}
\def\vb{{\mbox{\boldmath $v$}}}

\def\Ab{{\mbox{\boldmath $A$}}}
\def\Bb{{\mbox{\boldmath $B$}}}

\def\Db{{\mbox{\boldmath $D$}}}
\def\Fb{{\mbox{\boldmath $F$}}}
\def\Ib{{\mbox{\boldmath $I$}}}
\def\Kb{{\mbox{\boldmath $K$}}}
\def\Mb{{\mbox{\boldmath $M$}}}
\def\Nb{{\mbox{\boldmath $N$}}}
\def\Pb{{\mbox{\boldmath $P$}}}
\def\Rb{{\mbox{\boldmath $R$}}}
\def\Sb{{\mbox{\boldmath $S$}}}
\def\Tb{{\mbox{\boldmath $T$}}}
\def\Ub{{\mbox{\boldmath $U$}}}
\def\Vb{{\mbox{\boldmath $V$}}}
\def\Wb{{\mbox{\boldmath $W$}}}
\def\Omb{{\mbox{\boldmath $\Omega$}}}

\def\lambdab{{\mbox{\boldmath $\lambda$}}}
\def\mub{{\mbox{\boldmath $\mu$}}}
\def\xib{{\mbox{\boldmath $\xi$}}}

\def\Fbc{{\mbox{\boldmath $\cal{F}$}}}
\def\Tbc{{\mbox{\boldmath $\cal{T}$}}}

\begin{document}

\title{Retarded Many-Sphere Hydrodynamic Interactions in a Viscous Fluid}
\medskip
\author{P.~P.~J.~M.~Schram}
\address{Eindhoven University of Technology, P.~O. Box 513, MB 5600, Eindhoven,
The Netherlands}
\author{A.~S.~Usenko and I.~P.~Yakimenko}
\address{Bogolyubov Institute for Theoretical Physics, Ukrainian National
 Academy of Sciences, Metrologicheskaya Str. 14b, Kiev 03143, Ukraine}
\date{\today}

\maketitle
\bigskip

%\begin{abstract}

An alternative method is suggested for the description of the velocity and
pressure fields in an unbounded incompressible viscous fluid induced by an
arbitrary number of spheres moving and rotating in it. Within the framework of
this approach, we obtain the general relations for forces and torques exerted
by the fluid on the spheres. The behavior of the translational, rotational, and
coupled friction and mobility tensors in various frequency domains are analyzed
up to the terms of the third order in the dimensionless parameter $b$  equal to
the ratio of a typical radius of a sphere to the penetration depth of
transverse waves and a certain power of the dimensionless parameter $\sigma$
equal to the ratio of a typical radius of a sphere to the distance between the
centers of two spheres. We establish that the retardation effects can
essentially affect the character of the hydrodynamic interactions between the
spheres.

%\end{abstract}

\bigskip

%\pacs{PACS numbers: 47.10+g; 83.10-y} \vfill

%%--------------------------------Section 1------------------------------

\newpage

\section{Introduction}  \label{Introduction}

The theory of hydrodynamic interactions between particles of finite size
immersed in a viscous fluid is of great interest for numerous problems of the
modern physics and technology such as the theory of Brownian motion
\cite{ref.ChowHermans,ref.Chow,ref.Bedeaux,ref.Mazur,ref.Clercx1,ref.SchramYak}
stimulated by the results of numerical calculations on computers (including
"Brownian dynamics" in computational physics \cite{ref.ResLeener}),
sedimentation \cite{ref.Happel,ref.Rutgers}, etc. As a rule, the central
problem of the theory is to determine forces and torques exerted by the fluid
on particles moving and rotating in it with given translational and angular
velocities.  To this end, the linearized Navier--Stokes equation for the fluid
is usually used \cite{ref.Happel,ref.Kim}. Starting from this equation, a great
number of rigorous results has been obtained  for a single sphere in a fluid
\cite{ref.Lamb,ref.Milne,ref.Batch,ref.Landau,ref.Loitsyan,ref.Schram}. For
more than one sphere, there are only a few rigorous results obtained for two
spheres in some particular cases of their stationary motion and rotation
\cite{ref.Stimson,ref.Maude,ref.Wakiya,ref.Davis}. As for the nonstationary
case, any exact results for the fluid velocity and pressure induced by two and
more interacting spheres are absent.

The situation with finding approximate solutions of the linearized
Navier--Stokes equation is more advanced (first of all, at the expense of the
classical method of reflections proposed by Smoluchowski and its further
modifications). The comprehensive review of the results obtained in this way is
given in \cite{ref.Happel} including the cases of mixed slip-stick boundary
conditions at the surfaces of the spheres
\cite{ref.Feld1,ref.Feld2,ref.Schmitz1,ref.Schmitz2,ref.Schmitz3} and permeable
spheres \cite{ref.Jones1,ref.Jones2,ref.Jones3}. It follows that the majority
of these results have been obtained in the stationary case and mainly for two
spheres, and the possibility of an application of the corresponding methods to
nonstationary many-sphere problems is not clear.

In \cite{ref.Mazur,ref.MazurSaarl}, on the basis of the method of induced
forces proposed in \cite{ref.MazurBed}, it was developed a procedure of the
derivation of a hierarchy of equations for irreducible force multipoles induced
in spheres immersed in a fluid.  Using this procedure, the friction and
mobility tensors of a system of an arbitrary number of spheres are obtained in
the stationary case and the results are represented in the form of power series
expansions in the dimensionless parameter  $\sigma$  equal to the ratio of a
typical radius of a sphere to the distance between the centers of two spheres
retaining in the series terms up to a certain order in this parameter.  This
method essentially differs from the methods of reflections and gives one the
possibility to find the friction and mobility tensors without knowledge of the
fluid velocity and pressure in an explicit form. Furthermore, the method is
free from the restriction to the stationarity of the system. Its generalization
to the nonstationary case has been done in \cite{ref.Saarl} where the
expressions for the mobility tensors have been represented in the form of power
series expansions in two dimensionless parameters: $\sigma$, which is typical
of the stationary case, and the parameter $b$, which is the ratio of a typical
radius of a sphere to the penetration depth of transverse waves.  At finite
frequencies, the expressions for the mobility tensors are obtained in
\cite{ref.Saarl} up to the third order in these two dimensionless parameters,
namely, up to terms of the order of $b^n\sigma^m$ with  $n + m = 3$,  where $n,
m = 0, 1, 2, 3$.

There are some other methods
\cite{ref.Freed1,ref.Freed2,ref.Pien1,ref.Pien2,ref.Clercx2,ref.Clercx3,ref.Hofman}
proposed for determination of an approximate solution of problems of
many-sphere hydrodynamic interactions in a fluid in both the stationary and
nonstationary cases.  As compared with the method of induced forces used in
\cite{ref.Mazur,ref.MazurSaarl,ref.Saarl}, these methods enable one to
calculate not only forces and torques exerted by the fluid on spheres but also
the velocity field of the fluid induced by these spheres (since the latter
problem is much more complicated, usually only the hydrodynamic forces and
torques are calculated.)

The original analytic-numerical method has been used in \cite{ref.Clercx2} to
study the quasistatic hydrodynamic interactions in a system of $N$ spherical
particles.  Unlike the other methods, this method is not based on using the
conventional scheme of the representation of the quantities under study as
series expansions in powers of the parameter $\sigma$ but reduces the problem
to the specific solution of an infinite system of linear equations. With this
method, in \cite{ref.Clercx2}, the stationary mobility tensors for a system of
two interacting spheres of equal radii have been obtained and analyzed in
detail. In \cite{ref.Clercx3}, the method has been extended to the description
of the retarded hydrodynamic interaction between two equal spheres in an
unbounded fluid. The nonstationary mobility tensors of this system have been
calculated and plotted versus the variable  $b$ for several values of the
parameter $\sigma$. Note that these numerical results are not restricted to the
range of small values of  $b$ like in \cite{ref.Saarl} where the corresponding
analytic relations for the mobility tensors are deduced.

As for the method developed in \cite{ref.Pien1,ref.Pien2}, it should be noted
that the final results presented in these papers are expressed in terms of
multiindex tensors defined as tensors inverse to certain rather complicated
multiindex tensors but the explicit expressions for these inverse tensors are
not given. Naturally, without knowledge of the explicit form for the inverse
tensors, the expressions given in \cite{ref.Pien1,ref.Pien2} cannot be reduced
to the known results obtained by other methods. Moreover, it can be shown
\cite{ref.Usenko} that one of the inverse tensors defining in
\cite{ref.Pien1,ref.Pien2} the relations for the forces and torques exerted by
the fluid on the spheres and the velocity field of the fluid induced by the
spheres does not exist (even for noninteracting spheres) because the
corresponding direct tensor is singular.  In \cite{ref.Usenko}, the reasons for
this fact have been analyzed in detail, and the procedure has been proposed for
finding the time-dependent distributions of the velocity and pressure fields in
an unbounded incompressible viscous fluid in an arbitrary external force field
induced by any number of spheres immersed in the fluid as well the forces and
torques exerted by the fluid on the spheres.  In the stationary case, this
program has been performed for the derivation of the general relations for the
velocity and pressure fields of the fluid, the friction and mobility tensors
for any number of spheres in a fluid in given external force fields and these
results are represented in the form of power series expansions in the
dimensionless parameter  $\sigma$ retaining in the series terms of the second
iteration inclusive.

However, it is well known \cite{ref.Landau,ref.Schram} that even in the case of
a single sphere an account of time dependence can lead to significant changes
both in the fluid velocity induced by the sphere moving in the fluid with
time-dependent velocity and in the force exerted by the fluid on the sphere,
namely, the asymptotics of the fluid velocity, being of the order of  $1/r$,
where  $r$  is the distance from the center of the sphere to the point of
observation, for the stationary motion of the sphere, is replaced by  $1/r^3$
and the time-dependent force exerted by the fluid on the sphere, in addition to
the ordinary Stokes friction term, contains a term depending on the
acceleration of the sphere and a memory term associated with hydrodynamic
retardation effects. In the nonstationary case, hydrodynamic interactions
between spheres also change, which is especially significant for large
distances between the spheres.  For example, certain mobility tensors, which
are proportional to the inverse distance between spheres to certain powers in
the stationary case, exponentially vanish with increase in the distance between
the spheres in the nonstationary case \cite{ref.Saarl}.  The account of the
main frequency-dependent terms in the mobilities in the case of small
frequencies leads to long-time tails for the velocity correlation functions of
two different spheres \cite{ref.Saarl}.

Naturally, it is of interest to investigate the nonstationary velocity and
pressure fields of the fluid  induced by several spheres immersed in the fluid.
However, this problem cannot be solved with the use of the methods exploited in
the pioneering papers \cite{ref.Mazur,ref.MazurSaarl,ref.Saarl}. It is also of
interest to extend the analytic results for the nonstationary mobilities
obtained in \cite{ref.Saarl} retaining higher-order terms as well as to derive
frequency-dependent friction tensors for a system of spheres.  These problems
are considered in the present paper.

In Sec.~2, we give the general relations for the fluid velocity and pressure
induced by a system of arbitrary number of spheres and the total forces and
torques exerted by the fluid and external force fields on the spheres.  The
results are obtained in the explicit form without imposing any additional
restrictions on the size of spheres, distances between them, and frequency
range.  These relations are expressed in terms of the unknown induced surface
forces distributed over the surfaces of the spheres.  For the determination of
the unknown harmonics of the induced surface force densities, a closed infinite
system of linear algebraic equations is given.  On the basis of analysis of the
quantities contained in this system, we propose to seek its solution by the
method of successive approximations using the system of noninteracting spheres
as a zero approximation and formulate a rule for determination of the fluid
velocity and pressure fields as well as the forces and torques exerted by the
fluid on spheres up to terms of  $b^3$ and a given power of the parameter
$\sigma$.

In Sec.~3, in the approximation of noninteracting spheres, we obtain the fluid
velocity and pressure induced by a system of spheres and analyze their behavior
in the far zone.  The relations for harmonics of induced surface force
densities obtained for the first and second iterations are given in Secs.~4 and
5, respectively.  We derive the general relations for the forces and torques
exerted by the fluid on the spheres up to the terms of the order of  $b^3$  and
certain powers of the parameter  $\sigma$  corresponding to the second
iteration. We analyze the friction tensors in the particular cases: for
sufficiently large distances between the spheres and in a low-frequency range.
We show that the account of retardation effects essentially changes the
character of hydrodynamic interactions between the spheres.

In Sec.~6, the general relations for the translational, rotational, and coupled
mobilities are derived up to the terms of the order  $b^3$  and  $\sigma^2$,
$\sigma^4$, and  $\sigma^3$, respectively.  We estimate contributions of
three-sphere interactions to the mobility tensors and show the possibility of
coupling between translational and rotational motions of the same sphere.  We
investigate as well the influence of nonstationarity on the behavior of the
mobilities in the low-frequency range and in the case of large distances
between the spheres.

%%--------------------------------Section 2------------------------------

%\newpage

\section{General Relations}  \label{General}

We consider  $N$  homogeneous macroscopic spheres of radii $a_\alpha$ and
masses  $m_\alpha$,  where  $\alpha = 1, 2,\ldots,N$, immersed in an unbounded
incompressible viscous fluid at rest at infinity in an external force field and
determine the fluid velocity and pressure induced by these spheres as well as
the forces and torques exerted by the fluid on the spheres.

In the linear approximation, the fluid is described by the linearized
Navier--Stokes equation

\begin{equation}
 \rho \frac{\partial \vb(\rb,t)}{\partial t} + \dive \Pb(\rb,t) =
  \Fb^{ext}(\rb,t)
\end{equation}

\noindent  and the continuity equation

\begin{equation}
 \dive \vb(\rb,t) = 0.
\end{equation}

\noindent  Here,  $\Pb(\rb,t)$  is the stress tensor of the fluid with the
components

\begin{equation}
 P_{ij}(\rb,t) = \delta_{ij} \, p(\rb,t) - \eta\left(\frac{\partial
  v_i(\rb,t)}{\partial r_j}
  + \frac{\partial v_j(\rb,t)}{\partial r_i}\right),  \qquad  i,j = x,y,z,
\end{equation}

\noindent   $p(\rb,t)$  and  $\vb(\rb,t)$  are the pressure and velocity fields
of the fluid,  $\rho$  and  $\eta$  are its density and viscosity,
respectively,  $\delta_{ij}$  is the Kronecker symbol, and $\Fb^{ext}(\rb,t)$
is the external force acting on a unit volume of the fluid. In the present
paper, for simplicity, we consider the case of a conservative external force
field, i.e.,

\begin{equation}
 \Fb^{ext}(\rb,t) = -\rho \nablab \varphi(\rb,t),
\end{equation}

\noindent  where  $\varphi(\rb,t)$  is the potential of this force field. (The
corresponding results for the general case where the external force field has
both potential and solenoidal components are given in \cite{ref.Usenko})

The position of the center of sphere  $\alpha$  at the time  $t$  is defined by
the radius vector $\Rb_{\alpha}(t)$  relative to the fixed Cartesian coordinate
system with origin at the point  $O$  (in what follows, the system $O$). In
parallel with this fixed system, we also introduce  $N$ local moving Cartesian
coordinate systems with origins  $O_\alpha$  at the centers of spheres
coinciding with their centers of mass (in what follows, systems $O_\alpha$) so
that sphere $\alpha$  does not move relative to system $O_\alpha$.  Therefore,
the radius vector  $\rb$  of any point of the space can be represented in the
form  $\rb = \Rb_\alpha + \rb_\alpha$,  where $\rb_\alpha$  is the radius
vector of this point relative to the local system $O_\alpha$.  The motion of
the spheres is described by the equations

\begin{eqnarray}
 m_\alpha \frac{d\Ub_\alpha(t)}{dt}  &=& \Fb_\alpha^{tot}(t),  \\
 I_\alpha \frac{d\Omb_\alpha(t)}{dt} &=& \Tb_\alpha^{tot}(t),
\end{eqnarray}

\noindent  where

\begin{eqnarray}
 \Ub_\alpha(t) &=& \frac{d\Rb_\alpha(t)}{dt}
\end{eqnarray}

\noindent  and  $\Omb_\alpha (t)$  are, respectively, the translational and
angular velocities of sphere  $\alpha$,  $I_\alpha = (2/5) m_\alpha a_\alpha^2$
is its moment of inertia,  $\Fb_\alpha^{tot}(t)$  is the total force and
$\Tb_\alpha^{tot}(t)$  is the total torque (here and below, all torques
corresponding to sphere  $\alpha$  are considered relative to its center)
acting on sphere  $\alpha$

\begin{eqnarray}
 \Fb_\alpha^{tot}(t) &=& \Fb_\alpha^{ext}(t) + \Fb_\alpha^{f}(t),\\
 \Tb_\alpha^{tot}(t) &=& \Tb_\alpha^{ext}(t) + \Tb_\alpha^{f}(t),
\end{eqnarray}

\noindent  where  $\Fb_\alpha^{ext}(t)$  and  $\Tb_\alpha^{ext}(t)$  are the
force and torque acting on sphere  $\alpha$  due to the external force field
and $\Fb_\alpha^{f}(t)$  and  $\Tb_\alpha^{f}(t)$  are the force and torque
exerted by the fluid on sphere  $\alpha$

\begin{eqnarray}
 \Fb_\alpha^{f}(t) &=& -\int\limits_{S_\alpha(t)} \! \Pb(\rb,t)\cdot \nb_\alpha
  \, dS_\alpha,  \\
 \Tb_\alpha^{f}(t) &=& -\int\limits_{S_\alpha(t)} \! \Bigl(\bigl( \rb - \Rb_\alpha(t)
  \bigr) \times \Pb(\rb,t) \Bigr) \cdot \nb_\alpha \, dS_\alpha,
\end{eqnarray}

\noindent  where  $S_\alpha(t)$  is the surface of sphere  $\alpha$  at the
time $t$ and  $\nb_\alpha$  is the outward unit vector to this surface.

Equations (2.1)--(2.3) describing the motion of the fluid are defined in the
domain  $r_\alpha \geq a_\alpha$, \  $\alpha = 1, 2\ldots,N$,  where $r_\alpha
\equiv |\rb_\alpha| = |\rb - \Rb_\alpha (t)|$.  Following \cite{ref.Usenko}, we
extend the external force field $\Fb^{ext}(\rb,t)$  given in Eq.~(2.1) for
$r_\alpha \geq a_\alpha$  to the domains $r_\alpha < a_\alpha$ so that the
extended quantity is described by the same analytic expression in the entire
space as in the domain $r_\alpha \geq a_\alpha$ (the analytic extension) and
introduce new functions $\tilde{\vb}(\rb,t)$  and $\tilde{p}(\rb,t)$ that
coincide with the fluid velocity and pressure in the domain  $r_\alpha \geq
a_\alpha$ being defined in the entire space as follows:

\begin{eqnarray}
 \tilde{\vb}(\rb,t) &=& \sum\limits_{\alpha = 1}^N \Biggl \{
  \Theta(r_\alpha - a_\alpha) \vb(\rb,t) + \biggl [1
  - \Theta(r_\alpha - a_\alpha) \biggr ] \Ub_\alpha(\rb,t)\Biggr \}, \\
 \tilde{p}(\rb,t) &=& \sum\limits_{\alpha = 1}^N \Biggl \{
  \Theta(r_\alpha - a_\alpha) p(\rb,t) + \biggl [1
  - \Theta(r_\alpha - a_\alpha) \biggr ]\biggl \{ p^{(0)}(\rb,t)
  - \rho\, \biggl(\rb \cdot \frac{d \Ub_\alpha(t)}{d t}\biggr) \biggr \}
  \Biggr\},
\end{eqnarray}

\noindent  where

\begin{equation}
 p^{(0)}(\rb,t) = -\rho\varphi(\rb,t) + p^{inf}
\end{equation}

\noindent is the pressure of the unbounded fluid in the external force field
$\Fb^{ext}(\rb,t)$  in the absence of spheres  and the constant  $p^{inf}$  is
the fluid pressure at infinity where the external force field is absent,
$\Theta(x)$  is the Heaviside function, and

\begin{equation}
 \Ub_\alpha(\rb,t) = \Ub_\alpha(t) + \Bigl( \Omb_\alpha(t)\times\rb_\alpha \Bigr).
\end{equation}

As a result, the quantities  $\tilde{\vb}(\rb,t)$  and  $\tilde{p}(\rb,t)$
satisfy the continuity equation (2.2) and the equation

\begin{equation}
 \rho \frac{\partial \tilde{\vb}(\rb,t)}{\partial t} + \dive \tilde{\Pb}(\rb,t)
  = \Fb^{ext}(\rb,t) + \Fb^{ind}(\rb,t)
\end{equation}

\noindent  given in the entire space, where the quantity
$\tilde{P}_{ij}(\rb,t)$  coinciding with the stress tensor for  $r_\alpha \geq
a_\alpha$  is given in the entire space by relation (2.3) with
$\tilde{\vb}(\rb,t)$  and  $\tilde{p}(\rb,t)$  instead of $\vb(\rb,t)$ and
$p(\rb,t)$,  respectively; the quantity  $\Fb^{ext}(\rb,t)$  is given in the
entire space by the analytic expression for the external force field, and the
induced force density  $\Fb^{ind}(\rb,t)$  has the form

\begin{equation}
 \Fb^{ind}(\rb,t) = \sum\limits_{\alpha = 1}^{N} \Fb_\alpha^{ind}(\rb,t),
\end{equation}

\noindent  where  $\Fb_\alpha^{ind}(\rb,t)$  is the induced force density for
sphere $\alpha$  containing the volume,  $\Fb_\alpha^{(V)ind}(\rb,t)$, and
surface, $\Fb_\alpha^{(S)ind}(\rb,t)$, components

\begin{equation}
 \Fb_\alpha^{ind}(\rb,t) = \Fb_\alpha^{(V)ind}(\rb,t) +
  \Fb_\alpha^{(S)ind}(\rb,t)
\end{equation}

\noindent  given inside the volume  $V_\alpha = (4/3)\pi a_\alpha^3$  occupied
by the sphere  $\alpha$  and on its surface $S_{\alpha}(t)$, respectively, and
defined as follows \cite{ref.Usenko}:

\begin{eqnarray}
 \Fb_\alpha^{(V)ind}(\rb,t) &=& \Theta(a_\alpha - r_\alpha)\,
  \rho\frac{\partial}{\partial t}\Bigl(\Omb_\alpha(t) \times \rb_\alpha\Bigr), \\
 \Fb_\alpha^{(S)ind}(\rb,t) &=& \int \!\! d\Omega_\alpha \,
  \delta(\rb - \Rb_\alpha(t) - \ab_\alpha) \fb_\alpha(\ab_\alpha,t),
\end{eqnarray}

\noindent  where  $\fb_\alpha(\ab_\alpha,t)$  is a certain unknown density of
the induced surface force distributed over the surface $S_{\alpha}(t)$  and
$\ab_\alpha \equiv (a_\alpha, \theta_\alpha, \varphi_\alpha)$ is the vector
directed from the center of sphere  $\alpha$  to the point on its surface
characterized by the polar  $\theta_\alpha$  and azimuth $\varphi_\alpha$
angles in the local spherical coordinate system  $O_\alpha$, \  $d\Omega_\alpha
= \sin{\theta_\alpha} \, d\theta_\alpha d\varphi_\alpha$ \  is the solid angle.

Force (2.10) and torque (2.11) exerted by the fluid on sphere  $\alpha$ are
represented as follows:

\begin{eqnarray}
 \Fb_\alpha^{f}(t) &=& \Fb_\alpha (t) + \Fb_\alpha^{in}(t)
  - \tilde{\Fb}_\alpha^{ext}(t),  \\
 \Tb_\alpha^{f}(t) &=& \Tb_\alpha (t) - \tilde{\Tb}_\alpha^{ext}(t),
\end{eqnarray}

\noindent  where  $\Fb_\alpha (t)$  and  $\Tb_\alpha(t)$  are, respectively,
the force and the torque exerted on sphere  $\alpha$  due to the induced force
density distributed over its surface

\begin{eqnarray}
 \Fb_\alpha (t) &=& -\int \!\! d\rb \, \Fb_\alpha^{(S)ind}(\rb,t)
  = -\int \!\! d\Omega_\alpha \, \fb_\alpha(\ab_\alpha,t),  \\
 \Tb_\alpha (t) &=& -\int \!\! d\rb \, \left(\rb_\alpha \times
  \Fb_\alpha^{(S)ind}(\rb,t)\right) = -\int \!\! d\Omega_\alpha \,
  \Bigl ( \ab_\alpha \times \fb_\alpha(\ab_\alpha,t) \Bigr ),
\end{eqnarray}

\noindent  $\tilde{\Fb}_\alpha^{ext}(t)$  and $\tilde{\Tb}_\alpha^{ext}(t)$
are, respectively, the force and the torque acting on a fluid sphere occupying
the volume  $V_\alpha$  instead of sphere  $\alpha$  due the external force
field analytically extended to this domain,

\begin{eqnarray}
 \tilde{\Fb}_\alpha^{ext}(t) &=& \int\limits_{V_\alpha} \!\! d\rb \,
  \Fb^{ext}(\rb,t),  \\
 \tilde{\Tb}_\alpha^{ext}(t) &=& \int\limits_{V_\alpha} \!\! d\rb \,
  \left(\rb_\alpha \times \Fb^{ext}(\rb,t)\right),
\end{eqnarray}

\noindent  and

\begin{equation}
 \Fb_\alpha^{in}(t) = \tilde{m}_\alpha \frac{d\Ub_\alpha (t)}{dt}
\end{equation}

\noindent  is the inertial force that is necessary to be applied to a fluid
sphere of volume  $V_\alpha$  to ensure its movement with the acceleration
$d\Ub_\alpha(t)/dt$  \cite{ref.MazurBed},  where  $\tilde{m}_\alpha = \rho
V_\alpha$  is the mass of the fluid displaced by sphere  $\alpha$.

In what follows, we expand various quantities in the Fourier integral with
respect to the time of the form

\begin{equation}
 A(\rb,t) = \frac{1}{2\pi} \int\limits_{-\infty}^\infty \!\!d\omega \,
  A(\rb,\omega) \exp(-i \omega t).
\end{equation}

Within the framework of the linear approximation both with respect to the fluid
velocity and the velocities of the spheres \cite{ref.Saarl}, the solution of
problem (2.2), (2.3), (2.16) for the required Fourier transforms of the
quantities $\tilde{\vb}(\rb,\omega)$  and $\tilde{p}(\rb,\omega)$ can be
represented as

\begin{eqnarray}
 \tilde{\vb}(\rb,\omega) &=& \sum\limits_{\beta = 1}^N
  \vb_\beta^{ind}(\rb,\omega),  \\
 \tilde{p}(\rb,\omega) &=& p^{(0)}(\rb,\omega) + p^{(ind)}(\rb,\omega)
  = p^{(0)}(\rb,\omega) + \sum\limits_{\beta = 1}^N p_\beta^{ind}(\rb,\omega),
\end{eqnarray}

\noindent  where

\begin{equation}
 \vb_\beta^{ind}(\rb,\omega) = \vb_\beta^{(V)ind}(\rb,\omega)
  + \vb_\beta^{(S)ind}(\rb,\omega),
\end{equation}

\begin{equation}
 \vb_\beta^{(V)ind}(\rb,\omega) = -i \xi_\beta b_\beta^2 \frac{2}{3\eta}
 \frac{1}{(2\pi)^3} \int\limits_{-\infty}^\infty \!\! d\kb \,
 \frac{\exp{(i\kb \cdot \left(\rb-\Rb_\beta\right))}}{k^2 +\kappa^2}
 \frac{j_2(\kappa a_\beta)}{k^2} \Bigl (\Omb_\beta(\omega) \times \kb \Bigr ),
\end{equation}

\begin{eqnarray}
 \vb_\beta^{(S)ind}(\rb,\omega) &=& \frac{1}{(2 \pi)^3}
  \int\limits_{-\infty}^\infty \!\! d\kb \,
  \exp{(i\kb \cdot \left(\rb-\Rb_\beta\right))}
  \int \!\! d\Omega_\beta \, \exp{(-i\kb \cdot \ab_\beta)} \,
  \Sb(\kb,\omega) \cdot \fb_\beta(\ab_\beta,\omega), \nonumber  \\
  && \\
 p_\beta^{ind}(\rb,\omega)  &=& -\frac{i}{(2 \pi)^3}
  \int\limits_{-\infty}^\infty \!\! d\kb \,
  \frac{\exp{(i\kb \cdot \left(\rb-\Rb_\beta\right))}}{k^2}
  \int \!\! d\Omega_\beta \, \exp{(-i\kb \cdot \ab_\beta)} \,
  \Bigl(\kb \cdot \fb_\beta(\ab_\beta,\omega)\Bigr).
\end{eqnarray}

\noindent  Here,  $\kappa = \sqrt{\omega/(i\nu)} = (1 - i \sign
\omega)/\delta$,  $\delta = \sqrt{2\nu/|\omega|}$,  $\nu = \eta/\rho$ is the
kinematic viscosity of the fluid,  $\xi_\beta = 6 \pi \eta a_\beta$  is the
Stokes friction coefficient for a sphere of radius  $a_\beta$  uniformly moving
in the fluid, $b_\beta = \kappa a_\beta$  is the dimensionless parameter that
characterizes the ratio of the radius  $a_\beta$  of sphere  $\beta$  to the
depth of penetration   $\delta$  of a plane transverse wave of frequency
$\omega$ created by an oscillating solid surface into the fluid
\cite{ref.Landau}, ($|b_\beta | = \sqrt{2}\,a_\beta/\delta$),  $j_n(x)$ is the
spherical Bessel function of order  $n$, and

\begin{equation}
 \Sb(\kb,\omega) = \frac{1}{\eta \left(k^2 + \kappa^2 \right)}
 \left(\Ib - \nb_k \nb_k \right)
\end{equation}

\noindent  is the dynamic Oseen tensor ($\Ib$  is the unit tensor and
$\nb_k = \kb/k$  is the unit vector directed along the vector  $\kb$).

Note that Eqs.~(2.29)--(2.34) are valid for any  $\rb$.  For $r_\alpha
> a_\alpha$,  the quantities  $\vb_\beta^{(S)ind}(\rb,\omega)$  and
$p_\beta^{ind}(\rb,\omega)$  can be interpreted, respectively, as the fluid
velocity and pressure at the point  $\rb$  generated by the induced surface
force $\Fb_\beta^{(S)ind}(\rb,\omega)$ distributed over the surface of sphere
$\beta$,  whereas the quantity  $\vb_\beta^{(V)ind}(\rb,\omega)$  has the sense
of the fluid velocity at the point  $\rb$  generated by the induced volume
force $\Fb_\beta^{(V)ind}(\rb,\omega)$  given in the volume $V_\beta$ occupied
by sphere  $\beta$.

It is convenient to represent the quantities $\vb_\beta^{(V)ind}(\rb,\omega)$,
$\vb_\beta^{(S)ind}(\rb,\omega)$, and  $p_\beta^{ind}(\rb,\omega)$ as the
following expansions in the spherical harmonics:

\begin{eqnarray}
 \vb_\beta^{(V)ind}(\Rb_\alpha+\rb_\alpha,\omega) &=& 2\sqrt{\pi} \,
  \sum\limits_{lm} \vb_{\beta, lm}^{(V)ind}(\Rb_\alpha,r_\alpha,\omega)
  Y_{lm}(\theta_\alpha,\varphi_\alpha),  \\
 \vb_\beta^{(S)ind}(\Rb_\alpha+\rb_\alpha,\omega) &=& 2\sqrt{\pi} \,
  \sum\limits_{lm} \vb_{\beta, lm}^{(S)ind}(\Rb_\alpha,r_\alpha,\omega)
  Y_{lm}(\theta_\alpha,\varphi_\alpha), \\
 p_\beta^{ind}(\Rb_\alpha+\rb_\alpha,\omega) &=& 2\sqrt{\pi} \,
  \sum\limits_{lm} p_{\beta, lm}^{ind}(\Rb_\alpha,r_\alpha,\omega)
  Y_{lm}(\theta_\alpha,\varphi_\alpha).
\end{eqnarray}

\noindent  Here, the spherical harmonics $Y_{lm}(\theta,\varphi)$  are defined
as \cite{ref.Nikiforov}

\begin{eqnarray}
 Y_{lm}(\theta,\varphi) &=& \frac{\exp{(im\varphi)}}{\sqrt{2\pi}}\,
  \Theta_{lm}(\cos\theta), \qquad  l = 0, 1, \ldots, \qquad  -l \leq m \leq l,
  \nonumber \\
 \Theta_{lm}(x) &=& (-1)^{\frac{m - |m|}{2}} \, \sqrt{\frac{2l + 1}{2}
  \frac{(l - |m|)!}{(l + |m|)!}}\, P_l^{|m|}(x),
\end{eqnarray}

\noindent  and  $P_l^m(x)$  is the associated Legendre polynomial.  For short,
the symbol  $\sum\limits_{lm}$  means $\sum\limits_{l = 0}^\infty
\sum\limits_{m = -l}^l$.  As was mentioned above, the position of any point of
the space defined by the radius vector $\rb$ relative to the fixed system $O$
can be represented in the form $\rb = \Rb_\alpha + \rb_\alpha$, where
$\Rb_\alpha$ is the radius vector that defines the position of the center of
certain sphere $\alpha$  and $\rb_\alpha$ is the radius vector directed from
this center to the point at hand.  We take the quantity $\rb_\alpha$  such that
it corresponds to the minimum difference $r_\beta - a_\beta$, where $\beta = 1,
2, \ldots, N$, if the point of observation belongs to the domain occupied by
the fluid, which means the minimum distance from the point at hand to the
surface of sphere $\alpha$. If the point of observation lies inside the domain
occupied by sphere $\alpha$, then $\rb_\alpha$ is the radius vector directed
from the center of this sphere to this point; $\rb_\alpha \equiv
(r_\alpha,\theta_\alpha,\varphi_\alpha)$, where $\theta_\alpha$ and
$\varphi_\alpha$ are, respectively, the polar and azimuth angles of the vector
$\rb_\alpha$  in the local spherical coordinate system $O_\alpha$.

The expansion coefficients in the series (2.36)--(2.38) have the form

\begin{eqnarray}
 \vb_{\beta, l_1 m_1}^{(V)ind}(\Rb_\alpha,r_\alpha,\omega) &=&
  \frac{2}{3} \, \xi_\beta b_\beta^2 \sum\limits_{l_2 m_2} (-1)^{l_2}
  P_{l_1 2,l_2}(r_\alpha,a_\beta,R_{\alpha\beta},\omega)
  \left(\Omb_\beta(\omega) \times
  \Wb_{l_1 m_1,00}^{l_2 m_2} \right) \nonumber \\
  & \times & Y_{l_2 m_2}^*(\Theta_{\alpha\beta},\Phi_{\alpha\beta}),  \\
 \vb_{\beta, l_1 m_1}^{(S)ind}(\Rb_\alpha,r_\alpha,\omega) &=&
  \sum\limits_{l_2 m_2}
  \Tb_{l_1 m_1}^{l_2 m_2}(r_\alpha,a_\beta,\Rb_{\alpha\beta},\omega)
  \cdot \fb_{\beta,l_2 m_2}(\omega),  \\
 p_{\beta, l_1 m_1}^{ind}(\Rb_\alpha,r_\alpha,\omega) &=&
  \sum\limits_{l_2 m_2}
  \Db_{l_1 m_1}^{l_2 m_2}(r_\alpha,a_\beta,\Rb_{\alpha\beta})
  \cdot \fb_{\beta,l_2 m_2}(\omega)
\end{eqnarray}

\noindent  for  $\beta  \neq  \alpha$  and

\begin{eqnarray}
 \vb_{\alpha, l_1 m_1}^{(V)ind}(\Rb_\alpha,r_\alpha,\omega) &=&
  \delta_{l_1, 1} \, \frac{\xi_\alpha b_\alpha^2}{6\pi \sqrt{3}} \,
  P_{1,2}(r_\alpha,a_\alpha,\omega)
  \Bigl(\Omb_\alpha(\omega) \times \eb_{m_1}^* \Bigr),  \\
 \vb_{\alpha, l_1 m_1}^{(S)ind}(\Rb_\alpha,r_\alpha,\omega) &=&
  \sum\limits_{l_2 m_2}
  \Tb_{l_1 m_1}^{l_2 m_2}(r_\alpha,a_\alpha,\omega) \cdot
  \fb_{\alpha,l_2 m_2}(\omega),  \\
 p_{\alpha, l_1 m_1}^{ind}(\Rb_\alpha,r_\alpha,\omega) &=&
  \sum\limits_{l_2 m_2}
  \Db_{l_1 m_1}^{l_2 m_2}(r_\alpha,a_\alpha) \cdot
  \fb_{\alpha,l_2 m_2}(\omega)
\end{eqnarray}

\noindent  for  $\beta = \alpha$.

\noindent  Here, $\Rb_{\alpha\beta} = \Rb_\alpha - \Rb_\beta \equiv
(R_{\alpha\beta},\Theta_{\alpha\beta},\Phi_{\alpha\beta})$  is the vector
between the centers of spheres  $\alpha$  and  $\beta$  pointing from sphere
$\beta$  to sphere  $\alpha$  in the spherical coordinate system  $O$. The
quantities $\fb_{\beta, lm}(\omega)$  are the harmonics of the induced surface
force densities  $\fb_\beta(\ab_\beta,\omega)$ expanded in the spherical
harmonics

\begin{equation}
 \fb_\beta(\ab_\beta,\omega) = \frac{1}{2 \sqrt{\pi}} \sum\limits_{lm}
  \fb_{\beta, lm}(a_\beta,\omega) Y_{lm}(\theta_\beta,\varphi_\beta),
\end{equation}

\noindent  where, for short, the argument $a_\beta$ in  $\fb_{\beta,
lm}(a_\beta,\omega)$  is omitted.  The quantities involved in
Eqs.~(2.40)--(2.45) are written as

\begin{eqnarray}
 \Tb_{l_1 m_1}^{l_2 m_2}(r_\alpha,a_\beta,\Rb_{\alpha\beta},\omega) &=&
  \sum\limits_{lm}
  \Tb_{l_1 m_1,lm}^{l_2 m_2}(r_\alpha,a_\beta,R_{\alpha\beta},\omega)
  Y_{lm}(\Theta_{\alpha\beta},\Phi_{\alpha\beta}),  \\
 \Tb_{l_1 m_1,lm}^{l_2 m_2} (r_\alpha,a_\beta,R_{\alpha\beta},\omega) &=&
  F_{l_1 l_2,l}(r_\alpha,a_\beta,R_{\alpha\beta},\omega)
  \Kb_{l_1 m_1,lm}^{l_2 m_2}, \\
 F_{l_1 l_2,l}(r_\alpha,a_\beta,R_{\alpha\beta},\omega) &=&
  \frac{2}{\pi \eta} \int\limits_0^\infty \!\! dk\, \frac{k^2}{k^2 + \kappa^2}
  \, j_{l_1}(kr_\alpha) j_{l_2}(ka_\beta) j_l(kR_{\alpha\beta}),
\end{eqnarray}

\begin{equation}
 \Kb_{l_1 m_1, lm}^{l_2 m_2} =
  i^{l_1 - l_2 + l} \int \!\! d\Omega_k \, (\Ib - \nb_k \nb_k)\,
  Y_{l_1 m_1}^*(\theta_k,\varphi_k) Y_{l_2 m_2}(\theta_k,\varphi_k)
  Y_{lm}^*(\theta_k,\varphi_k),
\end{equation}

\begin{eqnarray}
 \Db_{l_1 m_1}^{l_2 m_2}(r_\alpha,a_\beta,\Rb_{\alpha\beta}) &=&
  \sum\limits_{lm} \Db_{l_1 m_1,lm}^{l_2 m_2}(r_\alpha,a_\beta,R_{\alpha\beta})
  Y_{lm}(\Theta_{\alpha\beta},\Phi_{\alpha\beta}),  \\
 \Db_{l_1 m_1,lm}^{l_2 m_2} (r_\alpha,a_\beta,R_{\alpha\beta}) &=&
  C_{l_1 l_2,l}(r_\alpha,a_\beta,R_{\alpha\beta}) \Wb_{l_1 m_1,lm}^{l_2 m_2}, \\
 C_{l_1 l_2,l}(r_\alpha,a_\beta,R_{\alpha\beta}) &=& \frac{2}{\pi}
  \int\limits_0^\infty \!\! dk\,
  k \, j_{l_1}(kr_\alpha) j_{l_2}(ka_\beta) j_l(kR_{\alpha\beta}), \\
 \Wb_{l_1 m_1,lm}^{l_2 m_2} &=&
  i^{l_1 - l_2 + l - 1} \int \!\! d\Omega_k \, \nb_k \,
  Y_{l_1 m_1}^*(\theta_k,\varphi_k) Y_{l_2 m_2}(\theta_k,\varphi_k)
  Y_{lm}^*(\theta_k,\varphi_k),  \\
 P_{l_1 l_2,l}(r_\alpha,a_\beta,R_{\alpha\beta},\omega) &=&
  \frac{2}{\pi \eta} \int\limits_0^\infty \!\! dk\, \frac{k}{k^2 + \kappa^2}
  \, j_{l_1}(kr_\alpha) j_{l_2}(ka_\beta) j_l(kR_{\alpha\beta}).
\end{eqnarray}

\begin{eqnarray}
 \Tb_{l_1 m_1}^{l_2 m_2}(r_\alpha,a_\alpha,\omega) &=&
  \frac{1}{2 \sqrt{\pi}} \, F_{l_1 l_2}(r_\alpha,a_\alpha,\omega)
  \Kb_{l_1 m_1,00}^{l_2 m_2}, \\
 F_{l_1 l_2}(r_\alpha,a_\alpha,\omega) &=&
  \frac{2}{\pi \eta} \, \int\limits_0^\infty \!\! dk\, \frac{k^2}{k^2 + \kappa^2}
  \, j_{l_1}(kr_\alpha) j_{l_2}(k a_\alpha), \\
 \Db_{l_1 m_1}^{l_2 m_2}(r_\alpha,a_\alpha) &=&
  \frac{1}{2 \sqrt{\pi}} \, C_{l_1 l_2}(r_\alpha,a_\alpha)
  \Wb_{l_1 m_1,00}^{l_2 m_2}, \\
 C_{l_1 l_2}(r_\alpha,a_\alpha) &=&
  \frac{2}{\pi} \, \int\limits_0^\infty \!\! dk\,
  k \, j_{l_1}(kr_\alpha) j_{l_2}(k a_\alpha), \\
 P_{l_1 l_2}(r_\alpha,a_\alpha,\omega) &=&
  \frac{2}{\pi \eta} \int\limits_0^\infty \!\! dk\, \frac{k}{k^2 + \kappa^2}
  \, j_{l_1}(kr_\alpha) j_{l_2}(ka_\alpha).
\end{eqnarray}

The explicit form for the quantities defined by (2.49), (2.50), (2.52)--(2.55),
and (2.57)--(2.60) has been found early in \cite{ref.Usenko} (see Appendix A).

The quantities  $\eb_m,$  where  $ m = 0, \pm1,$  are defined as follows:

\begin{equation}
 \eb_0 \equiv \eb_z,  \qquad  \eb_{\pm 1} = \frac{1}{\sqrt{2}}
 \left(i\eb_y \pm \eb_x \right),
\end{equation}

\noindent  i.e., up to the factor  $(-1)^m,$  are the cyclic covariant unit
vectors \cite{ref.Varshalovich}, and  $\eb_x,$  $\eb_y,$ and $\eb_z$  are the
Cartesian unit vectors.

Since the quantities  $\Kb_{l_1 m_1, lm}^{l_2 m_2}$  and  $\Wb_{l_1 m_1,
lm}^{l_2 m_2}$ are not equal to zero, respectively, only for  $l = l_1 + l_2 -
2p$, where  $p = -1,0,1,\ldots,p_{max},$  $p_{max} = \min \Bigl(\left[(l_1 +
l_2)/2\right], 1 + \min (l_1, l_2) \Bigr)$, where  $\left[a\right]$ is the
integer part of  $a$  and  $\min (a,b)$ means the smallest quantity of $a$ and
$b$, and for  $l = l_1 + l_2 - 2p + 1$,  where  $p =
0,1,\ldots,\tilde{p}_{max},$ $\tilde{p}_{max} = \min \Bigl(\left[(l_1 + l_2 +
1)/2\right], 1 + \min (l_1, l_2) \Bigr)$,  the infinite sums over  $l$  in
relations (2.47) and (2.51) are replaced by sums containing a finite number of
terms corresponding to these values of $l$. The quantity  $\Wb_{l_1 m_1,
00}^{l_2 m_2}  \neq 0$  only for $l_2 = l_1 \pm 1 \geq 0$.  Therefore, the
infinite sums over  $l_2$ in relation (2.40) and (2.45) are replaced by sums
containing only terms with $l_2 = l_1 \pm 1 \geq 0$. Taking into account that
$\Kb_{l_1 m_1, 00}^{l_2 m_2} \neq 0$ only for $l_2 = l_1 + 2p \geq 0$, where $p
= 0, \pm 1$, the infinite sum over $l_2$ in relation (2.44) is replaced by the
sum with $l_2 = l_1 + 2p \geq 0$.

According to (2.43), only one harmonic of $\vb_{\alpha, l_1
m_1}^{(V)ind}(\Rb_\alpha,r_\alpha,\omega)$  is not equal to zero.  Using this
observation, we can represent this component of the fluid velocity as follows:

\begin{equation}
 \vb_\alpha^{(V)ind}(\Rb_\alpha + \rb_\alpha,\omega) =
  \frac{\xi_\alpha b_\alpha^2}{6\pi} \, P_{1,2}(r_\alpha,a_\alpha,\omega)
  \Bigl( \Omb_\alpha(\omega) \times \nb_\alpha \Bigr),
\end{equation}

\noindent  where  $\nb_\alpha = \rb_\alpha/r_\alpha$.

Relations (2.29), (2.31)--(2.33), (2.36), (2.37), (2.40), (2.41), (2.43), and
(2.44) completely determine the function $\tilde{\vb}(\rb,\omega)$ in the
entire space provided that the induced surface force densities are known.
Analogously, relations (2.30), (2.34), (2.38), (2.42), and (2.45) uniquely
reproduce the original function $\tilde{p}(\rb,\omega)$  at any point of the
space with the exception of the points of the surfaces of the spheres
($r_\alpha = a_\alpha, \ \alpha = 1, 2,\ldots, N$) where the obtained quantity
is equal to the half-sum of the original function $\tilde{p}(\rb,\omega)$ given
at $\rb = \Rb_\alpha + \ab_\alpha + 0$ and $\rb = \Rb_\alpha + \ab_\alpha - 0$
because the original function $\tilde{p}(\rb,\omega)$  defined by relation
(2.13) is discontinuous at the surfaces $r_\alpha = a_\alpha$. The fluid
pressure at the surfaces of the spheres denoted as  $p(\Rb_\alpha + \ab_\alpha
+ 0,\omega)$  can be represented as follows:

\begin{equation}
 p(\Rb_\alpha + \ab_\alpha + 0,\omega) = p^{(0)}(\Rb_\alpha + \ab_\alpha,\omega)
  + p^{ind}(\Rb_\alpha + \ab_\alpha + 0,\omega),
\end{equation}

\noindent  where

\begin{equation}
 p^{ind}(\Rb_\alpha + \ab_\alpha + 0,\omega) = 2 p^{ind}(\Rb_\alpha + \ab_\alpha,\omega)
  - i\omega \rho \left( \Rb_\alpha + \ab_\alpha \right) \cdot
  \Ub_\alpha (\omega)
\end{equation}

\noindent  and the quantities  $p^{(0)}(\Rb_\alpha + \ab_\alpha,\omega)$  and
$p^{ind}(\Rb_\alpha + \ab_\alpha,\omega)$  are defined by the corresponding
expressions for  $p^{(0)}(\rb,\omega)$  and $p^{ind}(\rb,\omega)$
given at $\rb = \Rb_\alpha + \ab_\alpha$ derived above.

Using the expansion (2.46) together with Eqs.~(2.23) and (2.24), one can
represent the Fourier transforms of the force  $\Fb_\alpha(\omega)$  and the
torque $\Tb_\alpha(\omega)$ exerted by the fluid on sphere  $\alpha$ in terms
of Fourier harmonics of the induced surface force density  $\fb_{\alpha,
lm}(\omega)$  as follows \cite{ref.Pien1,ref.Pien2,ref.Usenko,ref.Yosh}:

\begin{eqnarray}
 \Fb_\alpha (\omega) &=& -\fb_{\alpha, 00}(\omega),  \\
 \Tb_\alpha (\omega) &=& -\frac{a_\alpha}{\sqrt{3}} \sum\limits_{m = -1}^1
  \left( \eb_m \times \fb_{\alpha, 1m}(\omega) \right).
\end{eqnarray}

Thus, both the fluid velocity and pressure induced by the spheres as well as
the forces and torques exerted by the fluid on the spheres are expressed in
terms of the harmonics of the induced surface force densities.  Note that these
expressions have been obtained without imposing any additional restrictions on
the size of spheres, distances between them, and the frequency range.  To
determine these harmonics, using the stick boundary conditions for the fluid
velocity at the surfaces of the spheres \cite{ref.Happel,ref.Batch,ref.Landau},
the original problem is reduced to the determination of the unknown quantities
$\fb_{\beta,lm}(\omega)$  from the following infinite system of the linear
algebraic equations \cite{ref.Usenko}:

\begin{eqnarray}
 \sum\limits_{l_2 m_2} \Tb_{\alpha, l_1 m_1}^{\alpha, l_2 m_2}(\omega)
  \cdot \fb_{\alpha, l_2 m_2}(\omega) + \vb_{\alpha, l_1 m_1}^{\alpha (V)}(\omega)
  - \Vb_{\alpha, l_1 m_1}(\omega)
  &=& -\sum\limits_{\beta \neq \alpha} \Biggl \{ \sum\limits_{l_2 m_2}
  \Tb_{\alpha, l_1 m_1}^{\beta, l_2 m_2}(\omega)
  \cdot \fb_{\beta, l_2 m_2}(\omega) \nonumber  \\
  &+& \vb_{\alpha, l_1 m_1}^{\beta (V)}(\omega) \Biggr \},
\end{eqnarray}

\noindent  where

\begin{equation}
 \Vb_{\alpha, l_1 m_1}(\omega) = \delta_{l_1, 0}\, \delta_{m_1, 0}\, \Ub_\alpha(\omega)
  + \delta_{l_1, 1}\, \frac{a_\alpha}{\sqrt{3}}\,
  \Bigl(\Omb_\alpha(\omega)\times\eb_{m_1}^*\Bigr),
\end{equation}

\noindent  and, to simplify the representation, we introduce the
following notation:

\[
\begin{array}{lrllrl}
  \Tb_{\alpha, l_1 m_1}^{\beta, l_2m_2}(\omega) & \equiv &
   \Tb_{l_1  m_1}^{l_2m_2}(a_\alpha,a_\beta,\Rb_{\alpha\beta},\omega),  \qquad
   & \Tb_{\alpha, l_1 m_1}^{\alpha, l_2m_2}(\omega) & \equiv &
   \Tb_{l_1  m_1}^{l_2m_2}(a_\alpha,a_\alpha,\omega),  \\
  \vb_{\alpha,l_1 m_1}^{\beta (V)}(\omega) & \equiv &
   \vb_{\beta,l_1 m_1}^{(V)ind}(\Rb_\alpha,a_\alpha,\omega), \qquad
   & \vb_{\alpha,l_1 m_1}^{\alpha (V)}(\omega) & \equiv &
   \vb_{\alpha,l_1 m_1}^{(V)ind}(\Rb_\alpha,a_\alpha,\omega),
\end{array}
\]

\noindent  where the quantities
$\Tb_{l_1m_1}^{l_2m_2}(a_\alpha,a_\beta,\Rb_{\alpha\beta},\omega)$, \
$\Tb_{l_1m_1}^{l_2m_2}(a_\alpha,a_\alpha,\omega)$, \
$\vb_{\beta,l_1m_1}^{(V)ind}(\Rb_\alpha,a_\alpha,\omega)$, \   and
$\vb_{\alpha,l_1 m_1}^{(V)ind}(\Rb_\alpha,a_\alpha,\omega)$, \  are defined,
respectively, by relations (2.47), (2.56), (2.40), and (2.43) at $r_\alpha =
a_\alpha$.  Up to the terms  $\vb_{\beta,l_1
m_1}^{(V)ind}(\Rb_\alpha,a_\alpha,\omega)$,  where  $\beta = 1,2,\ldots,N$,
Eqs.~(2.67) agree with the corresponding equations given in
\cite{ref.Pien1,ref.Pien2} (for the discussion, see \cite{ref.Usenko}).
Following \cite{ref.Pien1,ref.Pien2,ref.Yosh}, the quantities
$\Tb_{l_1m_1}^{l_2m_2}(a_\alpha,a_\beta,R_{\alpha\beta},\omega)$, where  $\beta
\neq \alpha$,  and  $\Tb_{l_1m_1}^{l_2m_2}(a_\alpha,a_\alpha,\omega)$, are
called hydrodynamic interaction tensors (mutual interaction tensors for $\beta
\neq \alpha$  and self-interaction tensors for $\beta = \alpha$). The
quantities $F_{l_1 l_2,l}(a_\alpha,a_\beta,R_{\alpha\beta},\omega)$ and $F_{l_1
l_2}(a_\alpha,a_\alpha,\omega)$  defining the hydrodynamic interaction tensors,
the quantities $C_{l_1 l_2,l}(a_\alpha,a_\beta,R_{\alpha\beta})$  and $C_{l_1
l_2}(a_\alpha,a_\alpha)$ defining, respectively, the tensors $\Db_{l_1
m_1}^{l_2 m_2}(a_\alpha,a_\beta,\Rb_{\alpha\beta})$  and  $\Db_{l_1 m_1}^{l_2
m_2}(a_\alpha,a_\alpha)$,  and the quantities  $P_{l_1
2,l}(a_\alpha,a_\beta,R_{\alpha\beta},\omega)$  and
$P_{1,2}(a_\alpha,a_\alpha,\omega)$  defining, respectively,  $\vb_{\alpha,l_1
m_1}^{\beta (V)}(\omega)$  and  $\vb_{\alpha,l_1 m_1}^{\alpha (V)}(\omega)$ are
determined by relations (A6), (A16), (A7), (A10), (A14), and (A18) at $r_\alpha
= a_\alpha$.  For  $\beta \neq \alpha$, these quantities have the form

\begin{eqnarray}
 F_{l_1 l_2,l}(a_\alpha,a_\beta,R_{\alpha\beta},\omega) &=&
  (-1)^p \frac{2\kappa}{\pi\eta} \left\{ \tilde{j}_{l_1}(b_\alpha)
  \tilde{j}_{l_2}(b_\beta) \tilde{h}_l(y_{\alpha\beta})
  - \delta_{l, l_1 + l_2 + 2}
  \frac{\pi^{3/2}}{2}
  \frac{\Gamma\left(l_1 + l_2 + \frac{5}{2} \right)}
  {\Gamma\left(l_1 + \frac{3}{2} \right)\Gamma\left(l_2 + \frac{3}{2} \right)}
  \right. \nonumber \\
  &\times &  \left. \frac{\sigma_{\alpha\beta}^{l_1} \sigma_{\beta\alpha}^{l_2}}
  {y_{\alpha\beta}^3} \right \},
  \quad  l = l_1 + l_2 -2p \geq 0,  \quad  p = -1, 0, 1,\ldots, p_{max},  \\
 C_{l_1 l_2,l}(a_\alpha,a_\beta,R_{\alpha\beta}) &=&
  \delta_{l, l_1 + l_2 + 1} \frac{\sqrt{\pi}}{2}
  \frac{\Gamma\left(l_1 + l_2 + \frac{3}{2} \right)}
  {\Gamma\left(l_1 + \frac{3}{2} \right)\Gamma\left(l_2 + \frac{3}{2} \right)}
  \frac{\sigma_{\alpha\beta}^{l_1} \sigma_{\beta\alpha}^{l_2}}{R_{\alpha\beta}^2},
  \nonumber \\
  && \qquad \qquad \qquad l = l_1 + l_2 - 2p + 1 \geq 0,
  \qquad  p = 0, 1,\ldots, \tilde{p}_{max},  \nonumber  \\
 P_{l_1 2,l}(a_\alpha,a_\beta,R_{\alpha\beta},\omega) &=& (-1)^p
  \frac{2}{\pi \eta} \, \tilde{j}_{l_1}(b_\alpha)
  \tilde{j}_2(b_\beta) \tilde{h}_l(y_{\alpha\beta}),
  \qquad \qquad l = l_1 \pm 1 \geq 0.
\end{eqnarray}

\noindent  Here,  $\tilde{j}_l(x) = \sqrt{\pi/(2x)} I_{l +\frac{1}{2}}(x)$ and
$\tilde{h}_l(x) = \sqrt{\pi/(2x)} K_{l +\frac{1}{2}}(x)$  are the modified
spherical Bessel functions of the first and third kind, respectively,
\cite{ref.Abram},  $\Gamma(z)$  is the gamma function, the dimensionless
parameters $\sigma_{\alpha\beta} = a_\alpha/R_{\alpha\beta}$  and
$\sigma_{\beta\alpha} = a_\beta/R_{\alpha\beta}$, which are the ratios of the
radius of a sphere to the distance between the centers of two spheres, is
always smaller than one (for spheres of equal radii, they are equal and cannot
be greater than $1/2$; in dilute suspensions,  $\sigma_{\alpha\beta},
\sigma_{\beta\alpha} \ll 1$), the dimensionless parameter $y_{\alpha\beta} =
\kappa R_{\alpha\beta}$ ($|y_{\alpha\beta}| =
\sqrt{2}\,R_{\alpha\beta}/\delta$) characterizes the ratio of the distance
between the centers of spheres  $\alpha$  and $\beta$  to the depth of
penetration $\delta$ of a plane transverse wave of frequency $\omega$  into the
fluid.  In certain frequency ranges, the quantity $|y_{\alpha\beta}|$  can be
both smaller (for short distances between spheres) and greater (for space-apart
spheres) than one.

For  $\beta \neq \alpha$,  we have

\begin{eqnarray}
 F_{l_1 l_2}(a_\alpha,a_\alpha,\omega) &=&
  (-1)^p \frac{2\kappa}{\pi \eta} \,
  \tilde{j}_{l_{max})}(b_\alpha) \tilde{h}_{l_{min}}(b_\alpha),
  \quad  l_2 = l_1 + 2p \geq 0,  \quad  p = 0, \pm 1,  \\
 C_{l_1 l_2}(a_\alpha,a_\alpha) &=&  \frac{1}{2 a_\alpha^2}
  \left( \delta_{l_2, l_1 + 1} + \delta_{l_2, l_1 - 1} \right),
  \quad \qquad  l_2 = l_1 \pm 1 \geq 0,
\end{eqnarray}

\begin{equation}
 P_{1,2}(a_\alpha,a_\alpha,\omega) =
  \frac{2}{\pi \eta} \, \tilde{h}_1(b_\alpha) \tilde{j}_2(b_\alpha),
\end{equation}

\noindent  where  $l_{max} = \max(l_1, l_2)$  and  $l_{min} = \min(l_1, l_2)$.

The agreement between the quantities  $F_{l_1
l_2,l}(a_\alpha,a_\beta,R_{\alpha\beta},\omega)$  and  $F_{l_1
l_2}(a_\alpha,a_\alpha,\omega)$  defined by relations (2.69) and (2.71) and the
corresponding results given in \cite{ref.Pien2} is discussed in
\cite{ref.Usenko}.  Here, we only note that, in the general case, according to
(2.71), the self-interaction hydrodynamic tensors $\Tb_{\alpha,l_1
m_1}^{\alpha,l_2 m_2}(\omega)$  are not equal to zero for three values of $l_2$
[$l_2 = l_1, \, l_1 + 2,$  and  $l_1 - 2$ (for $l_1 \geq 2)$] instead of two
values of  $l_2$  [$l_2 = l_1, \, l_1 + 2$] mentioned in \cite{ref.Pien2}.

In \cite{ref.Pien1,ref.Pien2}, the similar system of equations in the unknown
harmonics of the induced surface force densities is considered for equal
spheres ($a_\alpha \equiv a, \  \alpha = 1,2,\ldots,N$) for $b_\alpha \equiv b,
\ |b| \ll 1$  and the corresponding solution of the system is given as a series
in two dimensionless parameters ($\sigma$  and  $b$).  However, the expressions
for the harmonics of the induced surface force densities as well as for the
forces and torques exerted by the fluid on the spheres determined up to the
third order of these parameters are expressed in \cite{ref.Pien1,ref.Pien2} in
terms of the tensors $\tilde{\Tb}_{\alpha,l_1m_1}^{\alpha,l_2m_2}(\omega)$
inverse to the self-interaction hydrodynamic tensors
$\Tb_{\alpha,l_1m_1}^{\alpha,l_2m_2}(\omega)$  and defined by the condition

\begin{equation}
 \sum\limits_{m_3 = -l_3}^{l_3} \tilde{\Tb}_{\alpha, l_1 m_1}^{\alpha, l_3 m_3}(\omega)
  \cdot \Tb_{\alpha, l_3 m_3}^{\alpha, l_2 m_2}(\omega)
  = \delta_{l_1, l_2}\,\delta_{m_1, m_2}\, \Ib.
\end{equation}

In \cite{ref.Pien2}, several inverse tensors [e.g., $\tilde{\Tb}_{\alpha,1
m_1}^{\alpha,1 m_2}(\omega)$  and  $\tilde{\Tb}_{\alpha,1 m_1}^{\alpha,3
m_2}(\omega)$] are expressed in terms of the inverse tensor $\tilde{\Kb}_{1
m_1,00}^{1 m_2}$, where the tensors $\tilde{\Kb}_{l_1 m_1, 00}^{l_1 m_2}$
inverse to the tensors $\Kb_{l_1 m_1, 00}^{l_1 m_2}$  are defined by the
condition

\begin{equation}
 \sum\limits_{m_3 = -l_1}^{l_1} \tilde{\Kb}_{l_1 m_1, 00}^{l_1 m_3} \cdot
  \Kb_{l_1 m_3,00}^{l_1 m_2} = \delta_{m_1, m_2} \Ib,
  \qquad l_1 = 0, 1, 2,\ldots \,\, .
\end{equation}

However, the explicit form for the tensors  $\tilde{\Kb}_{l_1 m_1, 00}^{l_1
m_2}$ is not given except for  $l_1 = 0$.  At the same time, as is shown in
\cite{ref.Usenko},

\begin{equation}
 \det \Kb_{1 m_1,00}^{1 m_2} = 0 \, ,
\end{equation}

\noindent which means that  the tensor  $\Kb_{1 m_1,00}^{1 m_2}$  is singular,
and hence the inverse tensor $\tilde{\Kb}_{1m_1,00}^{1m_2}$  does not exist
(at least, in the conventional sense).  As is shown below, this difficulty remains
valid even in the case of one sphere.

In the present paper, we use the procedure proposed in \cite{ref.Usenko} which is free
of the above-mentioned difficulties.  According to this procedure, in addition to the
original system of equations (2.67), the following equations may be used:

\begin{equation}
 Y_\alpha(\omega) + i \omega \rho \, 4\pi \sqrt{3} \, a_\alpha^2
  \Bigl( \Rb_\alpha \cdot \Ub_\alpha(\omega) \Bigr )
  = 4\pi \sqrt{3} \, a_\alpha^2 \sum\limits_{\beta \neq \alpha}
  p_\beta^{ind}(\Rb_\alpha,\omega), \quad  \alpha = 1,2, \ldots,N,
\end{equation}

\noindent  where

\begin{equation}
 Y_\alpha(\omega) \equiv \sum\limits_{m = -1}^1 \eb_m \cdot \fb_{\alpha, 1m}(\omega),
  \quad  \alpha = 1,2, \ldots,N,
\end{equation}

\begin{eqnarray}
 p_\beta^{ind}(\Rb_\alpha,\omega) &=& \frac{1}{4\pi R_{\alpha\beta}^2} \biggl \{
  \left ( \nb_{\alpha\beta} \cdot \fb_{\beta,00}(\omega) \right )
  + 4 \pi \sum\limits_{l_2 = 1}^\infty \sum\limits_{m_2 = -l_2}^{l_2}
  \sigma_{\beta\alpha}^{l_2} \sum\limits_{m = -(l_2 + 1)}^{l_2 + 1}
  \Wb_{00, l_2 + 1,m}^{l_2 m_2}\cdot  \fb_{\beta, l_2 m_2}(\omega) \nonumber \\
   & \times & Y_{l_2 + 1,m}(\Theta_{\alpha\beta},\Phi_{\alpha\beta}) \biggr \},
\end{eqnarray}

\noindent  and  $\nb_{\alpha\beta} = \Rb_{\alpha\beta}/R_{\alpha\beta}$.

For the derivation of Eqs.~(2.77), (2.79), it is necessary to set $r_\alpha =
\varepsilon$,  where  $\varepsilon \to +0$,  in relations  (2.30), (2.34),
(2.38), (2.42), and (2.45), take into account the explicit form for the
function $\tilde{p}(\rb,\omega)$  defined by relation (2.13) and relations
(A19) and (A20) for the quantities $C_{l_1 l_2,l}(0,a_\beta,R_{\alpha\beta})$
and $\lim\limits_{r_\alpha \to 0} C_{l_1 l_2}(r_\alpha,a_\alpha)$ given in the
Appendix A.

We investigate the system under consideration assuming that  $|b_\alpha| \ll 1$
($a_\alpha  \ll \delta /\sqrt{2}\,$, i.e.,  $|\omega| \ll \nu/a_\alpha^2$) and
represent all required quantities [the harmonics  $\fb_{\beta, l_2
m_2}(\omega)$  of the induced surface densities, the fluid velocity and
pressure, the forces and torques exerted by the fluid on the spheres, and the
friction and mobility tensors] as power series in two dimensional parameters
($\sigma$  and $b$) retaining terms up to the order of $b^3$.

In this approximation, the quantities  $F_{l_1
l_2,l}(a_\alpha,a_\beta,R_{\alpha\beta},\omega)$  and  $F_{l_1
l_2}(a_\alpha,a_\alpha,\omega)$  defining the hydrodynamic interaction tensors
are simplified to the form

\begin{eqnarray}
 F_{l_l l_2,l}(a_\alpha,a_\beta,R_{\alpha\beta},\omega) &=&
  \frac{(-1)^p}{\eta R_{\alpha\beta}}
  \frac{\sigma_{\alpha\beta}^{l_1}\sigma_{\beta\alpha}^{l_2}}
  {\Gamma\left(l_1+\frac{3}{2}\right) \Gamma\left(l_2+\frac{3}{2}\right)}
   \, \Biggl \{
  \frac{y_{\alpha\beta}^{l_1 + l_2 + 1}}{\sqrt\pi \, 2^{l_1 + l_2 + 1}}
  \Biggl [ 1 + \frac{1}{4} \Biggl (\frac{b_\alpha ^2}{l_1 + \frac{3}{2}}
  + \frac{b_\beta ^2}{l_2 + \frac{3}{2}} \Biggr) \nonumber  \\
  &+& \frac{1}{32} \Biggl ( \frac{b_\alpha ^4}{\Bigl(l_1 + \frac{3}{2}\Bigr)
  \Bigl(l_1 + \frac{5}{2}\Bigr)}
  + \frac{b_\beta ^4}{\Bigl(l_2 + \frac{3}{2}\Bigr)\Bigl(l_2 +
  \frac{5}{2}\Bigr)} \Biggr )
  + {\cal O}(b^6) \Biggr ] \, \tilde{h}_l (y_{\alpha\beta}) \nonumber  \\
  &-& \delta_{p, -1} \, \frac{\Gamma\left(l_1 + l_2 + \frac{5}{2} \right)}
  {y_{\alpha\beta}^2} \Biggr \}, \,
  l = l_1 + l_2 - 2p,  \, p = -1,0,1, \ldots, p_{max},
\end{eqnarray}

\begin{equation}
 F_{l_1 l_2}(a_\alpha,a_\alpha,\omega) = F_{l_1}(a_\alpha) P_{l_1,n}(b_\alpha),
  \qquad  l_2 = l_1 + 2n \ge 0, \quad n = 0, \pm 1,
\end{equation}

\noindent  where the quantity  $F_{l_1}(a_\alpha)$  corresponding to the
stationary case is defined as follows:

\begin{equation}
 F_{l_1}(a_\alpha) = \frac{1}{(2l_1 + 1)\eta a_\alpha} \,
\end{equation}

\noindent  and

\begin{eqnarray}
 P_{l,0}(b_\alpha) &=& 1 - b_\alpha d_l (b_\alpha),  \\
 P_{l,1}(b_\alpha) &=& -b_\alpha ^2 \frac{1 - \delta_{l,0} \, b_\alpha}
  {(2l + 3) (2l + 5)},  \\
 P_{l,-1}(b_\alpha) &=& -b_\alpha ^2 \frac{1 - \delta_{l,2} \, b_\alpha}
  {(2l - 1) (2l - 3)},  \\
 d_l (b_\alpha) &=& \delta_{l,0} + b_\alpha \frac{2}{(2l - 1) (2l + 3)}
  + \frac{b_\alpha ^2}{3}\, (\delta_{l,0} - \delta_{l,1}), \qquad  l \ge 0.
\end{eqnarray}

We expand the quantities  $P_{1,2}(a_\alpha,a_\alpha,\omega)$  and  $P_{l_1
2,l}(a_\alpha,a_\beta,R_{\alpha\beta},\omega)$  in power series in the small
parameter $b$ retaining terms up to  $b^3$  and substitute these expansions
into relations (2.43) and (2.40) at  $r_\alpha = a_\alpha$.  As a results, we
get

\begin{equation}
 \vb_{\alpha, l_1 m_1}^{\alpha (V)}(\omega) = \delta_{l_1, 1} \,
  \frac{a_\alpha b_\alpha^2}{15 \sqrt{3}} \,
  \Bigl(\Omb_\alpha(\omega) \times \eb_{m_1}^* \Bigr) + {\cal O}(b_\alpha^4),
\end{equation}

\begin{eqnarray}
 \vb_{\alpha, l_1 m_1}^{\beta (V)}(\omega) &=& \frac{2^{2 - l_1}}{15}
  \sqrt{\pi} \, b_\beta^2 \frac{a_\beta \sigma_{\alpha\beta}^{l_1}
  \sigma_{\beta\alpha}^2} {\Gamma \left (l_1 + \frac{3}{2} \right )}\,(-1)^{l_1}
  \sum\limits_{l_2 = l_1 \pm 1 \ge 1} \, \sum\limits_{m_2 = -l_2}^{l_2}
  (-1)^{\frac{1 \pm 1}{2}} \,
  \left(\Omb_\beta(\omega) \times \Wb_{l_1 m_1,00}^{l_2 m_2} \right) \nonumber \\
  & \times &  \tilde{h}_{l_2}(y_{\alpha\beta})
  Y_{l_2 m_2}^*(\Theta_{\alpha\beta},\Phi_{\alpha\beta}) + {\cal O}(b^4).
\end{eqnarray}

For the solution of the system of equations (2.67), (2.77), we  take into
account relations (2.79), (2.87), and (2.88) and the following estimates for
the hydrodynamic interaction tensors following from (2.80) and (2.81) valid for
any $l_1, l_2 \geq 0$:

\begin{eqnarray}
 a_\alpha \Tb_{\alpha,l_1 m_1}^{\alpha, l_2 m_2}(\omega) & \sim &
  \sigma_{\alpha\beta}^0 \, \psi_{\alpha,l_1 l_2} (b_\alpha),
  \qquad  l_2 = l_1 + 2n \ge 0,  \qquad  n = 0, \pm 1, \nonumber  \\
 a_\alpha \Tb_{\alpha,l_1 m_1}^{\beta, l_2 m_2}(\omega) & \sim&
  \sigma_{\alpha\beta}^{l_1 + 1} \sigma_{\beta\alpha}^{l_2}
  \psi_{\alpha\beta,l_1 l_2} (b_\alpha,b_\beta,y_{\alpha\beta})\, ,
\end{eqnarray}

\noindent  where  $\psi_{\alpha,l_1 l_2} (b_\alpha)$  and
$\psi_{\alpha\beta,l_1 l_2} (b_\alpha,b_\beta,y_{\alpha\beta})$  are certain
finite functions of the small parameters $b_\alpha$  and  $b_\beta$  and the
parameter $y_{\alpha\beta}$ free of any restrictions imposing on it.
 The function $\psi_{\alpha,l_1 l_2} (b_\alpha)$ satisfies the relation

\begin{equation}
 \lim_{b_\alpha \to 0} \psi_{\alpha,l_1 l_2} (b_\alpha) = \delta_{l_2,l_1} \,
  a_\alpha \Tb_{\alpha,l_1 m_1}^{\alpha, l_1 m_2},
\end{equation}

\noindent  where

\begin{equation}
 \Tb_{\alpha,l_1 m_1}^{\alpha, l_1 m_2} = \frac{3\sqrt{\pi}}
  {(2l_1 + 1) \xi_\alpha} \, \Kb_{l_1 m_1,00}^{l_1 m_2}
\end{equation}

\noindent  is the static hydrodynamic self-interaction tensor
\cite{ref.Pien2,ref.Usenko,ref.Yosh}.

These estimates enable us to seek a solution of system (2.67), (2.77) by the
method of successive approximations using Eqs.~(2.67), (2.77) with the zero
right-hand sides as the zero iteration, which corresponds to the absence of
interactions between the spheres.  This is quite natural because the sum and
the second term on the right-hand sides of Eqs.~(2.67) correspond to the
harmonics of the fluid velocities at the surface of sphere $\alpha$ induced by
the surface and volume force densities distributed, respectively, over the
surface of sphere $\beta$ and inside its volume, whereas analogous quantities
on the left-hand sides of these equations correspond to the harmonics of the
fluid velocities at the surface of sphere $\alpha$ induced by the surface and
volume force densities distributed, respectively, over the surface of this
sphere and inside its volume.  Analogous reasonings may be used for
Eqs.~(2.77).

Next, we represent the unknown
harmonics $\fb_{\alpha, lm}(\omega)$ in the form

\begin{equation}
 \fb_{\alpha, lm}(\omega)
  = \sum\limits_{n = 0}^\infty \fb_{\alpha, lm}^{(n)}(\omega),
\end{equation}

\noindent  where  $\fb_{\alpha, l_1 m_1}^{(n)}(\omega)$  is a solution of
system (2.67), (2.77) corresponding the {\it n\/}th iteration.  As is shown in
what follows, these quantities can be represented in the form

\begin{equation}
 \fb_{\alpha, lm}^{(n)}(\omega) = \fb_{\alpha, lm}^{(t,n)}(\omega)
  + \fb_{\alpha, lm}^{(r,n)}(\omega)
  + \delta_{l,1} \, \fb_{\alpha, 1m}^{(p,n)}(\omega), \qquad n = 0,1,2,\ldots,
\end{equation}

\noindent  where  $\fb_{\alpha, lm}^{(t,n)}(\omega)$  and  $\fb_{\alpha,
lm}^{(r,n)}(\omega)$  are the components of the harmonic  $\fb_{\alpha,
lm}^{(n)}(\omega)$  associated, respectively, with the translational motion of
sphere  $\alpha$  and its rotation and $\fb_{\alpha, lm}^{(p,n)}(\omega)$  is
the harmonic of the potential component of the induced surface force density.
In what follows, we show that the last quantity has no effect on the fluid
velocity and the forces and torques exerted by the fluid on the spheres.

In the theory of hydrodynamic interactions between spheres,
the main problem is to derive the required quantities up to the terms of a given
order in certain small parameters (in the general nonstationary case, these are the
sets of parameters $\sigma$  and  $b$).  Using the estimates (2.89), it can be shown that,
for the $n\/$th iteration, the main
terms of the power expansions of the quantities $\fb_{\alpha,
lm}^{(t,n)}(\omega)$ and $\fb_{\alpha, lm}^{(r,n)}(\omega)$  in $\sigma$  are
proportional to $\sigma^{l + n}$ and $\sigma^{l + n + 1}$, respectively.  This
enables us to formulate the following procedure for determination of the
induced velocity and pressure of the fluid (as well well the forces and torques
exerted by the fluid on the spheres) up to terms proportional to $\sigma^p$,
where $p$ is a certain positive integer:

\noindent  (i) to carry out  $p$  iterations;

\noindent  (ii) for each  $s\/$th iteration, $s \leq p$,  to keep only
harmonics with  $l = 0, 1,\ldots,p - s$;

\noindent  (iii) for each keeping harmonic, to retain all terms of power
series up to terms proportional to  $\sigma^p$ inclusive.

\noindent  For the fluid velocity and pressure, further simplification of the
approximate results is possible depending on the choice of a point of
observation.

In the present paper, we restrict ourselves to the consideration of only the
first two iterations. For this, it is sufficient to retain the terms of at most
the second [for $\fb_{\alpha}^{(t)}(\omega)$] and the third [for
$\fb_{\alpha}^{(r)}(\omega)$] orders in  $\sigma$, respectively.

Using Eqs.~(2.8), (2.9), (2.21), (2.22), (2.27), (2.65), (2.66), (2.92), and
(2.93), we represent the total force  $\Fb_\alpha^{tot}(\omega)$ and torque
$\Tb_\alpha^{tot}(\omega)$  acting on sphere  $\alpha$  as follows:

\begin{eqnarray}
 \Fb_\alpha^{tot}(\omega) &=& \Fb_\alpha^{ext}(\omega)
  - \tilde{\Fb}_\alpha^{ext}(\omega) + \Fbc_\alpha(\omega),  \\
 \Tb_\alpha^{tot}(\omega) &=& \Tb_\alpha^{ext}(\omega)
  - \tilde{\Tb}_\alpha^{ext}(\omega) + \Tbc_\alpha(\omega),
\end{eqnarray}

\noindent  where

\begin{eqnarray}
 \Fbc_\alpha(\omega) &=& \Fb_\alpha^{(t)}(\omega)+\Fb_\alpha^{(r)}(\omega), \\
 \Tbc_\alpha(\omega) &=& \Tb_\alpha^{(t)}(\omega)+\Tb_\alpha^{(r)}(\omega)
\end{eqnarray}

\noindent  are the force and torque exerted by the fluid on sphere $\alpha$ due
to the translational motion and rotation of all spheres,

\begin{eqnarray}
 \Fb_\alpha^{(t)}(\omega) &=& \Fb_\alpha^{in}(\omega)
  + \sum\limits_{n = 0}^\infty \Fb_\alpha^{(t,n)}(\omega), \\
 \Fb_\alpha^{in}(\omega) &=& -i\omega \tilde{m_\alpha} \Ub_\alpha (\omega), \\
 \Fb_\alpha^{(r)}(\omega) &=&  \sum\limits_{n = 0}^\infty
  \Fb_\alpha^{(r,n)}(\omega),\\
 \Tb_\alpha^{(t)}(\omega) &=&  \sum\limits_{n = 0}^\infty
  \Tb_\alpha^{(t,n)}(\omega),\\
 \Tb_\alpha^{(r)}(\omega) &=&  \sum\limits_{n = 0}^\infty
  \Tb_\alpha^{(r,n)}(\omega),
\end{eqnarray}

\noindent  and $\Fb_\alpha^{(t,n)}(\omega)$,  $\Fb_\alpha^{(r,n)}(\omega)$,
$\Tb_\alpha^{(t,n)}(\omega)$,  and  $\Tb_\alpha^{(r,n)}(\omega)$ are the forces
and torques caused by the translational motion [superscripts $(t,n)$] and
rotation [superscripts $(r,n)$] of the spheres corresponding to the {\it n\/}th
iteration defined as follows:

\begin{eqnarray}
 \Fb_\alpha^{(\zeta,n)}(\omega) &=& -\fb_{\alpha, 00}^{(\zeta,n)}(\omega),
  \qquad  \qquad  \qquad  \qquad  \qquad  \zeta = t, r,  \\
 \Tb_\alpha^{(\zeta,n)}(\omega) &=& -\frac{a_\alpha}{\sqrt{3}} \sum\limits_{m = -1}^1
  \left( \eb_m \times \fb_{\alpha, 1m}^{(\zeta,n)}(\omega) \right),
  \qquad  \zeta = t, r.
\end{eqnarray}

The similar representations take place also for the induced velocity and pressure of
the fluid.

%%-------------------------------------Section 3------------------------------

\section{Noninteracting Spheres}  \label{n = 0}

In the approximation of noninteracting spheres (zero iteration), the infinite
system of equations (2.67) is splitted into the collection of independent systems
of equations for each sphere  $\alpha$,

\begin{eqnarray}
 \sum\limits_{l_2 m_2} \Tb_{\alpha, l_1 m_1}^{\alpha, l_2 m_2}(\omega)
  \cdot \fb_{\alpha, l_2 m_2}^{(0)}(\omega)
  - \delta_{l_1, 0}\, \delta_{m_1, 0}\, \Ub_\alpha(\omega)
  &-& \delta_{l_1, 1}\, \frac{a_\alpha}{\sqrt{3}}\, \biggl( 1 -
  \frac{b_\alpha^2}{15} \biggr )\,
  \Bigl(\Omb_\alpha(\omega)\times\eb_{m_1}^*\Bigr) = 0, \nonumber  \\
  && \qquad \qquad \qquad \qquad \alpha = 1,2,\ldots,N.
\end{eqnarray}
Note that, for each  $\alpha$, we still have the infinite system of linear
algebraic equations in the unknown quantities $\fb_{\alpha, lm}^{(0)}(\omega)$
(instead of a collection of systems for each value of  $l$  in the stationary
case \cite{ref.Usenko}).

The system (3.1) may be solved together with Eqs.~(2.77), which are rewritten
in the approximation of noninteracting spheres as
\begin{equation}
 Y_\alpha^{(0)}(\omega) = - i \omega \rho \, 4\pi \sqrt{3} \, a_\alpha^2
  \Bigl( \Rb_\alpha \cdot \Ub_\alpha(\omega) \Bigr ),
  \quad  \alpha = 1,2, \ldots,N,
\end{equation}

\noindent  where

\begin{equation}
 Y_\alpha^{(0)}(\omega) \equiv \sum\limits_{m = -1}^1 \eb_m \cdot
  \fb_{\alpha, 1m}^{(0)}(\omega), \qquad \qquad \  \alpha = 1,2, \ldots,N.
\end{equation}
Physically, the presence of the non-zero term on the right-hand side of this
equation corresponds to the appearance of the potential component of the
induced surface forces even in the approximation of noninteracting spheres (at
the expense of an account for the time dependence of the problem).

In the general case of interacting spheres, the conditions of the absence of
the potential components of the induced surface forces have the form

\begin{equation}
 Y_\alpha(\omega) = 0,  \qquad \qquad   \alpha = 1,2, \ldots,N.
\end{equation}

Solving the systems of equations (3.1) and (3.2) up to terms of $b_\alpha^3$,
we get

\begin{eqnarray}
 \fb_{\alpha, lm}^{(t,0)}(\omega) &=&
  \delta_{l,0} \, \fb_{\alpha, 00}^{(t,0)}(\omega) +
  \delta_{l,2} \, \fb_{\alpha, 2m}^{(t,0)}(\omega), \\
 \fb_{\alpha, lm}^{(r,0)}(\omega) &=&
  \delta_{l,1} \, \fb_{\alpha, 1m}^{(r,0)}(\omega) = \delta_{l,1} \,
  \frac{\sqrt{3}}{2 a_\alpha} \, \xi_\alpha^r(\omega)
  \Bigl(\Omb_\alpha(\omega)\times\eb_m^*\Bigr), \\
 \fb_{\alpha, 1m}^{(p,0)}(\omega) &=& \eb_m^* \frac{Y_\alpha^{(0)}(\omega)}{3},
  \qquad  m = 0, \pm 1,
 \end{eqnarray}

\noindent  where

\begin{eqnarray}
 \fb_{\alpha, 00}^{(t,0)}(\omega) &=& \xi_\alpha p(b_\alpha)\Ub_\alpha(\omega),
 \\
 \fb_{\alpha, 2m}^{(t,0)}(\omega) &=& \frac{b_\alpha^2 \xi_\alpha}{\sqrt{30}}\,
  \Kb_m \cdot \Ub_\alpha(\omega),  \qquad  m = 0, \pm 1, \pm 2,  \\
 p(b) &=& 1 + b + \frac{b^2}{3}, \\
 \xi_\alpha^r(\omega) &=& \xi_\alpha^r s^r(b_\alpha),
\end{eqnarray}
\noindent $\xi_\alpha^r$ being the Stokes stationary friction coefficient for a
sphere rotating in a viscous fluid ($\xi_\alpha^r = 8 \pi \eta a_\alpha^3$),
and

\begin{equation}
 s^r(b_\alpha) = 1 + \frac{b_\alpha^2}{3}\left( 1 - b_\alpha\right )
\end{equation}

\noindent  is the corresponding frequency factor. In deriving (3.9), the use of
the tensor  $\Kb_m \equiv 2 \sqrt{30 \pi} \, \Kb_{2m,00}^{00}$,  where  $m = 0,
\pm 1, \pm 2$ (see Appendix A), and the relations

\begin{equation}
 \sum\limits_{m_2 = -2}^{2}\tilde{\Kb}_{2 m_1, 00}^{2 m_2} \cdot \Kb_{2 m_2,00}^{00}
  = 6\sqrt{\pi} \, \Kb_{2 m_1,00}^{00}
\end{equation}
have been made.

Note that in the approximation of noninteracting spheres, the expression for
the harmonic $\fb_{\alpha, 1m}^{(r,0)}(\omega)$  given in \cite{ref.Pien2} is
proportional to the tensor  $\tilde{\Kb}_{1m_1,00}^{1m_2}$  that is inverse to
the singular tensor  $\Kb_{1m_1,00}^{1m_2}$,  whereas according to (3.6), the
harmonic $\fb_{\alpha, 1m}^{(r,0)}(\omega)$ contains no singularities.

Substituting the time-dependent solution (3.5)--(3.7) into Eqs.~(2.41), (2.42),
(2.44), and (2.45) and using (3.13), we obtain the following relations for the
harmonics of the fluid velocity and pressure induced by the surface forces in
the approximation of noninteracting spheres:

\begin{eqnarray}
 \vb_{\alpha,l_1 m_1}^{(S,t,0)ind}(\Rb_\alpha,r_\alpha,\omega) &=&
  \frac{a_\alpha}{r_\alpha} \, p(b_\alpha)\biggl \{\delta_{l_1,0} \,\delta_{m_1,0} \,
  \tilde{j}_0(b_\alpha) \exp(-x_\alpha)\Ub_\alpha(\omega)  \nonumber \\
  &+& \delta_{l_1,2}\, \frac{3}{2} \, \sqrt{\frac{3}{10}} \frac{1}{x_\alpha^2} \biggl \{1 -
  \frac{2x_\alpha^3}{3\pi}\, \tilde{h}_2(x_\alpha)\tilde{j}_0(b_\alpha)\biggr \}
  \Kb_{m_1} \cdot \Ub_\alpha(\omega) \biggr \},  \\
 \vb_{\alpha,l_1 m_1}^{(S,r,0)ind}(\Rb_\alpha,r_\alpha,\omega) &=&
  \delta_{l_1,1} \, \frac{\sqrt{3}}{8 \pi a_\alpha} \, \xi_\alpha^r(\omega)
  F_{1,1}(r_\alpha,a_\alpha,\omega)
  \left( \Omb_\alpha(\omega) \times \eb_{m_1}^* \right),  \\
 \vb_{\beta,l_1 m_1}^{(S,t,0)ind}(\Rb_\alpha,r_\alpha,\omega) &=&
  \xi_\beta p(b_\beta)
  \Tb_{l_1 m_1}^{00}(r_\alpha,a_\beta,\Rb_{\alpha\beta},\omega) \cdot
  \Ub_\beta(\omega),  \qquad \beta \ne \alpha, \\
 \vb_{\beta,l_1 m_1}^{(S,r,0)ind}(\Rb_\alpha,r_\alpha,\omega) &=&
  \frac{3}{2 a_\beta} \, \xi_\beta^r(\omega) \sum\limits_{l = l_1 \pm 1 \ge 1}\,\,
  \sum\limits_{m = -l}^l
  F_{l_1 1,l}(r_\alpha,a_\alpha,R_{\alpha\beta},\omega)
  \left( \Omb_\beta(\omega) \times \Wb_{l_1 m_1,lm}^{00} \right) \nonumber  \\
  &\times & Y_{lm}(\Theta_{\alpha\beta},\Phi_{\alpha\beta}),
  \qquad \qquad \qquad  \beta \ne \alpha,  \\
 \vb_{\beta,l_1 m_1}^{(S,p,0)ind}(\Rb_\alpha,r_\alpha,\omega) &=& 0,
  \qquad  \beta = 1,2,\ldots,N,  \\
 p_{\alpha,l_1 m_1}^{(t,0)ind}(\Rb_\alpha,r_\alpha,\omega) &=&
  \delta_{l_1,1} \,\frac{\sqrt{3}}{2} \eta \frac{a_\alpha}{r_\alpha^2}\,
  \Biggl \{ p(b_\alpha) - \frac{1}{2} \left( 1 + b_\alpha + b_\alpha^2 \right )\,
  \delta_{r_\alpha,a_\alpha} \Biggr \}
  \biggl ( \eb_{m_1}^* \cdot \Ub_\alpha (\omega) \biggr ),  \\
 p_{\beta,l_1 m_1}^{(t,0)ind}(\Rb_\alpha,r_\alpha,\omega) &=& \xi_\beta
  p(b_\beta) \Db_{l_1 m_1}^{00}(r_\alpha,a_\beta,\Rb_{\alpha\beta}) \cdot
  \Ub_\beta(\omega),  \qquad  \beta \neq \alpha, \\
 p_{\beta,l_1 m_1}^{(r,0)ind}(\Rb_\alpha,r_\alpha,\omega) &=& 0,
  \qquad  \beta = 1,2,\ldots,N,  \\
 p_{\beta,l_1 m_1}^{(p,0)ind}(\Rb_\alpha,r_\alpha,\omega) &=& \delta_{\beta,\alpha} \,
  \delta_{l_1, 0} \, \delta_{r_\alpha,a_\alpha} \, \frac{i}{2} \, \omega \rho \,
  \Bigl( \Rb_\alpha \cdot \Ub_\alpha(\omega) \Bigr ),
  \qquad  \beta = 1,2,\ldots,N.
\end{eqnarray}

Note that, by virtue of (3.21), the rotation of the spheres has
no effect on the fluid pressure,

\begin{equation}
 p_\beta^{(r,0)ind}(\rb,\omega) = 0,  \qquad  \beta = 1,2,\ldots,N.
\end{equation}

In view of (3.7), the potential component
$\fb_\alpha^{(p,0)}(\ab_\alpha,\omega)$  of the induced surface force density
$\fb_\alpha^{(0)}(\ab_\alpha,\omega)$  in the approximation of noninteracting
spheres has the form

\begin{equation}
 \fb_\alpha^{(p,0)}(\ab_\alpha,\omega) = \frac{\ab_\alpha}{a_\alpha}
  \frac{Y_\alpha^{(0)}(\omega)}{4 \pi \sqrt{3}},
\end{equation}

\noindent  and, hence, has only the radial component. As a result, the potential component of
the induced surface force has no effect
on the forces and torques exerted by the fluid on the spheres immersed in it.
According to (3.18), the potential component
$\fb_\beta^{(p,0)}(\ab_\beta,\omega)$, where $\beta = 1,2,\ldots N$, of the
induced surface force density makes no contribution to the fluid velocity.

These conclusions concerning the influence of the potential component of the
induced surface force densities on the fluid velocity and on the forces and
torques exerted by the fluid on spheres also remain valid for any $n$\/th
iteration if the potential component $\fb_\alpha^{(p,n)}(\ab_\alpha,\omega)$ of
the induced surface force density $\fb_\alpha^{(n)}(\ab_\alpha,\omega)$
corresponding to this iteration has only harmonics with $l = 1$  defined by a
relation similar to (3.7).

According to relations (3.19) and (3.20), for the fluid pressure for $r_\alpha
> a_\alpha$,  the account of the time dependence leads only to the appearance of
the frequency-dependent factors  $p(b_\beta)$ equal to 1 in the stationary case
and the substitution of  $\Ub_\beta(\omega)$  for  $\Ub_\beta$.

Taking relations (2.87) and (3.15) into account, we obtain the following
expression for the fluid velocity:

\begin{eqnarray}
 \vb_\alpha^{(r,0)ind}(\Rb_\alpha + \rb_\alpha,\omega) &=&
  \vb_\alpha^{(S,r,0)ind}(\Rb_\alpha + \rb_\alpha,\omega) +
  \vb_\alpha^{(V)ind}(\Rb_\alpha + \rb_\alpha,\omega) \nonumber  \\
  &=& \biggl(\frac{a_\alpha}{r_\alpha} \biggr)^3\, q(b_\alpha) \psi(x_\alpha)
  \, \Bigl( \Omb_\alpha(\omega) \times \rb_\alpha \Bigr),
\end{eqnarray}

\noindent  where

\begin{eqnarray}
 q(b) &=&1 + \frac{b^2}{2} - \frac{b^3}{3},  \\
 \psi(x) &=& \frac{2}{\pi} \, x^2 \tilde{h}_1(x),
\end{eqnarray}

\noindent at an arbitrary point  $\rb = \Rb_\alpha + \rb_\alpha$  induced due
to the rotation of sphere  $\alpha$,  which agrees with well-known result
\cite{ref.Landau} for the fluid velocity induced by a rotating sphere for
$|b_\alpha| \ll 1$.

Using relation (3.19), we obtain the following expression for the fluid
pressure for  $r_\alpha > a_\alpha$  caused by the translational motion of
sphere $\alpha$:

\begin{equation}
 p_\alpha^{(t,0)ind}(\Rb_\alpha + \rb_\alpha,\omega) = \frac{3\eta}{2}
  \frac{a}{r_\alpha^2} \, p(b_\alpha) \, \Bigl(
  \nb_\alpha \cdot \Ub_\alpha(\omega) \Bigr),
\end{equation}

\noindent  which differs from the corresponding relation for the stationary
case \cite{ref.Landau} only by the frequency-dependent factor  $p(b_\alpha)$.

If the point of observation is far from sphere  $\alpha$  ($\Real x_\alpha \gg
1$), then using relations (3.14) and (3.25) and retaining the leading terms, we
obtain the following expressions for the fluid velocity induced by the motion
and rotation of sphere  $\alpha$:

\begin{eqnarray}
 \vb_\alpha^{(S,t,0)ind}(\Rb_\alpha + \rb_\alpha,\omega) & \approx & \frac{3}{2}
  \frac{a_\alpha}{r_\alpha} \frac{p(b_\alpha)}{x_\alpha^2}
  \left( 3 \nb_\alpha \nb_\alpha - \Ib \right) \cdot \Ub_\alpha(\omega), \\
 \vb_\alpha^{(r,0)ind}(\Rb_\alpha + \rb_\alpha,\omega) & \approx &
  \biggl(\frac{a_\alpha}{r_\alpha} \biggr)^2 b_\alpha \exp(-x_\alpha)
  \Bigl ( \Omb_\alpha(\omega) \times \rb_\alpha \Bigr ).
\end{eqnarray}

\noindent  Thus, the account of the time dependence in the Navier--Stokes
equation leads to the essential change in the asymptotics of the fluid velocity
in the far zone \cite{ref.Landau,ref.Schram}.  [Of course, here and below, it
is assumed that $\rb_\alpha$ is such that the linearized Navier--Stokes
equation (2.1) can be used. For details, see Appendix B.]

Next, using relations (3.16) and (3.20), we obtain the following expressions
for the fluid velocity and pressure induced in the far zone ($r_\alpha \gg
R_{\alpha\beta}$) due to the motion of sphere  $\beta \neq \alpha$ with the
velocity  $\Ub_\beta(\omega)$  in the zero approximation with respect to the
small ratio $R_{\alpha\beta}/r_\alpha$:

\begin{eqnarray}
 \vb_\beta^{(S,t,0)ind}(\Rb_\alpha + \rb_\alpha,\omega) & \approx & \frac{3}{2}
  \frac{a_\beta}{r_\alpha} \frac{p(b_\beta)}{x_\alpha^2}
  \left( 3 \nb_\alpha \nb_\alpha - \Ib \right) \cdot \Ub_\beta (\omega),  \\
 p_\beta^{(t,0)ind}(\Rb_\alpha + \rb_\alpha,\omega) & \approx & \frac{3 \eta}{2}
  \, \frac{a_\beta}{r_\alpha^2} \, p(b_\beta) \, \Bigl(
  \nb_\alpha \cdot \Ub_\beta(\omega) \Bigr) + {\cal O}(\sigma_{\beta\alpha}^2).
\end{eqnarray}

The use of Eqs.~(3.28), (3.29), (3.31), and (3.32) results in the following
space distributions of the fluid velocity and pressure induced by the motion of
$N$ spheres far from them in the zero approximation with respect to the ratio
$R_{\alpha\beta}/r_\alpha$:

\begin{eqnarray}
 \vb^{(S,t,0)ind}(\Rb_\alpha + \rb_\alpha,\omega) & \approx &
  \frac{3}{2 r_\alpha x_\alpha^2}
  \left( 3 \nb_\alpha \nb_\alpha - \Ib \right) \cdot
  \sum\limits_{\beta = 1}^N a_\beta p(b_\beta)\Ub_\beta (\omega),  \\
  p^{(t,0)ind}(\Rb_\alpha + \rb_\alpha,\omega) & \approx &
  \frac{3 \eta}{2 r_\alpha^2} \, \sum\limits_{\beta = 1}^N a_\beta p(b_\beta)
  \Bigl( \nb_\alpha \cdot \Ub_\beta (\omega) \Bigr).
\end{eqnarray}

In the case of equal spheres  $a_\beta \equiv a$  ($b_\beta \equiv b$), these
relations are simplified to the form

\begin{eqnarray}
 \vb^{(S,t,0)ind}(\Rb_\alpha + \rb_\alpha,\omega) & \approx &
  \frac{3}{2 r_\alpha x_\alpha^2} \, a p(b) \,
  \left( 3 \nb_\alpha \nb_\alpha - \Ib \right) \cdot \Ub^{tot}(\omega),  \\
 p^{(t,0)ind}(\Rb_\alpha + \rb_\alpha,\omega) & \approx & \frac{3}{2} \,
  \frac{\eta}{r_\alpha^2} \, a p(b) \,
  \Bigl( \nb_\alpha \cdot \Ub^{tot}(\omega) \Bigr),
\end{eqnarray}

\noindent  where

\begin{equation}
 \Ub^{tot}(\omega) = \sum\limits_{\beta = 1}^N \Ub_\beta(\omega).
\end{equation}

Hence, the velocity and pressure fields of the fluid caused by the
translational motion of equal spheres in the far zone, in the zero
approximation with respect to the ratio of the typical distance between two
spheres to the distance to the point of observation, can be represented as the
velocity and pressure fields of the fluid induced by a single sphere of radius
$a$  moving in the fluid with the translational velocity $\Ub^{tot}(t)$.

If all spheres move with the same translational velocity  $\Ub_\beta(\omega) =
\Ub_0(\omega)$, we get

\begin{eqnarray}
 \vb^{(S,t,0)ind}(\Rb_\alpha + \rb_\alpha,\omega) & \approx &
  \frac{3}{2} \frac{a_{eff}}{r_\alpha x_\alpha^2}
  \left(3\nb_\alpha \nb_\alpha - \Ib \right) \cdot \Ub_0(\omega),  \\
 p^{(t,0)ind}(\Rb_\alpha + \rb_\alpha,\omega)) & \approx & \frac{3\eta}{2}
  \frac{a_{eff}}{r_\alpha^2} \, \Bigl( \nb_\alpha \cdot \Ub_0(\omega) \Bigr),
\end{eqnarray}

\noindent  where  $a_{eff} = \sum\limits_{\beta = 1}^N a_\beta p(b_\beta)$.
Thus, in the far zone, the action of the system of spheres moving with the
equal velocities is equivalent to the action of a single sphere of radius
$a_{eff}$.

Using (2.103), (3.5), (3.6), (3.8), and (3.9), in the approximation of
noninteracting spheres, we obtain the following relations for the Fourier
components of the force exerted by the fluid on sphere  $\alpha$ caused by
translational and rotational motions of the spheres:

\begin{eqnarray}
 \Fb_\alpha^{(t,0)}(\omega) &=& -\xi_\alpha p(b_\alpha) \Ub_\alpha(\omega),  \\
 \Fb_\alpha^{(r,0)}(\omega) &=& 0.
\end{eqnarray}

If we restrict our consideration to the approximation of noninteracting
spheres, then, using relations (2.98), (2.99), and (3.40), we obtain the
following expression for the force acting by the fluid on sphere $\alpha$
moving in it with the velocity  $\Ub_\alpha (\omega)$

\begin{equation}
 \Fbc_\alpha(\omega) = -\xi_\alpha(\omega) \Ub_\alpha(\omega),  \\
\end{equation}

\noindent where

\begin{equation}
 \xi_\alpha(\omega) = \xi_\alpha s^t(b_\alpha)
\end{equation}

\noindent  is the Stokes frequency-dependent friction coefficient for a sphere
of radius  $a_\alpha$  moving with the velocity  $\Ub_\alpha(\omega)$  and

\begin{equation}
 s^t(b_\alpha) = 1 + b_\alpha + \frac{b_\alpha^2}{9}
\end{equation}

\noindent  is the frequency factor corresponding to the translational motion.

Equation (3.42) defines nothing but the celebrated Boussinesq formula
\cite{ref.Landau,ref.Schram} for the force exerted by the fluid being at rest
at infinity on a single sphere moving with time-dependent velocity. Indeed, returning to
the time variable, we immediately obtain
\begin{equation}
\Fbc_\alpha(t) = -\xi_\alpha \Ub_\alpha(t) -\frac{\tilde{m}_\alpha}{2}
 \frac{d\Ub_\alpha(t)}{dt} - 6 \sqrt{\pi\eta\rho} \, a_\alpha^2
 \int\limits_{-\infty}^t \, \frac{d\tau}{\sqrt{t - \tau}}
 \frac{d\Ub_\alpha(\tau)}{d\tau}.
\end{equation}

To determine the spectral component of the torque exerted by the fluid on
sphere $\alpha$,  we use relations (2.104), (3.5), and (3.6) and obtain the
well-known results \cite{ref.Saarl}

\begin{eqnarray}
 \Tb_\alpha^{(t,0)}(\omega) &=& 0,  \\
 \Tb_\alpha^{(r,0)}(\omega) &=& -\xi_\alpha^r(\omega) \Omb_\alpha(\omega).
\end{eqnarray}

\noindent  In the time representation, this reads

\begin{equation}
 \Tbc_\alpha(t) = -\xi_\alpha^r \Omb_\alpha(t)
  - 5 \tilde{I}_\alpha \frac{d\Omb_\alpha(t)}{dt}
  - \frac{5}{2} \frac{\tilde{I}_\alpha a_\alpha}{\sqrt{\pi \nu}} \,
  \int\limits_{-\infty}^t \, \frac{d\tau}{(t - \tau)^{3/2}}
  \frac{d\Omb_\alpha(\tau)}{d\tau},
\end{equation}

\noindent  where  $\tilde{I}_\alpha = (2/5) \tilde{m}_\alpha a_\alpha^2$  is
the moment of inertia of a fluid sphere of radius  $a_\alpha$.  Analogously to
the Boussinesq force (3.45), torque (3.48) exerted by the fluid on a sphere
depends on the angular velocity of the sphere, its acceleration, and the
history of rotation of the sphere. Furthermore, the second term in (3.48) is
independent of the fluid viscosity.

Finally, using the relations (2.5), (2.6), (2.94), and (2.95), we can represent the equations of
motion of sphere  $\alpha$ as follows:

\begin{eqnarray}
 \Bigl \{ -i \omega m_\alpha^{eff} + \xi_\alpha (1 + b_\alpha) \Bigr \}
  \Ub_\alpha (\omega) &=& \Fb_\alpha^{ext}(\omega)
  - \tilde{\Fb}_\alpha^{ext}(\omega),  \\
 \biggl \{ -i \omega \Bigl \{ I_\alpha + 5 (1 - b_\alpha)\tilde{I}_\alpha\Bigr \}
  + \xi_\alpha^r \biggr \} \Omb_\alpha (\omega) &=&
  \Tb_\alpha^{ext}(\omega) - \tilde{\Tb}_\alpha^{ext}(\omega),
\end{eqnarray}

\noindent  where  $m_\alpha^{eff} = m_\alpha + \tilde{m}_\alpha/2$  is the
effective mass of sphere  $\alpha$  \cite{ref.Landau,ref.Schram}.

%%-----------------------------------==Section 4------------------------------

\section{The First Iteration}  \label{n = 1}

For the first iteration, the system of equations (2.67), (2.77) is reduced to
the form

\begin{equation}
 \sum\limits_{l_2 m_2} \Tb_{\alpha, l_1 m_1}^{\alpha, l_2 m_2}(\omega)
  \cdot \fb_{\alpha, l_2 m_2}^{(1)}(\omega) = -\sum\limits_{\beta \neq \alpha}
  \biggl \{ \sum\limits_{l_2 m_2} \Tb_{\alpha, l_1 m_1}^{\beta, l_2 m_2}(\omega)
  \cdot \fb_{\beta, l_2 m_2}^{(0)}(\omega)
  + \vb_{\alpha, l_1 m_1}^{\beta (V)}(\omega) \biggr \},
\end{equation}

\begin{equation}
 Y_\alpha^{(1)}(\omega) \equiv \sum\limits_{m = -1}^1
  \eb_m \cdot \fb_{\alpha, 1m}^{(1)}(\omega)
  = 4\pi \sqrt{3} \, a_\alpha^2 \sum\limits_{\beta \neq \alpha}
  p_\beta^{(0)ind}(\Rb_\alpha,\omega), \quad  \alpha = 1,2, \ldots,N,
\end{equation}

\noindent  where  $p_\beta^{(0)ind}(\Rb_\alpha,\omega)$  is defined by relation
(2.79) with  $\fb_{\beta,l_2 m_2}(\omega) \rightarrow \fb_{\beta,l_2
m_2}^{(0)}(\omega)$,  which gives

\begin{equation}
 Y_\alpha^{(1)}(\omega) = \sqrt{3} \, \sum\limits_{\beta \neq \alpha}
  \xi_\beta p(b_\beta) \sigma_{\alpha\beta}^2 \Bigl(
  \nb_{\alpha\beta} \cdot \Ub_\beta(\omega) \Bigr), \quad  \alpha = 1,2,
  \ldots,N.
\end{equation}

As before, the necessity in Eq.~(4.3) is caused by (2.76). Solving Eqs.~(4.1)
and (4.2) up to the terms of the order $b^3$ and $\sigma^2$  [for
$\fb_{\alpha,l_2 m_2}^{(t,1)}(\omega)$] and $\sigma^3$ [for $\fb_{\alpha,l_2
m_2}^{(r,1)}(\omega)$] and using Eq.~(3.13), we obtain

\begin{eqnarray}
 \fb_{\alpha, lm}^{(\zeta,1)}(\omega) &=&
  \delta_{l,0} \, \fb_{\alpha, 00}^{(\zeta,1)}(\omega) +
  \delta_{l,1} \, \fb_{\alpha, 1m}^{(\zeta,1)}(\omega) +
  \delta_{l,2} \, \fb_{\alpha, 2m}^{(\zeta,1)}(\omega) +
  \delta_{l,3} \, \fb_{\alpha, 3m}^{(\zeta,1)}(\omega), \qquad \zeta = t, r, \\
 \fb_{\alpha, 1m}^{(p,1)}(\omega) &=& \eb_m^* \, \frac{Y_\alpha^{(1)}(\omega)}{3},
  \qquad \qquad  m = 0, \pm 1.
 \end{eqnarray}

\noindent  Here,

\begin{eqnarray}
 \fb_{\alpha,00}^{(t,1)}(\omega) &=& -\xi_\alpha p(b_\alpha)
  \sum\limits_{\beta \neq \alpha} \xi_\beta p(b_\beta)
  \Tb_M(\Rb_{\alpha\beta},\omega) \cdot \Ub_\beta(\omega) + {\cal O}(\sigma^3),  \\
 \fb_{\alpha,00}^{(r,1)}(\omega) &=& \xi_\alpha \tau(b_\alpha)
  \sum\limits_{\beta \neq \alpha} a_\beta \sigma_{\beta\alpha}^2 q(b_\beta)
  \psi(y_{\alpha\beta}) \Bigl( \nb_{\alpha\beta} \times \Omb_\beta(\omega) \Bigr)
  + {\cal O}(\sigma^4),  \\
\fb_{\alpha,1m}^{(\zeta,1)} &=& -\frac{2}{3} \, \xi_\alpha \left \{
  4\bb_{\alpha,m}^{(\zeta,1)}(\omega)+\left( \bb_{\alpha,m}^{(\zeta,1)}(\omega)\right)^T
  \right \},  \qquad  \qquad \zeta = t, r,  \qquad  m = 0, \pm 1, \\
\fb_{\alpha,2m}^{(t,1)}(\omega) &=&
  -\frac{b_\alpha^2 \xi_\alpha}{\sqrt{30}} \, \Kb_m \cdot
  \sum\limits_{\beta \neq \alpha} (1 + b_\beta) \xi_\beta
  \Tb_M(\Rb_{\alpha\beta},\omega) \cdot \Ub_\beta(\omega),
  \qquad  m = 0, \pm 1, \pm 2, \\
 \fb_{\alpha,2m}^{(r,1)}(\omega) &=&
  \frac{b_\alpha^2 \xi_\alpha}{\sqrt{30}} \, \Kb_m
  \cdot \sum\limits_{\beta \neq \alpha} a_\beta \sigma_{\beta\alpha}^2
  \psi(y_{\alpha\beta}) \Bigl( \nb_{\alpha\beta} \times \Omb_\beta(\omega)
  \Bigr),  \qquad  m = 0, \pm 1, \pm 2,  \\
 \fb_{\alpha,3m}^{(t,1)}(\omega) &=&
  - \frac{b_\alpha^2 \xi_\alpha}{15 \sqrt{21}} \, \sum\limits_{m_1 = -3}^3 \,
  \Nb_{m m_1} \cdot \biggl \{4\cb_{\alpha,m_1}^{(t,1)}(\omega)
  + \Bigl( \cb_{\alpha,m_1}^{(t,1)}(\omega)\Bigr)^T  \biggr \}, \nonumber  \\
  && \qquad \qquad \qquad \qquad \qquad \qquad \qquad \qquad \qquad \qquad
  m = 0, \pm 1, \pm 2, \pm 3, \\
 \fb_{\alpha,3m}^{(r,1)}(\omega) &=&
  - \frac{b_\alpha^2 \xi_\alpha}{15 \sqrt{21}} \, \sum\limits_{m_1 = -3}^3 \,
  \Nb_{m m_1}  \cdot \biggl \{4\bb_{\alpha,m_1}^{(r,1,0)}(\omega)
  + \left( \bb_{\alpha,m_1}^{(r,1,0)}(\omega)\right)^T  \biggr \}, \nonumber  \\
  && \qquad \qquad \qquad \qquad \qquad \qquad \qquad \qquad \qquad \qquad
  m = 0, \pm 1, \pm 2, \pm 3, \\
 \bb_{\alpha,m}^{(t,1)}(\omega) &=& \bb_{\alpha,m}^{(t,1,0)}(\omega) +
  \bb_{\alpha,m}^{(t,1,1)}(\omega) + \bb_{\alpha,m}^{(t,1,2)}(\omega),
  \qquad \qquad \quad \,  m = 0, \pm 1,  \\
 \bb_{\alpha,m}^{(r,1)}(\omega) &=& \bb_{\alpha,m}^{(r,1,0)}(\omega) +
  \bb_{\alpha,m}^{(r,1,1)}(\omega),
  \qquad \qquad  \qquad \qquad \qquad \  m = 0, \pm 1,  \\
 \bb_{\alpha,m}^{(\zeta,1,0)}(\omega) &=& \sum\limits_{\beta \neq \alpha}
  \bb_{\alpha,m}^{\beta(\zeta,1,0)}(\omega),
  \qquad \qquad \qquad \zeta = t, r, \qquad \qquad  m = 0, \pm 1, \\
 \bb_{\alpha,m}^{\beta(t,1,0)}(\omega) &=& \xi_\beta
  \Tb_{\alpha,1m}^{\beta,00}(\omega) \cdot \Ub_\beta(\omega),
  \qquad \qquad \qquad \qquad \qquad \quad  m = 0, \pm 1,\\
 \bb_{\alpha,m_1}^{\beta (r,1,0)}(\omega) &=& 2 \xi_\beta a_\beta \sum\limits_{l = 0, 2}
  \sum\limits_{m = -l}^l F_{1,1,l}(a_\alpha,a_\beta,R_{\alpha\beta},\omega)
  \biggl ( \Omb_\beta(\omega) \times \Wb_{1m_1,lm}^{00} \biggr)
  Y_{lm}(\Theta_{\alpha\beta},\Phi_{\alpha\beta}), \nonumber  \\
  && \qquad \qquad \qquad \qquad \qquad \qquad \qquad \qquad \qquad \qquad
  m_1 = 0, \pm 1, \\
 \bb_{\alpha,m}^{(t,1,1)}(\omega) &=& \sum\limits_{\beta \neq \alpha}
  b_\beta \left ( 1 + \frac{b_\beta}{3} \right)
  \bb_{\alpha,m}^{\beta(t,1,0)}(\omega) + b_\alpha^2 \frac{4\sqrt{\pi}}{15} \,
  \left( 1 - \frac{5}{6} \, b_\alpha \right)
  \sum\limits_{m_1 = -1}^1 \Kb_{1m,00}^{1m_1} \cdot \biggl \{
  4\bb_{\alpha,m_1}^{(t,1,0)}(\omega) \nonumber  \\
  &+& \left( \bb_{\alpha,m_1}^{(t,1,0)}(\omega)\right)^T  \biggr \},
  \qquad \qquad \qquad \qquad \qquad \qquad \quad \ m = 0, \pm 1,  \\
 \bb_{\alpha,m}^{(t,1,2)}(\omega) &=& b_\alpha^2 \frac{4\sqrt{\pi}}{15} \,
  \sum\limits_{m_1 = -1}^1 \Kb_{1m,00}^{1m_1} \cdot \biggl \{4\lb_{\alpha,m_1}^{(t,1)}(\omega)
  + \left( \lb_{\alpha,m_1}^{(t,1)}(\omega)\right)^T  \biggr \}, \qquad  m = 0, \pm 1, \\
 \lb_{\alpha,m}^{(t,1)}(\omega) &=& \sum\limits_{\beta \neq \alpha} b_\beta
  \bb_{\alpha,m}^{\beta(t,1,0)}(\omega), \qquad  m = 0, \pm 1,\\
 \bb_{\alpha,m_1}^{(r,1,1)}(\omega) &=& \frac{1}{3}
  \sum\limits_{\beta \neq \alpha} b_\beta^2 \, \biggl \{ (1 - b_\beta)
  \bb_{\alpha,m_1}^{\beta (r,1,0)}(\omega) \nonumber  \\
  &+& 2 \xi_\beta \sum\limits_{l = 0, 2} \sum\limits_{m = -l}^l
  P_{1,2,l}(a_\alpha,a_\beta,R_{\alpha\beta},\omega)
  \biggl ( \Omb_\beta(\omega) \times \Wb_{1m_1,00}^{lm} \biggr)
  Y_{lm}^*(\Theta_{\alpha\beta},\Phi_{\alpha\beta}) \biggr \} \nonumber  \\
  &+& b_\alpha^2 \frac{4\sqrt{\pi}}{15} \, \left( 1 - \frac{5}{6} \, b_\alpha \right)
  \sum\limits_{m_2 = -1}^1 \Kb_{1m_1,00}^{1m_2} \cdot
  \biggl \{4\bb_{\alpha,m_2}^{(r,1,0)}(\omega)
  + \left( \bb_{\alpha,m_2}^{(r,1,0)}(\omega)\right)^T  \biggr \}, \nonumber  \\
  && \qquad \qquad \qquad \qquad \qquad \qquad \qquad \qquad \qquad \qquad \qquad
  m_1 = 0, \pm 1, \\
 \cb_{\alpha,m}^{(t,1)}(\omega) &=& \bb_{\alpha,m}^{(t,1,0)}(\omega)
  + \lb_{\alpha,m}^{(t,1)}(\omega) = \sum\limits_{\beta \neq \alpha} (1 + b_\beta)
  \bb_{\alpha,m}^{\beta(t,1,0)}(\omega),  \qquad \  m = 0, \pm 1,
\end{eqnarray}

\begin{equation}
 \tau(b) = 1 + b + \frac{b^2}{2} + \frac{b^3}{6},
\end{equation}

\begin{eqnarray}
 \Tb_M(\Rb_{\alpha\beta},\omega) & \equiv &
  \Tb_{\alpha,00}^{\beta,00}(\Rb_{\alpha\beta},\omega) = \frac{1}{6\pi}\, \biggl \{
  F_{0,0,0}(a_\alpha,a_\beta,R_{\alpha\beta},\omega) \Ib \nonumber  \\
  &+& \frac{1}{2} \, F_{0,0,2}(a_\alpha,a_\beta,R_{\alpha\beta},\omega)
  \Bigl(3 \nb_{\alpha\beta} \nb_{\alpha\beta} - \Ib\Bigr) \biggr \}
\end{eqnarray}

\noindent  is the modified dynamic Oseen tensor
\cite{ref.Pien1,ref.Pien2,ref.Usenko}. This tensor is symmetric both in the
coordinate indices ($i, j = x,y,z$) and the indices of spheres ($\alpha,
\beta$).  Note that in the case under consideration, only terms corresponding
to the approximation of the problem with respect to the small parameters   $b$
and  $\sigma$  should be retained in  $\Tb_M(\Rb_{\alpha\beta},\omega)$,
namely,

\begin{eqnarray}
 F_{0,0,0}(a_\alpha,a_\beta,R_{\alpha\beta},\omega) & \approx &
  \frac{1}{\eta R_{\alpha\beta}} \left( 1 + \frac{b_\alpha^2 + b_\beta^2}{6}
  \right) \exp(-y_{\alpha\beta}),  \\
 F_{0,0,2}(a_\alpha,a_\beta,R_{\alpha\beta},\omega) & \approx &
  \frac{1}{\eta R_{\alpha\beta} y_{\alpha\beta}^2} \left \{3 -
  \left [ 1 + \frac{b_\alpha^2 + b_\beta^2}{6} + \frac{b_\alpha^4 +
  b_\beta^4}{120}\right ] \frac{2}{\pi} y_{\alpha\beta}^3
  \tilde{h}_2(y_{\alpha\beta}) \right \}.
\end{eqnarray}

In deriving relations (4.11) and (4.12), after the determination of the inverse
tensor  $\tilde{\Kb}_{3m_1, 00}^{3m_2}$, the tensor

\begin{equation}
\Nb_{m_1 m_2} = 2 \, \sqrt{\frac{7}{3}} \sum\limits_{m_3 = -3}^3
 \tilde{\Kb}_{3m_1, 00}^{3m_3}\cdot \Kb_{3m_3, 00}^{1m_2}, \qquad
 m_1 = 0, \pm 1, \pm 2, \pm 3,  \qquad m_2 = 0, \pm 1,
\end{equation}

\noindent  has been introduced. Its components are expressed in terms of the
tensor $\Kb_m$ with  $ m = 0, \pm1, \pm 2$  given by the relations (A2) in
Appendix A as follows:

\begin{eqnarray}
 \Nb_{mm} &=& \biggl \{ \delta_{m,0} \, \sqrt{\frac{3}{2}} + \Bigl( 1 -
  \delta_{m,0} \biggr \} \Kb_0, \qquad  m = 0, \pm 1, \nonumber \\
 \Nb_{m,0} &=& \frac{2}{\sqrt{3}} \, \Kb_m, \  m = \pm 1, \quad
  \Nb_{0,m} =  \frac{1}{\sqrt{2}} \, \Kb_{-m}, \  m = \pm 1, \quad
  \Nb_{m,-m} = \frac{1}{\sqrt{6}} \, \Kb_{\pm 2}, \  m = \pm 1,
  \nonumber  \\
 \Nb_{\pm 2,0} &=& \sqrt{\frac{5}{6}} \, \Kb_{\pm 2}, \quad
  \Nb_{2,1} = \sqrt{\frac{5}{3}} \, \Kb_1, \quad
  \Nb_{-2,-1} = \sqrt{\frac{5}{3}} \, \Kb_{-1}, \quad
  \Nb_{3,1} = \sqrt{\frac{5}{2}} \, \Kb_2, \quad  \nonumber  \\
  \Nb_{-3,-1} &=& \sqrt{\frac{5}{2}} \, \Kb_{-2}, \qquad
  \Nb_{2, -1} = \Nb_{-2,1} = \Nb_{\pm 3,0} = \Nb_{3, -1} = \Nb_{-3, 1} = 0.
\end{eqnarray}

Here and below, the three superscripts in quantities  $b_{\alpha,
lm}^{(\zeta,n,k)}(\omega)$, where $\zeta = t, r$  and  $n, k =0, 1, 2,\ldots,$
mean the following: $\zeta$ shows that the induced surface force densities
are caused by the translational $\zeta = t$) or rotational ($\zeta = r$) motion
of the spheres,  and the subscripts $n$  and  $k$  stand for the corresponding
iterations with respect to the parameters  $\sigma$ and $b$, respectively.

All quantities of the type $\left(\bb_{\alpha, 1m} \right)^T$  are defined as
follows: Consider three vectors  $\bb_m \equiv (b_{m,x},b_{m,y},b_{m,z})$,
where  $m = 0, \pm 1$,  and decompose them into the independent unit vectors
$\eb_0$ and $\eb_{\pm 1}$ defined by relations (2.61) taking into account that
any vector $\ab \equiv (a_x, a_y, a_z)$ may be represented in terms of these
unit vectors in the form $\ab \equiv (a_{+1}, a_{-1},a_0)$,  i.e.,

\begin{equation}
 \ab =  -a_{-1} \eb_1 - a_{+1} \eb_{-1} + a_0 \eb_0,
\end{equation}

\noindent  where

\begin{equation}
 a_{\pm 1} = \frac{ia_y \pm a_x}{\sqrt{2}}.
\end{equation}

Accordingly, we introduce the vectors  $\bb_m \equiv (b_{m+1},b_{m-1},b_{m0})$
and represent them as

\begin{equation}
 \bb_{m_1} = \sum\limits_{m_2 = -1}^1 \Bb_{m_1 m_2} \cdot \eb_{m_2},
  \qquad  m_1 = 0, \pm 1,
\end{equation}

\noindent  where

\begin{equation}
 \Bb_{m_1 m_2} =
  \left (
   \begin{array}{lll}
    b_{00}  &  -b_{0-1}  &  -b_{0+1} \\
    b_{+10}  &  -b_{+1-1}  &  -b_{+1+1} \\
    b_{-10} &  -b_{-1-1} &  -b_{-1+1}
   \end{array}
  \right ),
  \qquad  m_1, m_2 = 0, \pm 1.
\end{equation}

\noindent  Then  $\bb_m^T$  is the vector defined as follows:

\begin{equation}
 \bb_{m_1}^T = \sum\limits_{m_1 = -1}^1 \Bb_{m_1 m_2}^T \cdot \eb_{m_2},
  \qquad  m_1 = 0, \pm 1,
\end{equation}

\noindent  where  $\Bb_{m_1 m_2}^T$  is the matrix transposed to the matrix
$\Bb_{m_1 m_2}$.

Substituting (4.6) and (4.7) into (2.103) and (4.8) into (2.104), we obtain the
following relations for the forces and torques for the first iteration:

\begin{eqnarray}
 \Fb_\alpha^{(t,1)}(\omega) &=& - \sum\limits_{\beta \neq \alpha}
  \xib_{\alpha\beta}^{tt(1)}(\omega) \cdot \Ub_\beta (\omega),  \\
 \Fb_\alpha^{(r,1)}(\omega) &=& -\sum\limits_{\beta \neq \alpha}
  \xib_{\alpha\beta}^{tr(1)}(\omega) \cdot \Omb_\beta(\omega),  \\
 \Tb_\alpha^{(t,1)}(\omega) &=& -\sum\limits_{\beta \neq \alpha}
  \xib_{\alpha\beta}^{rt(1)}(\omega) \cdot \Ub_\beta (\omega),  \\
 \Tb_\alpha^{(r,1)}(\omega) &=& -\sum\limits_{\beta \neq \alpha}
  \xib_{\alpha\beta}^{rr(1)}(\omega) \cdot \Omb_\beta(\omega),
\end{eqnarray}

\noindent  where  $\xib_{\alpha\beta}^{tt(1)}(\omega)$  and
$\xib_{\alpha\beta}^{rr(1)}(\omega)$ are, respectively, the translational and
rotational friction tensors corresponding to the first iteration and
$\xib_{\alpha\beta}^{tr(1)}(\omega)$ and $\xib_{\alpha\beta}^{rt(1)}(\omega)$
are the friction tensors that couple translational motion of spheres and their
rotation.  These quantities have the form

\begin{eqnarray}
 \xib_{\alpha\beta}^{tt(1)}(\omega) &=& -\xi_\alpha \xi_\beta p(b_\alpha)
  p(b_\beta) \Tb_M(\Rb_{\alpha\beta},\omega) + {\cal O}(\sigma^3),  \\
 \xib_{\alpha\beta}^{tr(1)}(\omega) &=& -\xi_\beta a_\beta \sigma_{\alpha\beta}
  \sigma_{\beta\alpha} \tau(b_\alpha) q(b_\beta) \psi(y_{\alpha\beta})
  \left(\eb \cdot \nb_{\alpha\beta} \right) + {\cal O}(\sigma^4),  \\
 \xib_{\alpha\beta}^{rt(1)}(\omega) &=& -\xi_\alpha a_\alpha \sigma_{\alpha\beta}
  \sigma_{\beta\alpha} \tau(b_\beta) q(b_\alpha) \psi(y_{\alpha\beta})
  \left(\eb \cdot \nb_{\alpha\beta} \right) + {\cal O}(\sigma^4),  \\
 \xib_{\alpha\beta}^{rr(1)}(\omega) &=& \frac{\xi_\alpha^r}{3\pi}
  \sigma_{\beta\alpha}^3 q(b_\alpha) q(b_\beta) y_{\alpha\beta}^3 \Bigl \{
  \tilde{h}_2(y_{\alpha\beta})
  \Bigl(\Ib - 3 \nb_{\alpha\beta}\nb_{\alpha\beta} \Bigr)
  + 2 \tilde{h}_0(y_{\alpha\beta}) \Ib \Bigr \} + {\cal O}(\sigma^5),
\end{eqnarray}

\noindent  where  $\eb$  is the completely antisymmetric unit tensor of the
third rank,  $(\eb \cdot \nb) = e_{ijk} n_k$.

It is easy to verify that these friction tensors satisfy the symmetry relations
\cite{ref.Happel}

\begin{eqnarray}
 \left(\xib_{\beta\alpha}^{tt(1)}(\omega)\right)^T = \xib_{\alpha\beta}^{tt(1)}(\omega), \qquad
 \left(\xib_{\beta\alpha}^{rr(1)}(\omega)\right)^T &=& \xib_{\alpha\beta}^{rr(1)}(\omega), \qquad
 \left(\xib_{\beta\alpha}^{rt(1)}(\omega)\right)^T = \xib_{\alpha\beta}^{tr(1)}(\omega),
 \nonumber  \\
 && \qquad\qquad \qquad \alpha, \beta = 1, 2, \ldots, N,
\end{eqnarray}

\noindent  where   $\left(\xib_{\beta\alpha}^{rt(1)}(\omega) \right)^T$  means
the transposition of the matrix  $\xib_{\beta\alpha}^{rt(1)}(\omega)$  with
respect to the space indices  $i,j = x,y,z$.

Furthermore, the tensors  $\xib_{\alpha\beta}^{tt(1)}(\omega)$  and
$\xib_{\alpha\beta}^{rr(1)}(\omega)$  are independently symmetric both in the
space indices ($i, j = x, y, z$) and in the indices of spheres  $\alpha,
\beta$, i.e.,

\begin{equation}
 \begin{array}{llllll}
  \xib_{\alpha\beta}^{tt(1)}(\omega) &=& \xib_{\beta\alpha}^{tt(1)}(\omega),
   \qquad  & \left(\xib_{\alpha\beta}^{tt(1)}(\omega)\right)^T  &=&
   \xib_{\alpha\beta}^{tt(1)}(\omega),  \\
  \xib_{\alpha\beta}^{rr(1)}(\omega) &=& \xib_{\beta\alpha}^{rr(1)}(\omega),
   \qquad  & \left(\xib_{\alpha\beta}^{rr(1)}(\omega)\right)^T  &=&
   \xib_{\alpha\beta}^{rr(1)}(\omega).
 \end{array}
\end{equation}

By virtue of relations (4.35), (4.36), (4.39) and (4.40), it follows that the
components of the velocities  $\Ub_\beta(\omega)$ and $\Omb_\beta(\omega)$
parallel or antiparallel to the vector $\Rb_{\alpha\beta}$ connecting the
centers of spheres $\alpha$ and $\beta$ give no contribution, respectively, to
the torque and force exerted by the fluid on sphere  $\alpha$.

%%-----------------------------------==Section 5------------------------------

\section{The Second Iteration}  \label{n = 2}

For the second iteration, the system of equations (2.67), (2.77) is reduced to
the form

\begin{equation}
 \sum\limits_{l_2 m_2} \Tb_{\alpha, l_1 m_1}^{\alpha, l_2 m_2}(\omega)
  \cdot \fb_{\alpha, l_2 m_2}^{(2)}(\omega) = -\sum\limits_{\beta \neq \alpha}
  \sum\limits_{l_2 m_2} \Tb_{\alpha, l_1 m_1}^{\beta, l_2 m_2}(\omega)
  \cdot \fb_{\beta, l_2 m_2}^{(1)}(\omega),
\end{equation}

\begin{equation}
 Y_\alpha^{(2)}(\omega) \equiv \sum\limits_{m = -1}^1
  \eb_m \cdot \fb_{\alpha, 1m}^{(2)}(\omega)
  = 4\pi \sqrt{3} \, a_\alpha^2 \sum\limits_{\beta \neq \alpha}
  p_\beta^{(1)ind}(\Rb_\alpha,\omega), \quad  \alpha = 1,2, \ldots,N,
\end{equation}

\noindent  where  $p_\beta^{(1)ind}(\Rb_\alpha,\omega)$  is defined by
Eq.~(2.79) with  $\fb_{\beta,l_2 m_2}(\omega) \rightarrow \fb_{\beta,l_2
m_2}^{(1)}(\omega)$  and the harmonics $\fb_{\beta,l_2 m_2}^{(1)}(\omega)$ are
determined by relations (4.4)--(4.12) obtained for the first iteration. Solving
Eqs.~(5.1) and (5.2) up to the terms of the order $b^3$  and $\sigma^2$ [for
$\fb_{\alpha,l_2 m_2}^{(t,2)}(\omega)$] and $\sigma^3$ [for $\fb_{\alpha,l_2
m_2}^{(r,2)}(\omega)$] and using Eqs.~(3.13) and (4.27), we obtain

\begin{eqnarray}
 \fb_{\alpha, lm}^{(\zeta,2)}(\omega) &=&
  \delta_{l,0} \, \fb_{\alpha, 00}^{(\zeta,2)}(\omega) +
  \delta_{l,1} \, \fb_{\alpha, 1m}^{(\zeta,2)}(\omega) +
  \delta_{l,2} \, \fb_{\alpha, 2m}^{(\zeta,2)}(\omega) +
  \delta_{l,3} \, \fb_{\alpha, 3m}^{(\zeta,2)}(\omega), \qquad  \zeta = t, r, \\
 \fb_{\alpha, 1m}^{(p,2)}(\omega) &=& \eb_m^* \frac{Y_\alpha^{(2)}(\omega)}{3},
  \qquad \qquad m = 0 \pm 1.
 \end{eqnarray}

\noindent  Here,

\begin{eqnarray}
 \fb_{\alpha,00}^{(t,2)}(\omega) &=& \xi_\alpha p(b_\alpha)
  \sum\limits_{\gamma \neq \alpha} \xi_\gamma p(b_\gamma)
  \Tb_M(\Rb_{\alpha\gamma},\omega) \cdot
  \sum\limits_{\beta \neq \gamma} \xi_\beta p(b_\beta)
  \Tb_M(\Rb_{\gamma\beta},\omega) \cdot \Ub_\beta(\omega) + {\cal O}(\sigma^3),  \\
 \fb_{\alpha,00}^{(r,2)}(\omega) &=& -\xi_\alpha p(b_\alpha)
  \sum\limits_{\gamma \neq \alpha} \tau(b_\gamma)
  \Tb_M(\Rb_{\alpha\gamma},\omega) \cdot
  \sum\limits_{\beta \neq \gamma} a_\beta \xi_\beta \sigma_{\beta\gamma}
  \sigma_{\gamma\beta} \, q(b_\beta) \psi(y_{\gamma\beta})
  \Bigl( \nb_{\gamma\beta} \times \Omb_\beta(\omega) \Bigr),  \\
 \fb_{\alpha,1m}^{(\zeta,2)}(\omega) &=& -\frac{2}{3} \, \xi_\alpha \left \{
  4\bb_{\alpha,m}^{(\zeta,2)}(\omega)+\left( \bb_{\alpha,m}^{(\zeta,2)}(\omega)\right)^T
  \right \},  \qquad  \qquad \zeta = t, r, \qquad  m= 0, \pm 1, \\
 \fb_{\alpha,2m}^{(t,2)}(\omega) &=&
  \frac{b_\alpha^2 \xi_\alpha}{\sqrt{30}} \, \Kb_m \cdot
  \sum\limits_{\gamma \neq \alpha} \xi_\gamma \Tb_M(\Rb_{\alpha\gamma},\omega) \cdot
  \sum\limits_{\beta \neq \gamma} \Bigl(1 + b_\beta + b_\gamma \Bigr) \xi_\beta
  \Tb_M(\Rb_{\gamma\beta},\omega) \cdot \Ub_\beta(\omega), \nonumber  \\
  && \qquad \qquad \qquad \qquad \qquad \qquad \qquad \qquad \qquad \qquad
  m = 0, \pm 1, \pm 2, \\
 \fb_{\alpha,2m}^{(r,2)}(\omega) &=&
  -\frac{b_\alpha^2 \xi_\alpha}{\sqrt{30}} \, \Kb_m \cdot
  \sum\limits_{\gamma \neq \alpha} \Bigl( 1 + b_\gamma \Bigr)
  \Tb_M(\Rb_{\alpha\gamma},\omega) \cdot
  \sum\limits_{\beta \neq \gamma} a_\beta \xi_\beta \sigma_{\beta\gamma}
  \sigma_{\gamma\beta} \, \psi(y_{\gamma\beta})
  \Bigl( \nb_{\gamma\beta} \times \Omb_\beta(\omega) \Bigr), \nonumber  \\
  && \qquad \qquad \qquad \qquad \qquad \qquad \qquad \qquad \qquad \qquad
  m = 0, \pm 1, \pm 2, \\
 \fb_{\alpha,3m}^{(\zeta,2)}(\omega) &=&
  - \frac{b_\alpha^2 \xi_\alpha}{15 \sqrt{21}} \, \sum\limits_{m_1 = -3}^3 \,
  \Nb_{m m_1} \cdot \biggl \{4\cb_{\alpha,m_1}^{(\zeta,2)}(\omega)
  + \Bigl( \cb_{\alpha,m_1}^{(\zeta,2)}(\omega)\Bigr)^T  \biggr \},
   \qquad \zeta = t, r, \nonumber  \\
  && \qquad \qquad \qquad \qquad \qquad \qquad \qquad \qquad \qquad
  m = 0, \pm 1, \pm 2, \pm 3, \\
 Y_\alpha^{(2)}(\omega) &=& \sqrt{3} \sum\limits_{\beta \neq \alpha}
  \sigma_{\alpha\beta}^2 \,
  \Bigl(\nb_{\alpha\beta} \cdot \fb_{\beta,00}^{(t,1)}(\omega) \Bigr), \\
 \bb_{\alpha,m}^{(\zeta,2)}(\omega) &=& \bb_{\alpha,m}^{(\zeta,2,0)}(\omega) +
  \bb_{\alpha,m}^{(\zeta,2,1)}(\omega) + \bb_{\alpha,m}^{(\zeta,2,2)}(\omega),
  \qquad  \qquad  \zeta = t, r,  \qquad  m = 0, \pm 1, \\
 \bb_{\alpha,m}^{(t,2,0)}(\omega) &=& -\sum\limits_{\gamma \neq \alpha} \xi_\gamma
  \Tb_{\alpha,1m}^{\gamma,00}(\omega) \cdot \sum\limits_{\beta \neq \gamma} \xi_\beta
  \Tb_M(\Rb_{\gamma\beta},\omega) \cdot \Ub_\beta(\omega),
  \qquad  \qquad \quad m = 0, \pm 1, \\
 \bb_{\alpha,m}^{(t,2,1)}(\omega) &=& -\sum\limits_{\gamma \neq \alpha}
  \xi_\gamma \Tb_{\alpha,1m}^{\gamma,00}(\omega) \cdot \sum\limits_{\beta \neq \gamma}
  \Bigl (p(b_\gamma) p(b_\beta) - 1 \Bigr) \xi_\beta
  \Tb_M(\Rb_{\gamma\beta},\omega) \cdot \Ub_\beta(\omega) + b_\alpha^2
  \frac{4\sqrt{\pi}}{15} \, \nonumber  \\
  &\times& \biggl( 1 - \frac{5}{6} \, b_\alpha \biggr)
  \sum\limits_{m_1 = -1}^1 \Kb_{1m,00}^{1m_1} \cdot \biggl
  \{4\bb_{\alpha,m_1}^{(t,2,0)}(\omega)
  + \left( \bb_{\alpha,m_1}^{(t,2,0)}(\omega)\right)^T  \biggr \},
  \qquad m = 0, \pm 1, \\
 \bb_{\alpha,m}^{(r,2,0)}(\omega) &=& \sum\limits_{\gamma \neq \alpha}
  \Tb_{\alpha,1m}^{\gamma,00}(\omega) \cdot \sum\limits_{\beta \neq \gamma} a_\beta \xi_\beta
  \sigma_{\beta\gamma} \sigma_{\gamma\beta} \, \psi(y_{\gamma\beta})
  \Bigl( \nb_{\gamma\beta} \times \Omb_\beta(\omega) \Bigr),  \qquad  m = 0, \pm 1, \\
 \bb_{\alpha,m}^{(r,2,1)}(\omega) &=& \sum\limits_{\gamma \neq \alpha}
  \Tb_{\alpha,1m}^{\gamma,00}(\omega) \cdot \sum\limits_{\beta \neq \gamma}
  \Bigl (\tau(b_\gamma) q(b_\beta) - 1 \Bigr ) a_\beta \xi_\beta
  \sigma_{\beta\gamma} \sigma_{\gamma\beta} \, \psi(y_{\gamma\beta})
  \Bigl( \nb_{\gamma\beta} \times \Omb_\beta(\omega) \Bigr) + b_\alpha^2
  \frac{4\sqrt{\pi}}{15} \, \nonumber  \\
  &\times& \biggl( 1 - \frac{5}{6} \, b_\alpha \biggr)
  \sum\limits_{m_2 = -1}^1 \Kb_{1m,00}^{1m_1} \cdot \biggl \{
  4\bb_{\alpha,m_1}^{(r,2,0)}(\omega)
  + \left( \bb_{\alpha,m_1}^{(r,2,0)}(\omega)\right)^T  \biggr \},
  \qquad m = 0, \pm 1,  \\
 \bb_{\alpha,m}^{(\zeta,2,2)}(\omega) &=& b_\alpha^2 \frac{4\sqrt{\pi}}{15} \,
  \sum\limits_{m_1 = -1}^1 \Kb_{1m,00}^{1m_1} \cdot \biggl \{
  4\lb_{\alpha,m_1}^{(\zeta,2)}(\omega)
  + \left( \lb_{\alpha,m_1}^{(\zeta,2)}(\omega)\right)^T  \biggr \},
  \  \zeta = t, r,  \ \   m = 0, \pm 1,\\
 \lb_{\alpha,m}^{(t,2)}(\omega) &=& -\sum\limits_{\gamma \neq \alpha} \xi_\gamma
  \Tb_{\alpha,1m}^{\gamma,00}(\omega) \cdot \sum\limits_{\beta \neq \gamma} (b_\gamma + b_\beta)
  \xi_\beta \Tb_M(\Rb_{\gamma\beta},\omega) \cdot \Ub_\beta(\omega), \qquad m = 0, \pm 1, \\
 \lb_{\alpha,m}^{(r,2)}(\omega) &=& \sum\limits_{\gamma \neq \alpha} b_\gamma
  \Tb_{\alpha,1m}^{\gamma,00}(\omega) \cdot \sum\limits_{\beta \neq \gamma}
   a_\beta \xi_\beta \sigma_{\beta\gamma} \sigma_{\gamma\beta} \psi(y_{\gamma\beta})
  \Bigl( \nb_{\gamma\beta} \times \Omb_\beta(\omega) \Bigr), \qquad m = 0, \pm 1, \\
 \cb_{\alpha,m}^{(t,2)}(\omega) &=& \bb_{\alpha,m}^{(t,2,0)}(\omega)
  + \lb_{\alpha,m}^{(t,2)}(\omega) \nonumber  \\
  &=& -\sum\limits_{\gamma \neq \alpha} \xi_\gamma
  \Tb_{\alpha,1m}^{\gamma,00}(\omega) \cdot \sum\limits_{\beta \neq \gamma} \Bigl( 1 +
  b_\beta + b_\gamma \Bigr) \xi_\beta \Tb_M(\Rb_{\gamma\beta},\omega) \cdot
  \Ub_\beta(\omega),  \  m = 0, \pm 1, \\
 \cb_{\alpha,m}^{(r,2)}(\omega) &=& \bb_{\alpha,m}^{(r,2,0)}(\omega)
  + \lb_{\alpha,m}^{(r,2)}(\omega) \nonumber  \\
  &=& \sum\limits_{\gamma \neq \alpha} \Bigl( 1 + b_\gamma \Bigr)
  \Tb_{\alpha,1m}^{\gamma,00}(\omega) \cdot \sum\limits_{\beta \neq \gamma}
   a_\beta \xi_\beta \sigma_{\beta\gamma} \sigma_{\gamma\beta} \psi(y_{\gamma\beta})
  \Bigl( \nb_{\gamma\beta} \times \Omb_\beta(\omega) \Bigr), \  m = 0, \pm 1.
 \end{eqnarray}

Note that in these relations only the terms up to the order $b^3$ might be
retained.

Next, we substitute relations (5.5) and (5.6) into Eq.~(2.103) and relation
(5.7) into Eq.~(2.104). As a result, we obtain the following expressions for
the force and torque exerted by the fluid on sphere  $\alpha$ for the second
iteration:

\begin{eqnarray}
 \Fb_\alpha^{(t,2)}(\omega) &=& - \sum\limits_{\beta = 1}^N
  \xib_{\alpha\beta}^{tt(2)}(\omega) \cdot \Ub_\beta (\omega),  \\
 \Fb_\alpha^{(r,2)}(\omega) &=& -\sum\limits_{\beta = 1}^N
  \xib_{\alpha\beta}^{tr(2)}(\omega) \cdot \Omb_\beta(\omega),  \\
 \Tb_\alpha^{(t,2)}(\omega) &=& -\sum\limits_{\beta = 1}^N
  \xib_{\alpha\beta}^{rt(2)}(\omega) \cdot \Ub_\beta (\omega),  \\
 \Tb_\alpha^{(r,2)}(\omega) &=& -\sum\limits_{\beta = 1}^N
  \xib_{\alpha\beta}^{rr(2)}(\omega) \cdot \Omb_\beta(\omega),
\end{eqnarray}

\noindent  where  $\xib_{\alpha\beta}^{\mu \nu (2)}(\omega), \ \mu, \nu = t,r,
\  \alpha, \beta = 1, 2, \ldots, N$, are the friction tensors for the second
iteration, which can be written in the form

\begin{eqnarray}
 \xib_{\alpha\beta}^{tt(2)}(\omega) &=& \xi_\alpha \xi_\beta p(b_\alpha)
  p(b_\beta) \sum\limits_{\gamma \neq \alpha, \beta} p(b_\gamma) \xi_\gamma
  \Tb_M(R_{\alpha\gamma},\omega) \cdot \Tb_M(R_{\beta\gamma},\omega),  \\
 \xib_{\alpha\beta}^{tr(2)}(\omega) &=& -\xi_\alpha \xi_\beta a_\beta p(b_\alpha)
  q(b_\beta) \sum\limits_{\gamma \neq \alpha, \beta} \tau (b_\gamma)
  \sigma_{\beta\gamma} \sigma_{\gamma\beta} \psi(y_{\beta\gamma})
  \Tb_M(R_{\alpha\gamma},\omega) \cdot \left(\eb \cdot \nb_{\beta\gamma} \right),  \\
 \xib_{\alpha\beta}^{rt(2)}(\omega) &=& \xi_\alpha \xi_\beta a_\alpha p(b_\beta)
  q(b_\alpha) \sum\limits_{\gamma \neq \alpha, \beta} \tau (b_\gamma)
  \sigma_{\alpha\gamma} \sigma_{\gamma\alpha} \psi(y_{\alpha\gamma})
  \left(\eb \cdot \nb_{\alpha\gamma}\right) \cdot \Tb_M(R_{\beta\gamma},\omega),  \\
 \xib_{\alpha\beta}^{rr(2)}(\omega) &=& \frac{3}{4}
  \Bigl( \frac{a_\beta}{a_\alpha} \Bigr)^2 \, \xi_\alpha^r q(b_\alpha) q(b_\beta)
   \sum\limits_{\gamma \neq \alpha, \beta} \chi (b_\gamma)
  \sigma_{\alpha\gamma}^2 \sigma_{\beta\gamma} \sigma_{\gamma\beta}
  \psi(y_{\alpha\gamma}) \psi(y_{\beta\gamma})
  \Bigl \{ \bigl (\nb_{\alpha\gamma} \cdot \nb_{\beta\gamma} \bigr)\, \Ib
  \nonumber  \\
  &-& \nb_{\beta\gamma} \nb_{\alpha\gamma} \Bigr \},
\end{eqnarray}

\noindent  where

\begin{equation}
 \chi (b) = 1 + b + \frac{2}{3} \, b^2 + \frac{b^3}{3}\, .
\end{equation}

Using the explicit form (4.24) for the modified Oseen tensor
$\Tb_M(\Rb_{\alpha\beta},\omega)$, it is useful to represent the tensors
$\Tb_M(\Rb_{\alpha\gamma},\omega) \cdot \left(\eb \cdot \nb_{\beta\gamma}
\right )$  and  $\left(\eb \cdot \nb_{\alpha\gamma}\right) \cdot
\Tb_M(\Rb_{\beta\gamma},\omega)$  in relations (5.27) and (5.28) as

\begin{eqnarray}
 \Tb_M(\Rb_{\alpha\gamma},\omega) \cdot \left(\eb \cdot \nb_{\beta\gamma} \right)
  &=& \frac{1}{6\pi} \Biggl \{ \biggl \{
  F_{0,0,0}(a_\alpha, a_\gamma,R_{\alpha\gamma},\omega) - \frac{1}{2} \,
  F_{0,0,2}(a_\alpha, a_\gamma,R_{\alpha\gamma},\omega) \biggr \}
  \left(\eb \cdot \nb_{\beta\gamma}\right) \nonumber  \\
  &-& \frac{3}{2} \, F_{0,0,2}(a_\alpha, a_\gamma,R_{\alpha\gamma},\omega)
  \, \nb_{\alpha\gamma}
  \Bigl( \nb_{\alpha\gamma} \times \nb_{\beta\gamma} \Bigr) \Biggr \},  \\
 \left(\eb \cdot \nb_{\alpha\gamma}\right) \cdot \Tb_M(\Rb_{\beta\gamma},\omega)
  &=& \frac{1}{6\pi} \Biggl \{ \biggl \{
  F_{0,0,0}(a_\beta, a_\gamma,R_{\beta\gamma},\omega) - \frac{1}{2} \,
  F_{0,0,2}(a_\beta, a_\gamma,R_{\beta\gamma},\omega) \biggr \}
  \left(\eb \cdot \nb_{\alpha\gamma}\right) \nonumber  \\
  &-& \frac{3}{2} \, F_{0,0,2}(a_\beta, a_\gamma,R_{\beta\gamma},\omega)
  \Bigl( \nb_{\alpha\gamma} \times \nb_{\beta\gamma} \Bigr)
  \nb_{\beta\gamma} \Biggr \}.
\end{eqnarray}

In the particular case of two spheres ($\alpha, \beta = 1, 2$), according to
relations (5.26)--(5.29), all friction tensors $\xib_{\alpha\beta}^{\mu \nu
(2)}(\omega) = 0, \ \mu, \nu = t, r$,  for  $\beta \neq \alpha$,  i.e., these
components are caused only by three-particle interactions.

Just as the friction tensors  $\xib_{\alpha\beta}^{\mu \nu (1)}(\omega)$  for
the first iteration, the friction tensors $\xib_{\alpha\beta}^{\mu \nu
(2)}(\omega)$ for the second iteration satisfy the symmetry relations
\cite{ref.Happel}

\begin{eqnarray}
 \left(\xib_{\beta\alpha}^{tt(2)}(\omega)\right)^T  = \xib_{\alpha\beta}^{tt(2)}(\omega), \qquad
 \left(\xib_{\beta\alpha}^{rr(2)}(\omega)\right)^T &=& \xib_{\alpha\beta}^{rr(2)}(\omega), \qquad
 \left(\xib_{\beta\alpha}^{rt(2)}(\omega)\right)^T = \xib_{\alpha\beta}^{tr(2)}(\omega),
 \nonumber  \\
 && \qquad\qquad\qquad \alpha, \beta = 1, 2, \ldots, N,
\end{eqnarray}

Substituting relations (2.99), (3.40), (4.34), and (5.22) into (2.98),
relations (3.41), (4.35), and (5.23) into (2.100), relations (3.46), (4.36),
and (5.24) into (2.101), and relations (3.47), (4.37), and (5.25) into (2.102),
we obtain the following expressions for the forces and torques exerted by the
fluid on the spheres up to the terms of the order $b^3$:

\begin{eqnarray}
 \Fb_\alpha^{(t)}(\omega) &=& -\sum_{\beta = 1}^N
  \xib_{\alpha\beta}^{tt}(\omega) \cdot \Ub_\beta (\omega),  \\
 \Fb_\alpha^{(r)}(\omega) &=& -\sum_{\beta = 1}^N
  \xib_{\alpha\beta}^{tr}(\omega) \cdot \Omb_\beta (\omega),  \\
 \Tb_\alpha^{(t)}(\omega) &=&  -\sum_{\beta = 1}^N
  \xib_{\alpha\beta}^{rt}(\omega) \cdot \Ub_\beta (\omega),  \\
 \Tb_\alpha^{(r)}(\omega) &=& -\sum_{\beta = 1}^N
  \xib_{\alpha\beta}^{rr}(\omega) \cdot \Omb_\beta (\omega).
\end{eqnarray}

\noindent  Here,

\begin{eqnarray}
 \xib_{\alpha\beta}^{tt}(\omega) &=& \xi_\alpha (\omega) \delta_{\alpha\beta} \,
  \Ib + \xi_\alpha \lambdab_{\alpha\beta}^{tt}(\omega),
  \qquad \qquad \alpha, \beta = 1, 2, \ldots, N,  \\
 \xib_{\alpha\beta}^{rr}(\omega) &=& \xi_\alpha^r (\omega) \delta_{\alpha\beta} \,
  \Ib + \xi_\alpha^r \lambdab_{\alpha\beta}^{rr}(\omega),
  \qquad \qquad \alpha, \beta = 1, 2, \ldots, N, \\
 \xib_{\alpha\alpha}^{tr}(\omega) &=& -\xi_\alpha^2 a_\alpha p(b_\alpha) q(b_\alpha)
  \sum\limits_{\gamma \neq \alpha} \tau (b_\gamma)
  \sigma_{\alpha\gamma} \sigma_{\gamma\alpha} \psi(y_{\alpha\gamma})
  \Tb_M(\Rb_{\alpha\gamma},\omega) \cdot \left(\eb \cdot \nb_{\alpha\gamma}\right),\\
 \xib_{\alpha\beta}^{tr}(\omega) &=& -\xi_\beta a_\beta \, q(b_\beta) \biggl \{
  \tau(b_\alpha) \sigma_{\alpha\beta} \sigma_{\beta\alpha} \psi(y_{\alpha\beta})
  \left(\eb \cdot \nb_{\alpha\beta}\right) \nonumber  \\
  &+& \xi_\alpha p(b_\alpha) \sum\limits_{\gamma \neq \alpha, \beta} \tau (b_\gamma)
  \sigma_{\beta\gamma} \sigma_{\gamma\beta} \psi(y_{\beta\gamma})
  \Tb_M(\Rb_{\alpha\gamma},\omega) \cdot \left(\eb \cdot \nb_{\beta\gamma} \right)
  \biggr \},  \qquad  \beta \neq \alpha,  \\
 \xib_{\alpha\alpha}^{rt}(\omega) &=& \xi_\alpha^2 a_\alpha p(b_\alpha) q(b_\alpha)
  \sum\limits_{\gamma \neq \alpha} \tau (b_\gamma)
  \sigma_{\alpha\gamma} \sigma_{\gamma\alpha} \psi(y_{\alpha\gamma})
  \left(\eb \cdot \nb_{\alpha\gamma}\right) \cdot \Tb_M(\Rb_{\alpha\gamma},\omega),  \\
  \xib_{\alpha\beta}^{rt}(\omega) &=& -\xi_\alpha a_\alpha  q(b_\alpha) \biggl \{
  \tau(b_\beta) \sigma_{\alpha\beta} \sigma_{\beta\alpha} \psi(y_{\alpha\beta})
  \left(\eb \cdot \nb_{\alpha\beta}\right) \nonumber  \\
  &-& \xi_\beta p(b_\beta) \sum\limits_{\gamma \neq \alpha, \beta} \tau (b_\gamma)
  \sigma_{\alpha\gamma} \sigma_{\gamma\alpha} \psi(y_{\alpha\gamma})
  \left(\eb \cdot \nb_{\alpha\gamma}\right) \cdot \Tb_M(\Rb_{\beta\gamma},\omega)
  \biggr \},  \qquad  \beta \neq \alpha,  \\
 \lambdab_{\alpha\alpha}^{tt}(\omega) &=& \xi_\alpha p^2(b_\alpha)
  \sum\limits_{\gamma \neq \alpha} p(b_\gamma) \xi_\gamma
  \biggl( \Tb_M(\Rb_{\alpha\gamma},\omega) \biggr)^2,  \\
 \lambdab_{\alpha\beta}^{tt}(\omega) &=& -\xi_\beta p(b_\alpha) p(b_\beta) \biggl \{
  \Tb_M(\Rb_{\alpha\beta},\omega) \nonumber \\
  &-& \sum\limits_{\gamma \neq \alpha, \beta} p(b_\gamma) \xi_\gamma
  \Tb_M(\Rb_{\alpha\gamma},\omega) \cdot \Tb_M(\Rb_{\beta\gamma},\omega) \biggr \},
  \qquad  \beta \neq \alpha,  \\
 \lambdab_{\alpha\alpha}^{rr}(\omega) &=& \frac{3}{4} \, q^2(b_\alpha)
  \sum\limits_{\gamma \neq \alpha} \chi (b_\gamma)
  \sigma_{\alpha\gamma}^3 \sigma_{\gamma\alpha} \psi^2(y_{\alpha\gamma})
  \bigl( \Ib - \nb_{\alpha\gamma} \nb_{\alpha\gamma} \bigr ), \\
 \lambdab_{\alpha\beta}^{rr}(\omega) &=& q(b_\alpha) q(b_\beta) \Biggl \{
  \frac{\sigma_{\beta\alpha}^3}{3 \pi} \, y_{\alpha\beta}^3 \ \biggl \{
  \tilde{h}_2(y_{\alpha\beta})
  \bigl( \Ib - 3 \nb_{\alpha\beta} \nb_{\alpha\beta} \bigr )
  + 2 \tilde{h}_0(y_{\alpha\beta}) \Ib \biggr \}
  + \frac{3}{4} \biggl( \frac{a_\beta}{a_\alpha} \biggr)^2 \nonumber  \\
  & \times & \sum\limits_{\gamma \neq \alpha, \beta} \chi (b_\gamma)
  \sigma_{\alpha\gamma}^2 \sigma_{\beta\gamma} \sigma_{\gamma\beta}
  \psi(y_{\alpha\gamma}) \psi(y_{\beta\gamma})
  \Bigl \{ \bigl (\nb_{\alpha\gamma} \cdot \nb_{\beta\gamma} \bigr)\, \Ib
  - \nb_{\beta\gamma} \nb_{\alpha\gamma} \Bigr \} \Biggr \},
  \quad  \beta \neq \alpha.
\end{eqnarray}

Note that in relations (5.41), (5.43), (5.45), and (5.47), the terms
proportional to $\sum\limits_{\gamma \neq \alpha, \beta}$  correspond to the
three-particle interactions and, hence, they are absent in the particular case
of two spheres ($\alpha, \beta = 1, 2$).

Note that the friction tensors for the first and second iterations satisfy the
symmetry relations given by Eqs.~(4.42) and (5.33), and, hence, these relations
also hold for the total friction tensors $\xib_{\alpha\beta}^{\mu \nu
}(\omega), \  \mu, \nu = t,r, \ $  i.e.,

\begin{eqnarray}
 \left(\xib_{\beta\alpha}^{tt}(\omega)\right)^T = \xib_{\alpha\beta}^{tt}(\omega), \quad
 \left(\xib_{\beta\alpha}^{rr}(\omega)\right)^T &=& \xib_{\alpha\beta}^{rr}(\omega), \quad
 \left(\xib_{\beta\alpha}^{rt}(\omega)\right)^T = \xib_{\alpha\beta}^{tr}(\omega),
 \nonumber \\
 && \qquad\qquad\qquad  \alpha, \beta = 1, 2, \ldots, N.
\end{eqnarray}

As was mentioned above, all required quantities (including the friction
tensors) are obtained without imposing any additional restrictions on the
dimensional parameter  $y_{\alpha\beta}$.  In particular cases, the expressions
for the friction tensors can be essentially simplified. We consider the
following two opposite modes:  $\Real y_{\alpha\beta} \gg 1$  and
$|y_{\alpha\beta}| \ll 1$, which is typical of the dilute suspensions and
colloidal crystals, respectively.

For the first mode,  the frequency belongs to the range

\begin{equation}
 \frac{\nu}{a_{max}^2} \gg |\omega| \gg \frac{2 \nu}{R_{min}^2},
\end{equation}

\noindent  where  $a_{max}$  is the radius of the largest sphere and  $R_{min}$
is the distance between the centers of two closest spheres.

Then we can use the asymptotics

\begin{equation}
 \tilde{h}_n(z) \approx \frac{\pi}{2z}\, \exp(-z), \qquad  n = 0,1,2,\ldots,
\end{equation}

\noindent valid for large  $z$, which enables us to simplify the friction tensors
(5.38)--(5.43) to the form

\begin{eqnarray}
 \xib_{\alpha\beta}^{tt}(\omega) &\approx & \xi_\alpha(\omega) \delta_{\alpha\beta} \,
  \Ib + \frac{3}{2} \, \xi_\alpha p(b_\alpha) p(b_\beta) \Biggl \{ (1 - \delta_{\alpha\beta})
  \frac{\sigma_{\beta\alpha}}{y_{\alpha\beta}^2} \,
  \Bigl( \Ib - 3\nb_{\alpha\beta}\nb_{\alpha\beta} \Bigr) \nonumber  \\
  &+& \frac{3}{2} \,  \sum\limits_{\gamma \neq \alpha,\beta} p(b_\gamma)
  \frac{\sigma_{\beta\gamma} \sigma_{\gamma\alpha}}{y_{\alpha\gamma}^2 y_{\beta\gamma}^2}
  \Bigl \{ \Ib + 9 (\nb_{\alpha\gamma} \cdot \nb_{\beta\gamma})
  \nb_{\alpha\gamma} \nb_{\beta\gamma} - 3 \nb_{\alpha\gamma} \nb_{\alpha\gamma}
  - 3 \nb_{\beta\gamma} \nb_{\beta\gamma} \Bigr \} \Biggr \},\nonumber  \\
  && \qquad \qquad \qquad \qquad \qquad \qquad \qquad \qquad \qquad \qquad
  \alpha, \beta = 1, 2, \ldots, N,
\end{eqnarray}

\begin{eqnarray}
 \xib_{\alpha\beta}^{rr}(\omega) &\approx & \xi_\alpha^r (\omega)
  \delta_{\alpha\beta} \,\Ib, \qquad \ \alpha, \beta = 1, 2, \ldots, N, \\
 \xib_{\alpha\beta}^{tr}(\omega) &\approx & 0,
  \qquad \qquad \qquad \ \alpha, \beta = 1, 2, \ldots, N, \\
 \xib_{\alpha\beta}^{rt}(\omega) &\approx & 0,
  \qquad \qquad \qquad \ \alpha, \beta = 1, 2, \ldots, N.
\end{eqnarray}

This mode is characterized by an exponentially small coupling between the
translational and rotational motions of the spheres proportional to
$\exp(-y_{\alpha\beta})$  for  $\beta \neq \alpha$  as well as between the
rotational motions of different spheres.  Taking into account that in the
stationary case,  $\xib_{\alpha\beta}^{tr}$,  $\xib_{\alpha\beta}^{rt}$,  and
$\xib_{\alpha\beta}^{rr}$,  for  $\beta \neq \alpha$,  are proportional to
$1/R_{\alpha\beta}^2$  and  $1/R_{\alpha\beta}^3$, respectively,
\cite{ref.MazurSaarl}, we may conclude that the retardation effects lead to an
essential decrease in hydrodynamic interactions between the spheres.  For the
mutual translational friction tensors, this influence is not so considerable
because the behavior of $\xib_{\alpha\beta}^{tt}$  as $1/R_{\alpha\beta}$  in
the stationary case \cite{ref.Mazur} is replaced by $1/R_{\alpha\beta}^3$  in
the nonstationary case for large distances between the spheres.

The second mode,  $|y_{\alpha\beta}| \ll 1$,  corresponds to the low
frequency domain.  In this case,

\begin{eqnarray}
 \psi(y_{\alpha\beta}) &\approx & 1 - \frac{y_{\alpha\beta}^2}{2}
 + \frac{y_{\alpha\beta}^3}{3} + {\cal O}(y_{\alpha\beta}^4),  \\
 \Tb_M(\Rb_{\alpha\beta},\omega) &\approx & \Tb_M(\Rb_{\alpha\beta})
 - \frac{y_{\alpha\beta}}{6\pi\eta R_{\alpha\beta}} \, \Ib
 + {\cal O}(y_{\alpha\beta}^2),
\end{eqnarray}

\noindent where

\begin{equation}
 \Tb_M(\Rb_{\alpha\beta}) = \frac{1}{8\pi\eta R_{\alpha\beta}}
  \biggl \{ \Bigl( \Ib + \nb_{\alpha\beta}\nb_{\alpha\beta} \Bigr)
  + \left( \sigma_{\alpha\beta}^2 + \sigma_{\beta\alpha}^2 \right)
  \left( \frac{1}{3} \, \Ib - \nb_{\alpha\beta}\nb_{\alpha\beta} \right) \biggr \}
\end{equation}

\noindent  is the modified static Oseen tensor \cite{ref.Yosh}.  Note that
within the framework of the second iteration, only terms proportional to
$1/R_{\alpha\beta}$  should be retained in (5.57), i.e.,

\begin{equation}
 \Tb_M(\Rb_{\alpha\beta}) \approx \frac{1}{8\pi\eta R_{\alpha\beta}}
  \Bigl( \Ib + \nb_{\alpha\beta}\nb_{\alpha\beta} \Bigr).
\end{equation}

Substituting (5.55) and (5.56) into (5.38)--(5.47) and retaining only the main
frequency-dependent terms, we represent the friction tensors in the form

\begin{equation}
 \xib_{\alpha\beta}^{\mu \nu}(\omega) = \xib_{\alpha\beta}^{\mu \nu}
  + \Ab_{\alpha\beta}^{\mu \nu}(\omega), \quad \mu, \nu = t,r, \quad
  \alpha, \beta = 1, 2, \ldots, N.
\end{equation}

\noindent  Here,  $\xib_{\alpha\beta}^{\mu \nu}, \ \mu, \nu = t,r, \  \alpha,
\beta = 1, 2, \ldots, N,$  are the stationary friction tensors determined up to
the terms corresponding the second iteration
\cite{ref.Mazur,ref.Jones2,ref.MazurBed,ref.Usenko}:

\begin{eqnarray}
 \xib_{\alpha\beta}^{tt} &=& \xi_\alpha \left(\delta_{\alpha\beta} \Ib
  + \lambdab_{\alpha\beta}^{tt} \right),
  \qquad \alpha, \beta = 1, 2, \ldots, N,  \\
 \xib_{\alpha\beta}^{rr} &=& \xi_\alpha^r \left(\delta_{\alpha\beta} \Ib
  + \lambdab_{\alpha\beta}^{rr} \right),
  \qquad \alpha, \beta = 1, 2, \ldots, N,  \\
 \xib_{\alpha\alpha}^{tr} &=& -\xi_\alpha a_\alpha \frac{3}{4}
  \sum\limits_{\gamma \neq \alpha} \sigma_{\alpha\gamma}^2 \sigma_{\gamma\alpha}
  \left(\eb \cdot \nb_{\alpha\gamma} \right),  \\
  \xib_{\alpha\alpha}^{rt} &=& -\xib_{\alpha\alpha}^{tr},
\end{eqnarray}

\begin{eqnarray}
 \xib_{\alpha\beta}^{tr} &=& -\xi_\beta a_\beta \Biggl \{ \sigma_{\alpha\beta}
  \sigma_{\beta\alpha} \left( \eb \cdot \nb_{\alpha\beta} \right)\nonumber  \\
  &+&  \frac{3}{4} \sum\limits_{\gamma \neq \alpha, \beta}
  \sigma_{\alpha\gamma} \sigma_{\gamma\beta} \sigma_{\beta\gamma}
  \Bigl \{ \left(\eb \cdot \nb_{\beta\gamma} \right)
  - \nb_{\alpha\gamma} \left( \nb_{\alpha\gamma} \times \nb_{\beta\gamma}
  \right) \Bigr \} \Biggr \},  \qquad \beta \neq \alpha,  \\
 \xib_{\alpha\beta}^{rt} &=& - \xi_\alpha a_\alpha \Biggl \{ \sigma_{\alpha\beta}
  \sigma_{\beta\alpha} \left( \eb \cdot \nb_{\alpha\beta} \right) \nonumber  \\
  &-& \frac{3}{4} \sum\limits_{\gamma \neq \alpha, \beta}
  \sigma_{\alpha\gamma} \sigma_{\gamma\alpha} \sigma_{\beta\gamma}
  \Bigl \{ \left(\eb \cdot \nb_{\alpha\gamma} \right)
  + \left( \nb_{\beta\gamma} \times \nb_{\alpha\gamma} \right)
  \nb_{\beta\gamma} \Bigr \} \Biggr \},  \qquad \beta \neq \alpha, \\
 \lambdab_{\alpha\alpha}^{tt} &=& \frac{9}{16} \sum\limits_{\gamma \neq \alpha}
  \sigma_{\alpha\gamma} \sigma_{\gamma\alpha} \left( \Ib + 3 \nb_{\alpha\gamma}
  \nb_{\alpha\gamma} \right),  \\
 \lambdab_{\alpha\beta}^{tt} &=& -\frac{3}{4} \Biggl \{ \sigma_{\beta\alpha}
  \left(\Ib + \nb_{\alpha\beta} \nb_{\alpha\beta} \right) \nonumber  \\
  &-& \frac{3}{4} \sum\limits_{\gamma \neq \alpha, \beta}
  \sigma_{\gamma\alpha} \sigma_{\beta\gamma}
  \left( \Ib + \nb_{\alpha\gamma} \nb_{\alpha\gamma} \right)
  \cdot \left( \Ib + \nb_{\beta\gamma} \nb_{\beta\gamma} \right) \Biggr \},
  \qquad \qquad  \beta \neq \alpha,  \\
 \lambdab_{\alpha\alpha}^{rr} &=& \frac{3}{4} \sum\limits_{\gamma \neq \alpha}
  \sigma_{\alpha\gamma}^3 \sigma_{\gamma\alpha} \left( \Ib - \nb_{\alpha\gamma}
  \nb_{\alpha\gamma} \right), \\
 \lambdab_{\alpha\beta}^{rr} &=& \frac{1}{2} \Biggl \{ \sigma_{\beta\alpha}^3
  \left(\Ib - 3 \nb_{\alpha\beta} \nb_{\alpha\beta} \right)  \nonumber  \\
  &+& \frac{3}{2} \, \left( \frac{a_\beta}{a_\alpha} \right)^2
  \sum\limits_{\gamma \neq \alpha, \beta}
  \sigma_{\beta\gamma} \sigma_{\gamma\beta} \sigma_{\alpha\gamma}^2
  \Bigl \{ \left( \nb_{\alpha\gamma} \cdot \nb_{\beta\gamma} \right) \Ib
  - \nb_{\beta\gamma} \nb_{\alpha\gamma} \Bigr \} \Biggr \},
  \quad  \beta \neq \alpha.
\end{eqnarray}

The frequency-dependent terms $\Ab_{\alpha\beta}^{\mu \nu}(\omega), \ \mu, \nu
= t,r, \ \  \alpha, \beta = 1, 2, \ldots, N,$  which vanish as  $\omega \to 0$,
have the form

\begin{eqnarray}
 \Ab_{\alpha\beta}^{tt}(\omega) &\approx & \xi_\alpha \Biggl \{ b_\beta \Ib
  - \Bigl( 1 - \delta_{\alpha\beta} \Bigr) \frac{3}{4} \, b_\beta
  \Bigl( \sigma_{\alpha\beta} + \sigma_{\beta\alpha} \Bigr)
  \Bigl( \Ib + \nb_{\alpha\beta} \nb_{\alpha\beta} \Bigr)
  - \frac{3}{4} \sum\limits_{\gamma \neq \alpha,\beta}
  \biggl \{ \sigma_{\beta\gamma} b_\gamma \Bigl( \Ib  \nonumber  \\
  &+& \nb_{\beta\gamma} \nb_{\beta\gamma} \Bigr) + \sigma_{\gamma\alpha} b_\beta
  \Bigl( \Ib + \nb_{\alpha\gamma} \nb_{\alpha\gamma} \Bigr) \Biggr \}, \\
 \Ab_{\alpha\beta}^{rr}(\omega) &\approx & \xi_\alpha^r \Biggl \{
  \frac{b_\beta^2}{3} \Bigl(\delta_{\alpha\beta} - b_\beta \Bigr) \, \Ib
  + \Bigl( 1 - \delta_{\alpha\beta} \Bigr) \frac{b_\beta^2}{4} \,
  \sigma_{\beta\alpha} \Bigl( \Ib + \nb_{\alpha\beta} \nb_{\alpha\beta} \Bigr)
  + \frac{3}{8} \biggl( \frac{a_\beta}{a_\alpha} \biggr)^2  \nonumber  \\
  && \times \sum\limits_{\gamma \neq \alpha,\beta} \biggl \{2 \sigma_{\alpha\gamma}^2
  \sigma_{\beta\gamma} \sigma_{\gamma\beta} b_\gamma
  - \Bigl( \sigma_{\beta\gamma} \sigma_{\gamma\beta} b_\alpha^2 +
  \sigma_{\alpha\gamma}^2 b_\beta b_\gamma \Bigr) \Bigl( 1 + b_\gamma \Bigr) +
  \frac{2}{3} \Bigl( \frac{\sigma_{\beta\gamma} \sigma_{\gamma\beta}}{\sigma_{\alpha\gamma}}
  \, b_\alpha^3  \nonumber  \\
  &+& \frac{{\sigma_{\alpha\gamma}^2}}{\sigma_{\beta\gamma}} \, b_\beta^2
  b_\gamma \Bigr) \biggr \}
  \biggl \{ \Bigl( \nb_{\alpha\gamma} \cdot \nb_{\beta\gamma} \Bigr) \Ib
  - \nb_{\beta\gamma} \nb_{\alpha\gamma} \biggr \} \Biggr \},  \\
 \Ab_{\alpha\beta}^{tr}(\omega) &\approx & \xi_\beta a_\beta b_\alpha \Biggl \{
  \Bigl( 1 - \delta_{\alpha\beta} \Bigr)
  \biggl(\frac{b_\beta}{2} - \sigma_{\alpha\beta} \sigma_{\beta\alpha} \biggr)
  \Bigl( \eb \cdot \nb_{\alpha\beta} \Bigr) +
  \sum\limits_{\gamma \neq \alpha,\beta}
  \sigma_{\beta\gamma} \sigma_{\gamma\beta}
  \Bigl( \eb \cdot \nb_{\beta\gamma} \Bigr) \Biggr \}, \\
 \Ab_{\alpha\beta}^{rt}(\omega) &\approx & \xi_\alpha a_\alpha b_\beta \Biggl \{
  \Bigl( 1 - \delta_{\alpha\beta} \Bigr)
  \biggl(\frac{b_\alpha}{2} - \sigma_{\alpha\beta} \sigma_{\beta\alpha} \biggr)
  \Bigl( \eb \cdot \nb_{\alpha\beta} \Bigr) -
  \sum\limits_{\gamma \neq \alpha,\beta}
  \sigma_{\alpha\gamma} \sigma_{\gamma\alpha}
  \Bigl( \eb \cdot \nb_{\alpha\gamma} \Bigr) \Biggr \}.
\end{eqnarray}

According to (5.70)--(5.73), in the low-frequency range, the main
frequency-dependent components of the friction tensors are proportional to
$\omega^{1/2}$.  For the rotational and coupled friction tensors, we also
retain terms proportional to $\omega$  (for the rotational friction tensors,
even terms proportional to $\omega^{3/2}$) because they are of the order
$\sigma^n$  [$n = 0$  for $\Ab_{\alpha\alpha}^{rr}(\omega)$  and
$\Ab_{\alpha\beta}^{tr}(\omega)$  and $\Ab_{\alpha\beta}^{rt}(\omega)$  for
$\beta\neq \alpha$ and  $n  = 1$ for $\Ab_{\alpha\beta}^{rr}(\omega)$ for
$\beta \neq \alpha$], whereas the terms proportional to $\omega^{1/2}$ are of
the order of $\sigma^m$  [$m = 2$  for $\Ab_{\alpha\beta}^{tr}(\omega)$  and
$\Ab_{\alpha\beta}^{rt}(\omega)$, $\alpha, \beta = 1, 2, \ldots, N$, and $m =
4$  for $\Ab_{\alpha\beta}^{rr}(\omega)$  for  $\alpha, \beta = 1, 2, \ldots,
N$] with $ m > n$.

%%-----------------------------------==Section 6------------------------------

\section{Mobility Tensors}  \label{Mobilities}

Substituting Eqs.~(5.34) and (5.35) into (2.96) and Eqs.~(5.36) and (5.37) into
(2.97) and solving the resulting system of equations for the quantities
$\Ub_\alpha (\omega)$ and $\Omb_\alpha(\omega)$,  we can represent these
quantities as follows:

\begin{eqnarray}
 \Ub_\alpha(\omega) &=& = -\sum\limits_{\beta = 1}^N \left \{
  \mub_{\alpha\beta}^{tt}(\omega) \cdot \Fbc_\beta(\omega)
  + \mub_{\alpha\beta}^{tr}(\omega) \cdot \Tbc_\beta(\omega) \right \},  \\
 \Omb_\alpha(\omega) &=& = -\sum\limits_{\beta = 1}^N \left \{
  \mub_{\alpha\beta}^{rt}(\omega) \cdot \Fbc_\beta(\omega)
  + \mub_{\alpha\beta}^{rr}(\omega) \cdot \Tbc_\beta(\omega) \right \},
\end{eqnarray}

\noindent  where $\mub_{\alpha\beta}^{tt}(\omega)$ and
$\mub_{\alpha\beta}^{rr}(\omega)$  are, respectively, the translational and
rotational mobility tensors, and  $\mub_{\alpha\beta}^{tr}(\omega)$  and
$\mub_{\alpha\beta}^{rt}(\omega)$  are the mobility tensors that couple
translational and rotational motions of the spheres.  Up to the third order in
the parameter $b$  and the same orders in the parameter $\sigma$ that for the
corresponding friction tensors, i.e., up to terms of  $\sigma^2$,  $\sigma^4$,
and $\sigma^3$, for the translational, rotational, and coupled mobility
tensors, respectively, the final expressions for the mobilities can be
represented in the form

\begin{eqnarray}
 \mub_{\alpha\alpha}^{tt}(\omega) &=& \mu_\alpha^t(\omega) \Ib
  + \frac{2}{9\xi_\alpha} \sum\limits_{\gamma \neq \alpha} b_\gamma^2
  \sigma_{\alpha\gamma} \sigma_{\gamma\alpha}
  \biggl \{ \exp(-2y_{\alpha\gamma}) \Ib + \frac{w^2(y_{\alpha\gamma})}{16} \,
  \Bigl( \Ib + 3 \nb_{\alpha\gamma}\nb_{\alpha\gamma} \Bigr) \nonumber  \\
  &-& \frac{w(y_{\alpha\gamma})}{2} \, \exp(-y_{\alpha\gamma})
  \Bigl( \Ib - 3 \nb_{\alpha\gamma}\nb_{\alpha\gamma} \Bigr) \biggr \}, \\
 \mub_{\alpha\beta}^{tt}(\omega) &=& \biggl \{ 1 + \frac{2}{9} \, \Bigl [
  b_\alpha^2 (1-b_\alpha) + b_\beta^2 (1-b_\beta) \Bigr ] \biggr \}
  \Tb_M (\Rb_{\alpha\beta},\omega) + \frac{2}{9\xi_\alpha}
  \sum\limits_{\gamma \neq \alpha, \beta} b_\gamma^2
  \sigma_{\alpha\gamma} \sigma_{\gamma\beta}
  \biggl \{ \exp(-y_{\alpha\gamma}) \Ib \nonumber  \\
  &-& \frac{w(y_{\alpha\gamma})}{4} \,
  \Bigl( \Ib - 3 \nb_{\alpha\gamma}\nb_{\alpha\gamma} \Bigr) \biggr \}
  \cdot
  \biggl \{ \exp(-y_{\beta\gamma}) \Ib - \frac{w(y_{\beta\gamma})}{4} \,
  \Bigl( \Ib - 3 \nb_{\beta\gamma}\nb_{\beta\gamma} \Bigr) \biggr \},
  \, \, \beta \neq \alpha,  \\
 \mub_{\alpha\alpha}^{rr}(\omega) &=& \mu_\alpha^r(\omega) \Ib
  + \frac{1}{6\xi_\alpha^r}   \sum\limits_{\gamma \neq \alpha} b_\gamma^2
  \sigma_{\alpha\gamma}^3 \sigma_{\gamma\alpha} \psi^2(y_{\alpha\gamma})
  \Bigl( \Ib - \nb_{\alpha\gamma}\nb_{\alpha\gamma} \Bigr), \\
 \mub_{\alpha\beta}^{rr}(\omega) &=& \frac{b_\alpha^3}{3 \pi \xi_\alpha^r}
  \biggl \{ \tilde{h}_2(y_{\alpha\beta}) \Bigl(
  3 \nb_{\alpha\beta}\nb_{\alpha\beta} - \Ib \Bigr)
  - 2 \tilde{h}_0(y_{\alpha\beta}) \Ib \biggr\} \nonumber \\
  &+& \frac{1}{6 \xi_\alpha^r} \Bigl( \frac{a_\alpha}{a_\beta} \Bigr)^2
  \sum\limits_{\gamma \neq \alpha, \beta} b_\gamma^2
  \sigma_{\alpha\gamma} \sigma_{\gamma\alpha} \sigma_{\beta\gamma}^2
  \psi(y_{\alpha\gamma}) \psi(y_{\beta\gamma})
  \Bigl \{ \Bigl( \nb_{\alpha\gamma} \cdot \nb_{\beta\gamma} \Bigr) \Ib
  - \nb_{\beta\gamma}\nb_{\alpha\gamma} \Bigr \} \Biggr \},
  \  \beta \neq \alpha,  \\
 \mub_{\alpha\alpha}^{tr}(\omega) &=& -\frac{1}{36 \pi \eta}
  \sum\limits_{\gamma \neq \alpha} b_\gamma^2
  \frac{\sigma_{\gamma\alpha}}{R_{\alpha\gamma}^2} \, \psi(y_{\alpha\gamma})
  \biggl \{ \exp(-y_{\alpha\gamma}) -\frac{w(y_{\alpha\gamma})}{4}
  \biggr \} \Bigl( \eb \cdot \nb_{\alpha\gamma} \Bigr),  \\
 \mub_{\alpha\beta}^{tr}(\omega) &=& \frac{k(b_\alpha)}{8\pi \eta R_{\alpha\beta}^2}
  \Bigl( 1 + \frac{b_\beta^2}{6} \Bigr)
  \psi(y_{\alpha\beta}) \Bigl( \eb \cdot \nb_{\alpha\beta} \Bigr)
  - \frac{1}{36\pi \eta} \sum\limits_{\gamma \neq \alpha, \beta} b_\gamma^2
  \frac{\sigma_{\gamma\beta}}{R_{\alpha\gamma}R_{\beta\gamma}} \,
  \psi(y_{\beta\gamma}) \nonumber  \\
  &\times& \Biggl \{ \biggl \{ \exp(-y_{\alpha\gamma})
  - \frac{w(y_{\alpha\gamma})}{4}
  \biggr \} \Bigl( \eb \cdot \nb_{\beta\gamma} \Bigr) - \frac{3}{4} \,
  w(y_{\alpha\gamma}) \nb_{\alpha\gamma}
  \Bigl( \nb_{\alpha\gamma} \times \nb_{\beta\gamma} \Bigr) \Biggr \},
  \quad \beta \neq \alpha,  \\
 \mub_{\alpha\alpha}^{rt}(\omega) &=&
  \Bigl( \mub_{\alpha\alpha}^{tr}(\omega) \Bigr)^T = \frac{1}{36 \pi \eta}
  \sum\limits_{\gamma \neq \alpha} b_\gamma^2
  \frac{\sigma_{\gamma\alpha}}{R_{\alpha\gamma}^2} \, \psi(y_{\alpha\gamma})
  \biggl \{ \exp(-y_{\alpha\gamma}) -\frac{w(y_{\alpha\gamma})}{4}
  \biggr \} \Bigl( \eb \cdot \nb_{\alpha\gamma} \Bigr),  \\
 \mub_{\alpha\beta}^{rt}(\omega) &=& \frac{k(b_\beta)}{8 \pi \eta R_{\alpha\beta}^2}
  \Bigl( 1 + \frac{b_\alpha^2}{6} \Bigr)
  \psi(y_{\alpha\beta}) \Bigl( \eb \cdot \nb_{\alpha\beta} \Bigr)
  + \frac{1}{36\pi \eta} \sum\limits_{\gamma \neq \alpha, \beta} b_\gamma^2
  \frac{\sigma_{\gamma\alpha}}{R_{\alpha\gamma}R_{\beta\gamma}} \,
  \psi(y_{\alpha\gamma}) \nonumber  \\
  &\times& \Biggl \{ \biggl \{ \exp(-y_{\beta\gamma})
  - \frac{w(y_{\beta\gamma})}{4}
  \biggr \} \Bigl( \eb \cdot \nb_{\alpha\gamma} \Bigr) + \frac{3}{4} \,
  w(y_{\beta\gamma})
  \Bigl( \nb_{\beta\gamma} \times \nb_{\alpha\gamma} \Bigr)
  \nb_{\beta\gamma} \Biggr \},  \quad \beta \neq \alpha.
\end{eqnarray}

\noindent  Here,

\begin{eqnarray}
 \mu_\alpha^t(\omega) &=& \frac{1}{\xi_\alpha} \biggl( 1 - b_\alpha +
  \frac{8}{9} \, b_\alpha^2 - \frac {7}{9}\, b_\alpha^3 \biggr),  \\
 \mu_\alpha^r(\omega) &=& \frac{1}{\xi_\alpha^r} \biggl[ 1 - \frac{b_\alpha^2}{3}
  \Bigl( 1 - b_\alpha \Bigr) \biggr ]
\end{eqnarray}

\noindent  are, respectively, the frequency-dependent translational and
rotational  mobility tensors of a single sphere calculated up to the terms of order $b^3$
\cite{ref.Saarl}, and

\begin{eqnarray}
 w(x) &=& \frac{6}{x^2} - \frac{4}{\pi} \, x \tilde{h}_2(x),  \\
 k(b) &=& \frac{\tau(b)}{s^t (b)} \approx 1 + \frac{b^2}{9}
 \biggl( \frac{7}{2} - 2 b \biggr).
\end{eqnarray}

Note that the obtained mobility tensors defined by relations
(6.3)--(6.10) satisfy the symmetry relations \cite{ref.Happel}

\begin{equation}
 \left(\mub_{\beta\alpha}^{tt}(\omega) \right)^T = \mub_{\alpha\beta}^{tt}(\omega), \
 \left(\mub_{\beta\alpha}^{rr}(\omega) \right)^T = \mub_{\alpha\beta}^{rr}(\omega), \
 \left(\mub_{\beta\alpha}^{rt}(\omega) \right)^T = \mub_{\alpha\beta}^{tr}(\omega), \
 \   \alpha, \beta = 1, 2, \ldots, N.
\end{equation}

In \cite{ref.Saarl}, the frequency-dependent mobility tensors are derived up to
the terms of the third order in two parameters  $b$  and  $\sigma$,  i.e. up to
the terms of  $b^n\sigma^m$,  where  $n + m = 3$ ("$n + m=3$ approximation").
According to one of the main conclusions in \cite{ref.Saarl}, the
frequency-dependent mobility tensors determined in this approximation contain
only the terms corresponding to the two-particle interactions, which is similar
to the behavior of the frequency-independent mobility tensors determined up to
the terms $\sigma^3$ ( or  $\sigma^4$  for $\mub_{\alpha\beta}^{rr}$)
\cite{ref.Mazur,ref.MazurSaarl}. Moreover, the diagonal ($\beta = \alpha$)
translational and rotational mobility tensors determined in \cite{ref.Saarl}
are independent of the hydrodynamic interactions between the spheres, i.e.,

\begin{equation}
 \mub_{\alpha\alpha}^{tt}(\omega) = \mu_\alpha^t(\omega) \Ib,  \qquad
 \mub_{\alpha\alpha}^{rr}(\omega) = \mu_\alpha^r(\omega) \Ib,
\end{equation}

\noindent  and, hence, the self-interaction of a sphere due the action of the
fluid induced by this sphere and reflected from the rest spheres is not
described within the framework of approximation  $n + m = 3$. Furthermore,
there is no coupling between translational and rotational motions of the same
sphere, i.e.,

\begin{equation}
 \mub_{\alpha\alpha}^{tr}(\omega) = 0,  \qquad \qquad \qquad
 \mub_{\alpha\alpha}^{rt}(\omega) = 0.
\end{equation}

Unlike \cite{ref.Saarl}, expressions (6.3)--(6.10) represented in the form of
expansions in the parameters  $b$  and  $\sigma$ contain the terms up to
$b^3\sigma^m$,  where $m = 2$  for  $\mub_{\alpha\beta}^{tt}(\omega)$,  $m = 3$
for $\mub_{\alpha\beta}^{tr}(\omega)$  and $\mub_{\alpha\beta}^{rt}(\omega)$,
and $m = 4$  for $\mub_{\alpha\beta}^{rr}(\omega)$, i.e., $n + m > 3$.  An
account of such terms in the mobility tensors leads to some new effects, caused
by the contribution of the three-particle hydrodynamic interactions to the
mobility tensors. This contribution is described by the terms proportional to
the sum $\sum\limits_{\gamma \neq \alpha, \beta}$ in Eqs.~(6.4), (6.6), (6.8),
and (6.10). It follows that the main terms corresponding to the three-particle
hydrodynamic interactions are proportional to $b^2\sigma^2$  for
$\mub_{\alpha\beta}^{tt}(\omega)$,  $b^2\sigma^4$  for
$\mub_{\alpha\beta}^{rr}(\omega)$,  and $b^2\sigma/R_{\alpha\beta}^2$  for
$\mub_{\alpha\beta}^{tr}(\omega)$  and  $\mub_{\alpha\beta}^{rt}(\omega)$. In
addition, the terms proportional to $\sum\limits_{\gamma \neq \alpha}$ in the
diagonal mobility tensors $\mub_{\alpha\alpha}^{tt}(\omega)$ and
$\mub_{\alpha\alpha}^{rr}(\omega)$  defined by relations (6.3) and (6.5)
describe the self-interaction of sphere  $\alpha$  due to the action of the
fluid induced by this sphere and reflected from the rest ones. This effect is
of the order of $b^2\sigma^2$ for $\mub_{\alpha\alpha}^{tt}(\omega)$ and
$b^2\sigma^4$ $\mub_{\alpha\alpha}^{rr}(\omega)$.  According to Eqs.~(6.7) and
(6.9) for the mobility tensors $\mub_{\alpha\alpha}^{tr}(\omega)$ and
$\mub_{\alpha\alpha}^{rt}(\omega)$, there is a coupling between translational
and rotational motions of the same sphere, which is of the order of
$b^2\sigma/R_{\alpha\gamma}^2$.

Now, let us examine  the mobility tensors defined by Eqs.~(6.3)--(6.10) in the
approximation $n + m = 3$. We see that the relations for
$\mub_{\alpha\alpha}^{tt}(\omega)$, $\mub_{\alpha\alpha}^{rr}(\omega)$ and
$\mub_{\alpha\beta}^{tr}(\omega)$, $\mub_{\alpha\beta}^{rt}(\omega)$, where
$\alpha, \beta = 1, 2, \ldots, N$,  coincide with the corresponding relations
given in \cite{ref.Saarl}.  For the rotational mobility tensors
$\mub_{\alpha\beta}^{rr}(\omega)$,  where $\beta \neq \alpha$, we get

\begin{equation}
 \mub_{\alpha\beta}^{rr}(\omega) = -\frac{\kappa^3}{24 \pi^2 \eta}
  \biggl \{ \tilde{h}_2(y_{\alpha\beta})
  \Bigl( \Ib - 3 \nb_{\alpha\beta}\nb_{\alpha\beta} \Bigr)
  + 2 \tilde{h}_0(y_{\alpha\beta}) \Ib \biggr\},
\end{equation}

\noindent which differs from relation (4.5) in \cite{ref.Saarl} by the second
term in the braces. [In terms of the notation used in the present paper,
relation (4.5) in \cite{ref.Saarl} is reduced to the form (6.18) with the
substitution of  $\ (1/2) \tilde{h}_0(y_{\alpha\beta}) \Bigl( 7 \Ib - 9
\nb_{\alpha\beta}\nb_{\alpha\beta} \Bigr) \ $  for $\
2\tilde{h}_0(y_{\alpha\beta})\Ib$.]

The translational mobility tensors  $\mub_{\alpha\beta}^{tt}(\omega)$,  where
$\beta \neq \alpha$, in approximation  $n + m = 3$ is reduced to the form

\begin{equation}
 \mub_{\alpha\beta}^{tt}(\omega) = \Tb_M (\Rb_{\alpha\beta},\omega)
  + \frac{2}{9} \, \Bigl( b_\alpha^2 + b_\beta^2 \Bigr)
  \Tb_M (R_{\alpha\beta}),
\end{equation}
which essentially differs from the translational mobility tensor given by
Eq.~(4.3) in \cite{ref.Saarl} because of the presence of the terms necessary
for holding the symmetry relations \cite{ref.Happel}.

In the particular case of two spheres ($\alpha, \beta = 1, 2$), Eqs.~(6.4),
(6.6), (6.8), and (6.10) for  $\beta \neq \alpha$ are simplified to the form

\begin{eqnarray}
 \mub_{12}^{tt}(\omega) &=& \mub_{21}^{tt}(\omega) = \biggl \{ 1 + \frac{2}{9}
  \, \Bigl [ b_1^2 (1 - b_1) + b_2^2 (1 - b_2) \Bigr ]\biggr \}
  \Tb_M (\Rb_{12},\omega),  \\
 \mub_{12}^{rr}(\omega) &=& \mub_{21}^{rr}(\omega) =
  -  \frac{\kappa^3}{24 \pi^2 \eta} \biggl \{\tilde{h}_2(y_{12})
  \Bigl( \Ib - 3 \nb_{12}\nb_{12} \Bigr)
  + 2 \tilde{h}_0(y_{12}) \Ib \biggr\},  \\
 \mub_{12}^{tr}(\omega) &=& \frac{k(b_1)}{8 \pi \eta R_{12}^2}
  \biggl( 1 + \frac{b_2^2}{6} \biggr)
  \psi(y_{12}) \Bigl( \eb \cdot \nb_{12} \Bigr),  \\
 \mub_{12}^{rt}(\omega) &=& \frac{k(b_2)}{8 \pi \eta R_{12}^2}
  \biggl( 1 + \frac{b_1^2}{6} \biggr)
  \psi(y_{12}) \Bigl( \eb \cdot \nb_{12} \Bigr),
\end{eqnarray}

\noindent  and the mobility tensors  $\mub_{21}^{tr}(\omega)$  and
$\mub_{21}^{rt}(\omega)$  are defined, respectively, by relations (6.22) and
(6.23) by interchanging  the indices 1 and 2.

Thus, in the case of two spheres, the translational and mobility tensors are
independently symmetric in the space indices ($i, j = x, y, z$) and in the
indices of spheres  $\alpha,\beta = 1, 2$, i.e.,

\begin{equation}
 \begin{array}{lllllll}
  \mub_{\alpha\beta}^{tt}(\omega) &=& \mub_{\beta\alpha}^{tt}(\omega), \qquad
   & \left(\mub_{\alpha\beta}^{tt}(\omega)\right)^T &=& \mub_{\alpha\beta}^{tt}(\omega),
   \qquad  & \alpha,\beta = 1, 2,  \\
  \mub_{\alpha\beta}^{rr}(\omega) &=& \mub_{\beta\alpha}^{rr}(\omega),  \qquad
   & \left(\mub_{\alpha\beta}^{rr}(\omega)\right)^T &=& \mub_{\alpha\beta}^{rr}(\omega),
   \qquad  & \alpha,\beta = 1, 2.
 \end{array}
\end{equation}

Analogously to the friction tensors, we investigate the behavior of the
mobility tensors in two special opposite cases.  First, we consider the case
$\Real y_{\alpha\beta} \gg 1$ corresponding to large distances between the
spheres. Using asymptotics (5.50) and

\begin{equation}
 w(y_{\alpha\beta}) \approx \frac{6}{y_{\alpha\beta}^2},
\end{equation}

\noindent we reduce the mobility tensors (6.3)--(6.10) to the form

\begin{eqnarray}
 \mub_{\alpha\beta}^{tt}(\omega) &\approx &  \mu_\alpha(\omega) \delta_{\alpha\beta} \,
  \Ib + \frac{1}{2 \xi_\alpha} \, \Biggl \{ \Bigl( 1 - \delta_{\alpha\beta} \Bigr)
  \frac{3 \sigma_{\alpha\beta}}{y_{\alpha\beta}^2} \, \biggl \{ 1 + \frac{2}{9}
  \Bigl [ b_\alpha^2 (1 - b_\alpha) + b_\beta^2 (1 - b_\beta) \Bigr ]\biggr \}
  \Bigl( 3\nb_{\alpha\beta}\nb_{\alpha\beta}  \nonumber  \\
  &-& \Ib \Bigr) + \sum\limits_{\gamma \neq \alpha,\beta} \frac{\sigma_{\alpha\gamma}
  \sigma_{\gamma\beta}^3}{y_{\alpha\gamma}^2} \, \Bigl \{ \Ib
  + 9 (\nb_{\alpha\gamma} \cdot \nb_{\beta\gamma})
  \nb_{\alpha\gamma} \nb_{\beta\gamma} - 3 \nb_{\alpha\gamma} \nb_{\alpha\gamma}
  - 3 \nb_{\beta\gamma} \nb_{\beta\gamma} \Bigr \} \Biggr \}, \nonumber  \\
  && \qquad \qquad \qquad \qquad \qquad \qquad \qquad \quad \ \
  \alpha, \beta = 1, 2, \ldots, N,
\end{eqnarray}

\begin{eqnarray}
 \mub_{\alpha\beta}^{rr}(\omega) &\approx & \mu_\alpha^r (\omega)
  \delta_{\alpha\beta} \,\Ib,
  \qquad \quad \ \ \alpha, \beta = 1, 2, \ldots, N, \\
 \mub_{\alpha\beta}^{tr}(\omega) &\approx & 0,
  \qquad \qquad \qquad \qquad  \alpha, \beta = 1, 2, \ldots, N, \\
 \mub_{\alpha\beta}^{rt}(\omega) &\approx & 0,
  \qquad \qquad \qquad \qquad \alpha, \beta = 1, 2, \ldots, N.
\end{eqnarray}

Note that the nonstationary mobility tensors  $\mub_{\alpha\beta}^{tr}(\omega)$
and  $\mub_{\alpha\beta}^{rt}(\omega)$,  $\  \alpha, \beta = 1, 2, \ldots, N$,
and the mutual rotational mobility tensors $\mub_{\alpha\beta}^{rr}(\omega)$,
$\beta \neq \alpha$,  exponentially vanish as  $\exp(-y_{\alpha\beta})$,
whereas in the stationary case, for $\beta\neq \alpha$,  these quantities
decrease as $1/R_{\alpha\beta}^2$  and $1/R_{\alpha\beta}^3$, respectively,
\cite{ref.MazurSaarl}.

Relations (6.27)--(6.29) coincide with the corresponding relations given in
\cite{ref.Saarl}, whereas the translational mobility tensor defined by (6.26)
differs from the corresponding result given in \cite{ref.Saarl}.  Indeed, in
approximation $n + m = 3$, relation (6.26) is simplified to the form

\begin{equation}
 \mub_{\alpha\beta}^{tt}(\omega) \approx  \mu_\alpha(\omega) \delta_{\alpha\beta} \,
  \Ib - \frac{1}{4\pi \eta R_{\alpha\beta} y_{\alpha\beta}^2} \,
  \biggl [ 1 + \frac{2}{9} \Bigl( b_\alpha^2  + b_\beta^2  \Bigr) \biggr ]
  \Bigl( \Ib - 3\nb_{\alpha\beta}\nb_{\alpha\beta} \Bigr),
\end{equation}

\noindent  which differs from Eq.~(4.6) in \cite{ref.Saarl}. Indeed, in terms
of notation used in the present paper, the latter takes the form

\begin{equation}
 \mub_{\alpha\beta}^{tt}(\omega) \approx  \mu_\alpha(\omega) \delta_{\alpha\beta} \,
  \Ib - \frac{1}{4\pi \eta R_{\alpha\beta} y_{\alpha\beta}^2} \,
  \biggl ( 1 + \frac{2}{9}\,  b_\beta^2 \frac{a_\beta}{a_\alpha} \biggr)
  \Bigl( \Ib - 3\nb_{\alpha\beta}\nb_{\alpha\beta} \Bigr),
\end{equation}

\noindent  which does not satisfy the symmetry relation
\cite{ref.Happel}.

Finally, consider  the mobility tensors in the low-frequency domain
$|y_{\alpha\beta}| \ll 1$.  Using Eqs.~(5.55), (5.56) and the series
\begin{equation}
 w(y_{\alpha\beta}) \approx  1 - \frac{y_{\alpha\beta}^2}{4}
 + \frac{2}{15}y_{\alpha\beta}^3 + ...,
\end{equation}

\noindent and retaining only the main frequency-dependent terms, we simplify
the set of the mobility tensors given by Eqs.~(6.3)--(6.10) to the form

\begin{equation}
 \mub_{\alpha\beta}^{\mu \nu}(\omega) = \mub_{\alpha\beta}^{\mu \nu}
  + \Mb_{\alpha\beta}^{\mu \nu}(\omega), \quad \mu, \nu = t,r, \quad
  \alpha, \beta = 1, 2, \ldots, N.
\end{equation}

\noindent  Here,  $\mub_{\alpha\beta}^{\mu \nu}, \ \mu, \nu = t,r, \ \alpha,
\beta = 1, 2, \ldots, N,$  are the stationary mobility tensors calculated
within the framework of the second iteration
\cite{ref.Mazur,ref.MazurSaarl,ref.Usenko}:

\begin{eqnarray}
 \mub_{\alpha\beta}^{tt} &=& \frac{1}{\xi_\alpha} \left \{ \delta_{\alpha\beta}
  \Ib + \left( 1 - \delta_{\alpha\beta} \right) \frac{3}{4} \, \sigma_{\alpha\beta}
  \left( \Ib + \nb_{\alpha\beta} \nb_{\alpha\beta} \right) \right \},  \\
 \mub_{\alpha\beta}^{rr} &=& \frac{1}{\xi_\alpha^r} \left \{ \delta_{\alpha\beta}
  \Ib + \left( 1 - \delta_{\alpha\beta} \right) \frac{\sigma_{\alpha\beta}^3}{2}
  \left(3 \nb_{\alpha\beta} \nb_{\alpha\beta} - \Ib \right) \right \},  \\
 \mub_{\alpha\beta}^{tr} &=& \left( 1 - \delta_{\alpha\beta} \right)
  \frac{1}{8\pi \eta R_{\alpha\beta}^2}
  \left( \eb \cdot \nb_{\alpha\beta} \right),  \\
 \mub_{\alpha\beta}^{rt} &=& \mub_{\alpha\beta}^{tr}.
\end{eqnarray}

The frequency-dependent terms $\Mb_{\alpha\beta}^{\mu \nu}(\omega), \ \mu, \nu
= t,r, \ \ \alpha, \beta = 1, 2, \ldots, N,$  which vanish as  $\omega \to 0$,
have the form

\begin{eqnarray}
 \Mb_{\alpha\beta}^{tt}(\omega) &\approx & -\frac{b_\alpha}{\xi_\alpha} \, \Ib,
  \\
 \Mb_{\alpha\beta}^{rr}(\omega) &\approx & -\frac{b_\alpha^2}{3 \xi_\alpha^r}
  \Bigl \{ \delta_{\alpha\beta} \, \Ib + \Bigl( 1 - \delta_{\alpha\beta} \Bigr)
  \frac{3}{4} \, \sigma_{\alpha\beta}
  \Bigl( \Ib + \nb_{\alpha\beta} \nb_{\alpha\beta} \Bigr) \Bigr \}  \nonumber  \\
  &+& \frac {1}{6 \xi_\alpha^r} \biggl( \frac{a_\alpha}{a_\beta} \biggr)^2
  \sum\limits_{\gamma \neq \alpha,\beta}
  b_\gamma^2 \sigma_{\alpha\gamma} \sigma_{\gamma\alpha} \sigma_{\beta\gamma}^2
  \biggl \{ \Bigl( \nb_{\alpha\gamma} \cdot \nb_{\beta\gamma} \Bigr) \Ib
  - \nb_{\beta\gamma} \nb_{\alpha\gamma} \biggr \},  \\
 \Mb_{\alpha\beta}^{tr}(\omega) &\approx & - \frac{1}{48\pi \eta} \Biggl \{
  \Bigl( 1 - \delta_{\alpha\beta} \Bigr)
   \frac{3 y_{\alpha\beta}^2}{R_{\alpha\beta}^2}
  \Bigl( \eb \cdot \nb_{\alpha\beta} \Bigr) \nonumber  \\
  &+&  \sum\limits_{\gamma \neq \alpha, \beta} b_\gamma^2
  \frac{\sigma_{\gamma\beta}}{R_{\alpha\gamma} R_{\beta\gamma}}
  \biggl \{ \Bigl( \eb \cdot \nb_{\beta\gamma} \Bigr)
  - \nb_{\alpha\gamma}
  \Bigl( \nb_{\alpha\gamma} \times \nb_{\beta\gamma} \Bigr) \biggl \}\Biggr \}, \\
 \Mb_{\alpha\beta}^{rt}(\omega) &\approx & -\frac{1}{48\pi \eta} \Biggl \{
  \Bigl( 1 - \delta_{\alpha\beta} \Bigr)
  \frac{3 y_{\alpha\beta}^2}{R_{\alpha\beta}^2}
  \Bigl( \eb \cdot \nb_{\alpha\beta} \Bigr) \nonumber  \\
  &-&  \sum\limits_{\gamma \neq \alpha, \beta} b_\gamma^2
  \frac{\sigma_{\gamma\alpha}}{R_{\alpha\gamma} R_{\beta\gamma}}
  \biggl \{ \Bigl( \eb \cdot \nb_{\alpha\gamma} \Bigr)
  + \Bigl( \nb_{\beta\gamma} \times \nb_{\alpha\gamma} \Bigr)
  \nb_{\beta\gamma} \biggl \} \Biggr \}.
\end{eqnarray}

The frequency-dependent component  $\Mb_{\alpha\beta}^{tt}(\omega)$  of the
translational mobility tensor coincides with the corresponding result given in
\cite{ref.Saarl}.  In \cite{ref.Saarl}, the mobility tensors in the
low-frequency range are calculated to the first order in the parameters  $b$
and  $\sigma$, and, within the framework of this approximation, for  $\beta
\neq \alpha$,

\begin{eqnarray}
 \mub_{\alpha\beta}^{rt}(\omega) &=& 0, \qquad
  \mub_{\alpha\beta}^{tr}(\omega) = 0, \qquad  \alpha, \beta = 1, 2,\ldots,N,
  \nonumber  \\
 \mub_{\alpha\beta}^{rr}(\omega) &=& 0, \qquad \qquad \qquad \qquad \quad
  \beta \neq \alpha.
\end{eqnarray}

As is seen from relations (6.39)--(6.41) derived in the present paper within
the framework of the approximation discussed above, the mobility tensors
$\mub_{\alpha\beta}^{rr}(\omega)$, $\mub_{\alpha\beta}^{tr}(\omega)$, and
$\mub_{\alpha\beta}^{rt}(\omega)$  for $\alpha, \beta = 1, 2,\ldots,N$, are not
equal to zero but proportional to $\omega$ in the low-frequency range.

%%--------------------------------Section 7------------------------------

\section{Conclusion}  \label{Conclusion}

In the present paper, we calculate the distributions of the
velocity and pressure fields in an unbounded incompressible viscous fluid
induced in the nonstationary case by an arbitrary number of spheres moving and
rotating in it as well as the time-dependent forces and torques exerted by the
fluid on the spheres.  All the quantities of interest are expressed in terms of the
induced surface force densities without imposing any additional restrictions on the
size of spheres, distances between them, and frequency range.  Within the
framework of this approach, we derived the general relations for the
translational, rotational, and coupled friction and mobility tensors in the
case of an arbitrary number of spheres up to the terms of the order $b^3$ and $\sigma^2$,
$\sigma^4$, and   $\sigma^3$, respectively, and  analyzed the contributions to the mobility tensors
caused by the effects of the three-particle hydrodynamic interactions and the self-interaction
of spheres. We investigated as well the
influence of retardation effects on the behavior of the friction and mobility
tensors in the low-frequency range and for large distances between the spheres.

Finally, we formulated the procedure which allow one to calculate both the velocity and pressure
fields in the fluid and the forces and torques exerted by the fluid on the spherical
particles up to the terms
of  the order $\sigma^p$,  where  $p$  is any positive integer. With the use of this procedure,
the hydrodynamic interaction in a viscous fluid can be described analytically within to the
terms of any given order in the parameter $\sigma$.

%%--------------------------------Appendix------------------------------

%\newpage
\appendix
\section{}

The quantities  $\Kb_{l_1 m_1, lm}^{l_2 m_2}$  and  $\Wb_{l_1 m_1, lm}^{l_2
m_2}$,  where  $l_1, l_2, l \geq 0, \ l_1 \geq m_1 \geq -l_1, \ l_2 \geq m_2
\geq -l_2$,  and $\ l \geq m \geq -l$  defined by relations (2.50) and (2.54),
respectively, can be represented in the explicit form in terms of the Wigner
3j-symbols \cite{ref.Varshalovich,ref.Davydov} as follows (for details, see
\cite{ref.Usenko,ref.Yosh}):

\begin{eqnarray}
 \Kb_{l_1 m_1, lm}^{l_2 m_2} &=& \frac{i^{l_1 - l_2 + l}}{3\sqrt{\pi}}
  \sqrt{(2l_1 +1)(2l_2 +1)(2l + 1)} \,(-1)^{m_1 + m}
  \left \{
   \left ( \begin{array}{rrr}
   l_1  &  l_2  &  l  \\
   0  &  0  &  0  \\
   \end{array} \right)
  \left ( \begin{array}{rrr}
   l_1  &  l_2  &  l  \\
  -m_1  &  m_2  &  m  \\
  \end{array} \right)
 \Ib \right. \nonumber  \\
 &-& (-1)^m \frac{1}{2} \sqrt{\frac{3}{2}} \sum\limits_{k = -2}^2
 (-1)^k \Kb_k \sum\limits_{j = j_{min}}^1 (2L + 1)
  \left ( \begin{array}{rrr}
  2  &  l  &  L  \\
  0  &  0  &  0  \\
  \end{array} \right)
  \left ( \begin{array}{rrr}
  l_1  &  l_2  &  L  \\
  0    &  0    &  0  \\
  \end{array} \right)  \nonumber  \\
 & \times & \left.
  \left ( \begin{array}{rrl}
  2  &   l  &  L  \\
  k  &  -m  &  m - k  \\
  \end{array} \right)
  \left ( \begin{array}{rrl}
   l_1   &  l_2  &  L  \\
  -m_1   &  m_2  &  k - m  \\
  \end{array} \right)
 \right \},  \qquad \qquad  l_1,l_2,l \geq 0,
\end{eqnarray}

\noindent  where

\begin{eqnarray}
 L = l + 2j,  \qquad \qquad
  j_{min} = \left \{
  \begin{array}{rll}
   1,  \quad  &  \mbox{if}  &  \quad l =    0  \\
   0,  \quad  &  \mbox{if}  &  \quad l =    1  \\
  -1,  \quad  &  \mbox{if}  &  \quad l \geq 2, \\
  \end{array}  \right.  \nonumber
\end{eqnarray}

\begin{eqnarray}
 \Kb_0 &=& \sqrt{\frac{2}{3}} \left (-\eb_x \eb_x - \eb_y \eb_y + 2\eb_z\eb_z
  \right) = \sqrt{\frac{2}{3}} \left (\eb_1 \eb_{-1} + \eb_{-1} \eb_1
  + 2 \eb_0 \eb_0 \right),  \nonumber  \\
 \Kb_1 &=& \eb_x \eb_z + \eb_z \eb_x -i\left (\eb_y \eb_z + \eb_z \eb_y \right)
  = -\sqrt{2} \left (\eb_{-1} \eb_0 + \eb_0 \eb_{-1} \right), \nonumber  \\
 \Kb_{-1} &=& -\Kb_1^* = -\sqrt{2} \left (\eb_1 \eb_0 + \eb_0 \eb_1 \right), \\
  \nonumber
 \Kb_2 &=& \eb_x \eb_x - \eb_y \eb_y - i\left (\eb_x \eb_y + \eb_y \eb_x \right)
   = 2 \eb_{-1} \eb_{-1},  \nonumber  \\
 \Kb_{-2} &=& \Kb_2^* = 2 \eb_1 \eb_1,  \nonumber
\end{eqnarray}

\noindent  and

\begin{eqnarray}
 \Wb_{l_1 m_1, lm}^{l_2 m_2} &=& \frac{i^{l_1 - l_2 + l - 1}}{2\sqrt{\pi}}
  \sqrt{(2l_1 +1)(2l_2 +1)(2l + 1)} \,(-1)^{m - m_2}
  \sum\limits_{k = -1}^1 \eb_k^* \sum\limits_{j = j_{min}}^1 (2L + 1)
  \nonumber  \\
   & \times &\left ( \begin{array}{rrr}
   1  &  l  &  L  \\
   0  &  0  &  0  \\
   \end{array} \right)
   \left ( \begin{array}{rrr}
   l_1  &  l_2  &  L  \\
   0    &  0    &  0  \\
   \end{array} \right)
   \left ( \begin{array}{rrl}
   1  &   l  &  L  \\
   k  &  -m  &  m - k  \\
   \end{array} \right)
   \left ( \begin{array}{rrl}
    l_1  &  l_2  &  L  \\
   -m_1  &  m_2  &  k - m  \\
   \end{array} \right), \nonumber \\
   & & \qquad \qquad \qquad \qquad \qquad \qquad \qquad \qquad \qquad
   \qquad \qquad  l_1,l_2,l \geq 0.
\end{eqnarray}

\noindent  where

\begin{eqnarray}
 L = l + 2j -1,  \qquad \qquad
  j_{min} = \left \{
  \begin{array}{rll}
   1,  \quad  &  \mbox{if}  &  l =    0  \\
   0,  \quad  &  \mbox{if}  &  l \geq 1.  \\
  \end{array}  \right.  \nonumber
\end{eqnarray}

Note that  $\Kb_{l_1 m_1, lm}^{l_2 m_2} \neq 0$  only for

\begin{equation}
 l = l_1 + l_2 - 2p \geq 0,
\end{equation}

\noindent  where  $p = -1,0,1,\ldots,p_{max},$  $p_{max} = \min
\Bigl(\left[(l_1 + l_2)/2\right], 1 + \min (l_1, l_2) \Bigr)$, $\left[a\right]$
is the integer part of  $a$,  and  $\min (a,b)$ means the smallest quantity of
$a$ and  $b$, and  $ \Wb_{l_1 m_1, lm}^{l_2 m_2} \neq 0$  only for

\begin{equation}
 l = l_1 + l_2 - 2p + 1 \geq 0,
\end{equation}

\noindent  where  $p = 0,1,\ldots,\tilde{p}_{max},$  $\tilde{p}_{max} = \min
\Bigl(\left[(l_1 + l_2 + 1)/2\right], 1 + \min (l_1, l_2) \Bigr)$.

This means that the quantities $F_{l_1
l_2,l}(r_\alpha,a_\beta,R_{\alpha\beta},\omega)$  and $C_{l_1
l_2,l}(r_\alpha,a_\beta,R_{\alpha\beta})$  defined by relations (2.49) and
(2.53) for  $\beta \neq \alpha$  should be determined only for the values of
$l$ given by (A4) and (A5), respectively. The explicit form for these
quantities is as follows \cite{ref.Usenko}:

\begin{eqnarray}
 F_{l_1 l_2,l}(r_\alpha,a_\beta,R_{\alpha\beta},\omega) &=&
  (-1)^p \frac{2\kappa}{\pi\eta} \left\{ \tilde{j}_{l_1}(x_\alpha)
  \tilde{j}_{l_2}(b_\beta) \tilde{h}_l(y_{\alpha\beta})
  - \delta_{l, l_1 + l_2 + 2}
  \frac{\pi^{3/2}}{2}
  \frac{\Gamma\left(l_1 + l_2 + \frac{5}{2} \right)}
  {\Gamma\left(l_1 + \frac{3}{2} \right)\Gamma\left(l_2 + \frac{3}{2} \right)}
  \right. \nonumber \\
  &\times &  \left. \frac{r_\alpha^{l_1} \sigma_{\beta\alpha}^{l_2}}
  {y_{\alpha\beta}^3 R_{\alpha\beta}^{l_1}} \right \},
  \quad  l = l_1 + l_2 -2p \geq 0,  \quad  p = -1, 0, 1,\ldots, p_{max},  \\
 C_{l_1 l_2,l}(r_\alpha,a_\beta,R_{\alpha\beta}) &=&
  \delta_{l, l_1 + l_2 + 1} \frac{\sqrt{\pi}}{2}
  \frac{\Gamma\left(l_1 + l_2 + \frac{3}{2} \right)}
  {\Gamma\left(l_1 + \frac{3}{2} \right)\Gamma\left(l_2 + \frac{3}{2} \right)}
  \frac{r_\alpha^{l_1} \sigma_{\beta\alpha}^{l_2}}{R_{\alpha\beta}^{l_1 + 2}},
  \nonumber \\
  && \qquad \qquad \qquad l = l_1 + l_2 - 2p + 1 \geq 0,
  \qquad  p = 0, 1,\ldots, \tilde{p}_{max},
\end{eqnarray}

\noindent  for $r_\alpha \leq R_{\alpha\beta} - a_\beta$  and

\begin{eqnarray}
 F_{l_1 l_2,l}(r_\alpha,a_\beta,R_{\alpha\beta},\omega) &=&
  (-1)^{l_{2} - p} \, \frac{2\kappa}{\pi\eta} \Biggl \{ \tilde{h}_{l_1}(x_\alpha)
  \tilde{j}_{l_2}(b_\beta) \tilde{j}_l(y_{\alpha\beta})
  - \delta_{l, l_1 - l_2 - 2}
  \frac{\pi^{3/2}}{2}  \nonumber \\
  &\times & \frac{\Gamma\left(l_1 + \frac{1}{2} \right)}
  {\Gamma\left(l_2 + \frac{3}{2} \right)
  \Gamma\left(l_1 - l_2 - \frac{1}{2} \right)}
  \frac{\sigma_{\beta\alpha}^{l_2} R_{\alpha\beta}^{l_1 - 2}}
  {x_\alpha^3 r_\alpha^{l_1 - 2}} \Biggr \}, \nonumber \\
  && \qquad \qquad \qquad l = l_1 + l_2 - 2p \geq 0,
  \qquad  p = -1, 0, 1,\ldots, p_{max},
\end{eqnarray}

\begin{eqnarray}
 C_{l_1 l_2,l}(r_\alpha,a_\beta,R_{\alpha\beta}) &=&
  \frac{\sqrt{\pi}}{2} \frac{\Gamma\left(l_1 + \frac{1}{2} \right)}
  {\Gamma\left(l_2 + \frac{3}{2} \right)}
  \frac{\sigma_{\beta\alpha}^{l_2} R_{\alpha\beta}^{l_1 - 1}}
  {r_\alpha^{l_1 + 1}} \left \{ \delta_{l, l_1 - l_2 - 1} \,
  \frac{1}{\Gamma\left(l_1 - l_2 + \frac{1}{2} \right)}
  +  \delta_{l, l_1 - l_2 - 3} \right. \nonumber  \\
  &\times & \left. \frac{2}{\Gamma\left(l_1 - l_2 - \frac{3}{2} \right)}
  \left [ \frac{1}{2l_1 - 1} \left (\frac{r_\alpha}{R_{\alpha\beta}} \right )^2
  - \frac{\sigma_{\beta\alpha}^2}{2l_2 + 3}
  - \frac{1}{2l + 3} \right ] \right \}
  \nonumber \\
  && \qquad \qquad \qquad l = l_1 + l_2 - 2p + 1 \geq 0,
  \qquad  p = 0, 1,\ldots, \tilde{p}_{max},
\end{eqnarray}

\noindent  for  $r_\alpha \geq R_{\alpha\beta} + a_\beta$.

\noindent  Here,  $\tilde{j}_l(x) = \sqrt{\pi/(2x)} I_{l +\frac{1}{2}}(x)$ and
$\tilde{h}_l(x) = \sqrt{\pi/(2x)} K_{l +\frac{1}{2}}(x)$  are the modified
spherical Bessel functions of the first and third kind, respectively,
\cite{ref.Abram},  $\Gamma(z)$  is the gamma function, $\sigma_{\beta\alpha} =
a_\beta/R_{\alpha\beta}$, $x_\alpha = \kappa r_\alpha$, and $y_{\alpha\beta} =
\kappa R_{\alpha\beta}$.

In the particular case  $l = 0$,  the general relations (A1) and (A3) for the
quantities  $\Kb_{l_1 m_1, lm}^{l_2 m_2}$  and  $\Wb_{l_1 m_1, lm}^{l_2 m_2}$
are simplified to the form

\begin{eqnarray}
 \Kb_{l_1 m_1, 00}^{l_2 m_2} &=&
  \frac{i^{l_1 - l_2}}{2 \sqrt{\pi}} \int \!\! d\Omega_k \, (\Ib - \nb_k
  \nb_k)\,
  Y_{l_1 m_1}^*(\theta_k,\varphi_k) Y_{l_2 m_2}(\theta_k,\varphi_k) \nonumber \\
  &=& \frac{1}{3\sqrt{\pi}} \left \{ \delta_{l_1, l_2} \, \delta_{m_1, m_2} \, \Ib
  - \sqrt{\frac{3 (2l_1 + 1)(2l_2 + 1)}{8}} \, i^{l_1 - l_2} (-1)^{m_1}
   \left ( \begin{array}{rrr}
   l_1  &  l_2  &  2  \\
    0  &    0  &  0  \\
   \end{array} \right)
  \right.  \nonumber  \\
  &\times & \left. \sum\limits_{k = -2}^2 \delta_{k, m_1 - m_2}
   \left ( \begin{array}{rrr}
    l_1  &  l_2  &  2  \\
   -m_1  &  m_2  &  k  \\
   \end{array} \right)
  \Kb_k \right \}
\end{eqnarray}

\noindent  and

\begin{eqnarray}
 \Wb_{l_1 m_1, 00}^{l_2 m_2} &=&
  \frac{i^{l_1 - l_2 - 1}}{2 \sqrt{\pi}} \int \!\! d\Omega_k \, \nb_k \,
  Y_{l_1 m_1}^*(\theta_k,\varphi_k) Y_{l_2 m_2}(\theta_k,\varphi_k) \nonumber \\
  &=& \frac{i^{l_1 - l_2 - 1}}{2\sqrt{\pi}}
  \sqrt{(2l_1 + 1)(2l_2 + 1)} (-1)^{m_1}
   \left ( \begin{array}{rrr}
   l_1  &  l_2  &  1  \\
    0  &    0  &  0   \\
   \end{array} \right) \nonumber  \\
  &\times& \sum\limits_{k = -1}^1 \delta_{k, m_1 - m_2}
   \left ( \begin{array}{rrr}
    l_1  &  l_2  &  1  \\
   -m_1  &  m_2  &  k  \\
   \end{array} \right)
   \eb_k^*.
\end{eqnarray}

With regard for the properties of 3j-symbols, we have

\begin{equation}
 \Kb_{l_1 m_1, 00}^{l_2 m_2} \neq 0 \quad  \mbox{only for}  \quad
 l_2 = l_1 + 2p \geq 0, \quad \mbox{where} \quad
  p = \left \{
  \begin{array}{rrl}
   0, 1,     \quad  &  \mbox{if}  &  l_1 =    0,1  \\
   0, \pm 1, \quad  &  \mbox{if}  &  l_1 \geq 2,
  \end{array}  \right.
\end{equation}

\noindent  and

\begin{equation}
 \Wb_{l_1 m_1, 00}^{l_2 m_2} \neq 0 \quad  \mbox{only for}  \quad
 l_2 = l_1 + 2p +1 \geq 0,  \quad \mbox{where} \quad
  p = \left \{
  \begin{array}{rrl}
   0,     \quad  &  \mbox{if}  &  l_1 =    0  \\
   0, -1, \quad  &  \mbox{if}  &  l_1 \geq 1.
  \end{array}  \right.
\end{equation}

According to (A13), the quantity  $P_{l_1
2,l}(r_\alpha,a_\beta,R_{\alpha\beta},\omega)$  contained in relation (2.40)
with $l_2 = 2$  should be determined only for $l = l_1 \pm 1 \geq 0$.  For this
value of $l$, we have \cite{ref.Usenko}

\begin{equation}
 P_{l_1 2,l}(r_\alpha,a_\beta,R_{\alpha\beta},\omega) = (-1)^p
  \frac{2}{\pi \eta} \, \tilde{j}_{l_1}(x_\alpha)
  \tilde{j}_2(b_\beta) \tilde{h}_l(y_{\alpha\beta})
\end{equation}

\noindent  for  $r_\alpha \leq R_{\alpha\beta} - a_\beta$  and

\begin{equation}
 P_{l_1 2,l}(r_\alpha,a_\beta,R_{\alpha\beta},\omega) = (-1)^{p + 1}
  \frac{2}{\pi \eta} \, \tilde{h}_{l_1}(x_\alpha)
  \tilde{j}_2(b_\beta) \tilde{j}_l(y_{\alpha\beta})
\end{equation}

\noindent  for  $r_\alpha \geq R_{\alpha\beta} + a_\beta$.

According to (A12) and (A13), the quantities
$F_{l_1l_2}(r_\alpha,a_\alpha,\omega)$  and $C_{l_1l_2}(r_\alpha,a_\alpha)$,
should be determined only for $l_2 = l_1 + 2p \ge 0$,  where  $p = 0, \pm 1$,
and  $l_ 2 = l_1 \pm 1 \geq 0$, respectively.  As a result, for  $r_\alpha \geq
a_\alpha$,  we get \cite{ref.Usenko}

\begin{eqnarray}
 F_{l_1 l_2}(r_\alpha,a_\alpha,\omega) &=&
  (-1)^p \, \frac{2\kappa}{\pi \eta} \left \{
  \tilde{h}_{l_1}(x_\alpha) \tilde{j}_{l_2}(b_\alpha)
  - \delta_{l_2, l_1 -2} \left (l_1 - \frac{1}{2} \right) \frac{\pi}{x_\alpha^3}
  \left( \frac{a_\alpha}{r_\alpha} \right)^{l_1 -2} \right \}, \nonumber \\
  && \qquad \qquad \qquad \qquad l_2 = l_1 + 2p \geq 0,  \quad p = 0, \pm 1,
\end{eqnarray}

\begin{equation}
 C_{l_1 l_2}(r_\alpha,a_\alpha) = \delta_{l_2, l_1 - 1} \frac{1}{r_\alpha^2}
  \left( \frac{a_\alpha}{r_\alpha} \right)^{l_1 - 1}
  + \delta_{r_\alpha, a} \, \frac{1}{2 a_\alpha^2}
  \left( \delta_{l_2, l_1 + 1} - \delta_{l_2, l_1 - 1} \right),
  \quad  l_2 = l_1 \pm 1 \geq 0.
\end{equation}

\noindent  For $P_{1,2}(r_\alpha,a_\alpha,\omega)$,  where   $r_\alpha \geq
a_\alpha$,  we have

\begin{equation}
 P_{1,2}(r_\alpha,a_\alpha,\omega) =
  \frac{2}{\pi \eta} \, \tilde{h}_1(x_\alpha) \tilde{j}_2(b_\alpha).
\end{equation}

In the particular case  $r_\alpha = 0$, we have

\begin{equation}
 C_{l_1 l_2,l}(0,a_\beta,R_{\alpha\beta}) = \delta_{l_1,0} \,
  \delta_{l,l_2 + 1}\,
  \frac{\sigma_{\beta\alpha}^{l_2}}{R_{\alpha\beta}^2} \, ,
  \qquad l = l_2 \pm 1 \geq 0,
\end{equation}

\begin{equation}
 \lim\limits_{r_\alpha \to 0} C_{l_1 l_2}(r_\alpha,a_\alpha)
  =  \delta_{l_1,0}\, \delta_{l_2,1} \, \frac{1}{a_\alpha^2}\, .
\end{equation}

\section{On the Validity of Solutions of the Linearized Navier--Stokes Equation}

All relations obtained in the present paper are true within the framework of
the linearized Navier--Stokes equation.  Using the explicit form for the
velocity and pressure fields of the fluid induced by a single sphere or a
system of noninteracting spheres, we consider several problems of the
correctness of the linearization of the Navier--Stokes equation in the
nonstationary case.  The Fourier transform of the linearized Navier--Stokes
equation (2.1) has the form

\begin{equation}
 \nablab p(\rb,\omega) - \bigl( i \omega \rho + \eta \triangle \bigr) \vb(\rb,\omega)
  = -\rho \nablab \varphi(\rb,\omega),
\end{equation}

\noindent  where  $\triangle$  is the Laplace operator.  The linearized
equation (B1) is valid
\cite{ref.Happel,ref.Lamb,ref.Milne,ref.Batch,ref.Landau,ref.Loitsyan} provided
that the force acting on a unit volume of the fluid due to the fluid viscosity

\begin{equation}
 \Fb^{vis}(\rb,t) = \eta \triangle \, \vb(\rb,t)
\end{equation}

\noindent  is much greater than the nonlinear (inertial) term

\begin{equation}
 \Fb^{nl}(\rb,t) = -\rho \Bigl ( \vb(\rb,t) \cdot \nablab \Bigr ) \, \vb(\rb,t).
\end{equation}

\noindent  To estimate the possibility to neglect the nonlinear term, we use
the following condition:

\begin{equation}
 \overline{|\Fb^{vis}(\rb,t)|} \gg \overline{|\Fb^{nl}(\rb,t)|},
\end{equation}

\noindent  where the bar in (B4) means averaging over a certain time interval.

Since in the approximation of noninteracting spheres, the rotation of spheres
does not make a contribution to the fluid pressure [relation (3.21)], we
rewrite Eq.~(B1) in terms of the notation used in the present paper as follows:

\begin{equation}
 \nablab p^{(t,0)ind}(\rb,\omega) - \bigl( i \omega \rho + \eta \triangle \bigr)
  \Bigl( \vb^{(t,0)ind}(\rb,\omega) + \vb^{(r,0)ind}(\rb,\omega) \Bigr) = 0,
\end{equation}

\noindent  where

\begin{eqnarray}
 \vb^{(t,0)ind}(\rb,\omega) &=& \sum\limits_{\beta = 1}^N
  \vb^{(S,t,0)ind}_\beta (\rb,\omega),  \\
 \vb^{(r,0)ind}(\rb,\omega) &=& \vb^{(S, r,0)ind}(\rb,\omega) + \vb^{(V)ind}(\rb,\omega)
  = \sum\limits_{\beta = 1}^N \vb^{(r,0)ind}_\beta (\rb,\omega),  \\
 p^{(t,0)ind}(\rb,\omega) &=& \sum\limits_{\beta = 1}^N
  p^{(t,0)ind}_\beta (\rb,\omega).
\end{eqnarray}

We represent the Fourier transform of the viscous force $\Fb^{vis}(\rb,\omega)$
as a sum of two components

\begin{equation}
 \Fb^{vis}(\rb,\omega) = \Fb^{vis (t)}(\rb,\omega) + \Fb^{vis (r)}(\rb,\omega),
\end{equation}

\noindent  where

\begin{eqnarray}
 \Fb^{vis (t)}(\rb,\omega) &=& \eta \triangle \, \vb^{(t,0)ind}(\rb,\omega), \\
 \Fb^{vis (r)}(\rb,\omega) &=& \eta \triangle \, \vb^{(r,0)ind}(\rb,\omega)
  = - i \omega \rho \vb^{(r,0)ind}(\rb,\omega)
\end{eqnarray}

\noindent  are the viscous forces due to the translational motion of the
spheres and their rotation, respectively.  As is seen from relation (B11), the
nonstationarity of the problem leads to the appearance of the component of the
viscous force induced by rotation of spheres.

First, we consider the case of a single sphere  $\alpha$.  To verify condition
(B4), the general relations for the fluid velocities
$\vb^{(t,0)ind}(\rb,\omega)$  and   $\vb^{(r,0)ind}(\rb,\omega)$  in the entire
frequency range are necessary instead of particular relations (3.29) and
(3.25). To this end, we use the well-known results for the fluid velocity
induced by a single sphere moving with arbitrary velocity  $\Ub_\alpha(t)$
\cite{ref.Schram}

\begin{eqnarray}
 \vb^{(t,0)ind}(\rb,\omega) &=& \frac{3}{2 x_\alpha^2} \frac{a_\alpha}{r_\alpha}
  \biggl \{ p(b_\alpha) \Bigl ( 3 \nb_\alpha \nb_\alpha - \Ib \Bigr)
  + \Bigl \{ \Bigl( 1  +  x_\alpha + x_\alpha^2 \Bigr) \Ib  \nonumber  \\
  &-& \Bigl( 3 +  3 x_\alpha + x_\alpha^2 \Bigr) \nb_\alpha \nb_\alpha \Bigr \}
  \exp(-(x_\alpha - b_\alpha)) \biggr \} \Ub_\alpha (\omega)
\end{eqnarray}

\noindent  or rotating with angular velocity  $\Omb_\alpha(t)$
\cite{ref.Landau}

\begin{equation}
 \vb^{(r,0)ind}(\rb,\omega) = \biggl( \frac{a_\alpha}{r_\alpha} \biggr)^3
  \frac{1 + x_\alpha}{1 + b_\alpha} \exp(-(x_\alpha - b_\alpha)) \,
  \Bigl( \Omb_\alpha(\omega) \times \rb_\alpha \Bigr).
\end{equation}

Substituting (B12) and (B13), respectively, into (B10) and (B11), after certain
transformations, we obtain the following relations for the Fourier transforms
of the translational and rotational components of the viscous force:

\begin{eqnarray}
 \Fb^{vis(t)}(\rb,\omega) &=& \eta \frac{3 a_\alpha}{2 r_\alpha^3}
  \biggl \{ \Bigl( 1  +  x_\alpha + x_\alpha^2 \Bigr) \Ib
  \nonumber  \\
  &-& \Bigl( 3 +  3 x_\alpha + x_\alpha^2 \Bigr) \nb_\alpha \nb_\alpha\biggr \}
  \exp(-(x_\alpha - b_\alpha)) \, \Ub_\alpha (\omega),  \\
 \Fb^{vis (r)}(\rb,\omega) &=& \eta \frac{b_\alpha^2 a_\alpha}{r_\alpha^3} \,
  \frac{1 + x_\alpha}{1 + b_\alpha} \exp(-(x_\alpha - b_\alpha)) \,
  \Bigl( \Omb_\alpha(\omega) \times \rb_\alpha \Bigr),
\end{eqnarray}

\noindent  which are valid for any  $r_\alpha \ge a_\alpha$.

It should be noted that the first term in (B12), being the main term of the
fluid velocity induced by the moving sphere for distances  $\Real x_\alpha \gg
1$, gives no contribution to the viscous force.  As a result, the quantity
$\Fb^{vis(t)}(\rb,\omega)$  exponentially decreases in this range despite the
fact that  $\vb^{(t,0)ind}(\rb,\omega)$  is proportional to  $1/r_\alpha^3$.

Lets us investigate the validity of solutions (B12) and (B13) of the linear
Navier--Stokes equation in two particular cases corresponding to a sphere
oscillating in a fluid with translational  $\Ub_\alpha(t) = \Ub_\alpha \cos
\omega t $  or rotational  $\Omb_\alpha(t) = \Omb_\alpha \cos \omega t $
velocities, where the frequency of oscillations  $\omega$  is such that
$|b_\alpha| \ll 1$. We investigate these problems in the domain far from the
sphere: $\Real x_\alpha \gg 1$.  To this end, using (B14) and (B15), we
represent $\Fb^{vis(\zeta)}(\rb,t)$,  where  $\zeta = t, r$,  as follows:

\begin{equation}
 \Fb^{vis(\zeta)}(\rb,t) = \Real\biggl( \Fb^{vis(\zeta)}(\rb,\omega)
  \exp(-i \omega t) \biggr), \qquad \qquad \zeta = t, r,
\end{equation}

\noindent  where

\begin{eqnarray}
 \Fb^{vis(t)}(\rb,\omega) &\approx& \frac{3 \eta \kappa b_\alpha}{2 r_\alpha}
  \exp(-x_\alpha)\Bigl ( \Ib - \nb_\alpha \nb_\alpha \Bigr)\, \Ub_\alpha,  \\
 \Fb^{vis (r)}(\rb,\omega) &\approx& \eta \frac{b_\alpha^3}{r_\alpha^2}
  \exp(-x_\alpha) \, \Bigl( \Omb_\alpha(\omega) \times \rb_\alpha \Bigr).
\end{eqnarray}

Substituting the fluid velocities defined by relations (3.29) and (3.30) for
$\Real x_\alpha \gg 1$  into (B3) and averaging the estimated forces, we obtain
that in this space domain, inequality (B4) is true for a sphere rotating with
oscillating angular velocity for any distances $r_\alpha$  (except for the
particular case where $\rb_\alpha \parallel \Omb_\alpha(\omega)$ for which both
$\Fb^{vis (r)}(\rb,\omega)$ and the corresponding nonlinear force are equal to
zero), whereas, for a sphere oscillating with translational velocity,
inequality (B4) is reduced to the form

\begin{equation}
 |\sin \psi| \exp(-\Real x_\alpha) \gg \Reyn \, \biggl( \frac{3}{2 |r_\alpha|^2}
  \biggr)^{5/2} \, |p(b_\alpha)|^2 \, \frac{a_\alpha}{r_\alpha}\,
  \Bigl(sin^2 \psi + 8 \cos^2 \psi (1 + \cos^2 \psi)\Bigr)^{1/2},
\end{equation}

\noindent  where

\begin{equation}
\Reyn = \frac {U_\alpha a_\alpha}{\nu}
\end{equation}

\noindent  is the Reynolds number,  $U_\alpha = |\Ub_\alpha|$,  and  $\psi$  is
the angle between the vectors  $\Ub_\alpha$  and  $\rb_\alpha$.

Since  $\Real x_\alpha \gg 1$,  inequality (B19) is not valid for any
$r_\alpha$ in this domain.  Therefore, expression (3.29) for the fluid velocity
induced by an oscillating sphere far from it obtained within the framework of
the linear approximation cannot be regarded as correct  in this space domain
for the problem considered because, in the general case, the nonlinear term is
much greater than the viscous force.

Using the results obtained above, we can make analogous conclusions for a
system of oscillating spheres in the approximation of noninteracting particles.
In particular, the fluid velocity and pressure induced in the far zone by
spheres oscillating with velocities  $\Ub_\beta(t) = \Ub_\beta \cos \omega t$,
where $\beta = 1, 2,\ldots,N$,  should be determined using the nonlinear
Navier--Stokes equation because the linear solutions defined by relations
(3.33) and (3.34) contradict the required condition (B4) of linearization of
the Navier--Stokes equation at any point in this domain.

%%-------------------------------References------------------------------


\newpage

\begin{thebibliography}{99}

\bibitem{ref.ChowHermans}
 T.\,S.~Chow and J.\,J.~Hermans, {\it Brownian motion of a spherical particle in a
 compressible fluid,\/} Physica, {\bf 65}, 156--162 (1973).

\bibitem{ref.Chow}
 T.\,S.~Chow, {\it Simultaneous translational and rotational Brownian movement of
 particles of arbitrary shape,\/} Phys. Fluids, {\bf 16}, No.~1, 31--34 (1973).

\bibitem{ref.Bedeaux}
 D.~Bedeaux and P.~Mazur, {\it Brownian motion and fluctuation hydrodynamics,\/}
 Physica, {\bf 76}, No.~2, 247--258 (1974).

\bibitem{ref.Mazur}
 P.~Mazur, {\it On the motion and Brownian motion of n spheres in a viscous
 fluid,\/} Physica, {\bf 110A}, 128--146 (1982).

\bibitem{ref.Clercx1}
 H.\,J.\,H.~Clercx and P.\,P.\,J.\,M.~Schram, {\it Brownian particles in shear
 flow and harmonic potentials: A study of long-time tails,\/} Phys. Rev. A,
 {\bf 46}, No.~4, 1942--1950 (1992).

\bibitem{ref.SchramYak}
 P.\,P.\,J.\,M.~Schram and I.\,P.~Yakimenko, {\it Hydrodynamic theory of Brownian
 motion in the compressible fluid,\/} Prikl. Hidromekh., {\bf 1~(73)}, No.~2, 53--63
 (1999).

\bibitem{ref.ResLeener}
 P.~R\'{e}sibois and M.\,~De~Leener, {\it Classical Kinetic Theory of Fluids\/}
 (Wiley, New York, 1977).

\bibitem{ref.Happel}
 J.~Happel and H.~Brenner, {\it Low Reynolds Number Hydrodynamics\/}
 (Kluwer Academic, Dordrecht, 1991).

\bibitem{ref.Rutgers}
 M.\,A.~Rutgers, J.-Z.~Xue, E.~Herbolzheimer, W.\,B.~Russel, and P.\,M.~Chaikin,
 {\it Crystalline fluidized beds,\/} Phys. Rev. E, {\bf 51}, No.~5, 4674--4678 (1995).

\bibitem{ref.Kim}
 S.~Kim and S.\,J.~Karrila, {\it Microhydrodynamics: Principles and Selected Applications\/}
 (Butterworh-Heinemann, Boston, 1991).

\bibitem{ref.Lamb}
 H.~Lamb, {\it Hydrodynamics\/} (Dover, New York, 1945).

\bibitem{ref.Milne}
 L.\,M.~Milne-Thomson, {\it Theoretical Hydrodynamics\/}
 (Macmillan, London, 1960).

\bibitem{ref.Batch}
 G.\,K.~Batchelor, {\it An Introduction to Fluid Dynamics\/}
 (Cambridge, 1970).

\bibitem{ref.Landau}
 L.\,D.~Landau and E.\,M.~Lifshitz, {\it Fluid Mechanics\/}
 (Pergamon, New York, 1978).

\bibitem{ref.Loitsyan}
 L.\,G.~Loitsyanskii, {\it Mechanics of Fluid and Gas\/}
 (Nauka, Moscow, 1987).

\bibitem{ref.Schram}
 P.\,P.\,J.\,M.~Schram, {\it Kinetic Theory of Gases and Plasmas\/}
 (Kluwer, Dordrecht, 1991).

\bibitem{ref.Stimson}
 M.~Stimson and G.\,B.~Jeffery, Proc. Roy. Soc. London, {\bf A111}, 110 (1926).

\bibitem{ref.Maude}
 A.\,D.~Maude, {\it End effects in a falling-sphere viscometer,\/}
 Brit. J. Apl. Phys., {\bf 12}, No.~6, 293--295 (1961).

\bibitem{ref.Wakiya}
 S.~Wakiya, {\it Slow motions of a viscous fluid around two spheres,\/}
 J. Phys. Soc. Japan, {\bf 22}, No.~4, 1101--1109 (1966).

\bibitem{ref.Davis}
 M.\,H.~Davis, {\it The slow translation and rotation of two unequal spheres in a
 viscous fluid,\/} Chem. Eng. Sci., {\bf 24}, No.~12, 1769--1776 (1969).

\bibitem{ref.Feld1}
 B.\,U.~Felderhof, {\it Force density induced on a sphere in linear
 hydrodynamics.  II.  Moving sphere, mixed boundary conditions,\/}
 Physica, {\bf 84A}, No.~3, 569--576 (1976).

\bibitem{ref.Feld2}
 B.\,U.~Felderhof, {\it Hydrodynamic interaction between two spheres,\/}
 Physica, {\bf 89A}, No.~2, 373--384 (1977).

\bibitem{ref.Schmitz1}
 R.~Schmitz and B.\,U.~Felderhof, {\it Creeping flow about a sphere,\/}
 Physica, {\bf 92A}, No.~3--4, 423--437 (1978).

\bibitem{ref.Schmitz2}
 R.~Schmitz and B.\,U.~Felderhof, {\it Creeping flow about a spherical
 particle,\/} Physica, {\bf 113A}, 90--102 (1982).

\bibitem{ref.Schmitz3}
 R.~Schmitz and B.\,U.~Felderhof, {\it Friction matrix for two spherical
 particles with hydrodynamic interaction,\/} Physica, {\bf 113A},
 103--116 (1982).

\bibitem{ref.Jones1}
 R.\,B.~Jones, {\it Hydrodynamic interaction of two permeable spheres.  I:
 The method of reflections,\/}
 Physica, {\bf 92A}, No.~3--4, 545--556 (1978).

\bibitem{ref.Jones2}
 R.\,B.~Jones, {\it Hydrodynamic interaction of two permeable spheres.  II:
 Velocity field and friction constants,\/}
 Physica, {\bf 92A}, No.~3--4, 557--570 (1978).

\bibitem{ref.Jones3}
 R.\,B.~Jones, {\it Hydrodynamic interaction of two permeable spheres.  III:
 Mobility tensors,\/} Physica, {\bf 92A}, No.~3--4, 571--583 (1978).

\bibitem{ref.MazurSaarl}
 P.~Mazur and W.~van~Saarloos, {\it Many-sphere hydrodynamic interactions and
 mobilities in a suspension,\/} Physica, {\bf 115A}, 21--57 (1982).

\bibitem{ref.MazurBed}
 P.~Mazur and D.~Bedeaux, {\it A generalization of Fax\'{e}n theorem to
 nonsteady motion of a sphere through an incompressible fluid in arbitrary
 flow,\/} Physica, {\bf 76}, No.~2, 235--246 (1974).

\bibitem{ref.Saarl}
 W.~van~Saarloos and P.~Mazur, {\it Many-sphere hydrodynamic interactions.  II.
 Mobilities at finite frequencies,\/}
 Physica, {\bf 120A}, 77--102 (1983).

\bibitem{ref.Freed1}
 K.\,F.~Freed and M.~Muthukumar, {\it On the Stokes problem for a suspension of
 spheres at finite concentrations,\/}
 J. Chem. Phys., {\bf 68}, No. 5, 2088--2096 (1978).

\bibitem{ref.Freed2}
 K.\,F.~Freed and M.~Muthukumar, {\it Dynamics and hydrodynamics of
 translational-rotational Brownian particles at finite concentrations,\/}
 J. Chem. Phys., {\bf 69}, No. 6, 2657--2671 (1978).

\bibitem{ref.Pien1}
 I.~Pie\'{n}kowska, {\it An unsteady Faxen's relation for the force including
 interaction effects,\/} Arch. Mech., {\bf 34}, No.~3, 297--306 (1982).

\bibitem{ref.Pien2}
 I.~Pie\'{n}kowska, {\it Unsteady friction and mobility relations for Stokes flow,\/}
 Arch. Mech., {\bf 36}, No.~5--6, 746--769 (1984).

\bibitem{ref.Clercx2}
 H.\,J.\,H.~Clercx and P.\,P.\,J.\,M.~Schram, {\it Quasistatic hydrodynamic
 interaction in suspensions,\/} Physica A, {\bf 174}, 293--324 (1991).

\bibitem{ref.Clercx3}
 H.\,J.\,H.~Clercx and P.\,P.\,J.\,M.~Schram, {\it Retarded hydrodynamic
 interaction in suspensions,\/} Physica A, {\bf 174}, 325--354 (1991).

\bibitem{ref.Hofman}
 J.\,M.\,A.~Hofman, H.\,J.\,H.~Clercx, and P.\,P.\,J.\,M.~Schram, {\it Effective
 viscosity of dense colloidal crystals,\/} Phys. Rev. E, {\bf 62}, No.~6,
 8212--8333 (2000).

\bibitem{ref.Usenko}
 A.\,S.~Usenko, {\it Velocity and Pressure Fields Induced by Spheres in an
 Unbounded Fluid,\/} Preprint physics/0204045 (2002).

\bibitem{ref.Nikiforov}
 A.\,F.~Nikiforov and V.\,B.~Uvarov, {\it Special Functions of Mathematical
 Physics\/} (Nauka, Moscow, 1978).

\bibitem{ref.Varshalovich}
 D.\,A~Varshalovich, A.\,N.~Moskalev, and V.\,K.~Khersonskii,
 {\it Quantum Theory of Angular Momentum\/} (Nauka, Leningrad, 1975).

\bibitem{ref.Yosh}
 T.~Yoshizaki and H.~Yamakawa, {\it Validity of the superposition approximation
 in an application of the modified Oseen tensor to rigid polymers,\/}
 J. Chem. Phys., {\bf 73},  No.~1, 578--582 (1980).

\bibitem{ref.Abram} M.~Abramowitz and I.\, A.~Stegun (editors), {\it Handbook of
 Mathematical Functions with Formulas, Graphs and Mathematical Tables\/},
 (National Bureau of Standards, Appl. Math., Ser. 55, 1964).

\bibitem{ref.Davydov}
 A.\,S.~Davydov, {\it Quantum Mechanics\/} (Macmillan, London, 1960).

\end{thebibliography}
\end{document}